\documentclass[aps,prd,superscriptaddress,nofootinbib,11pt]{revtex4}
\usepackage[english]{babel}
\usepackage[utf8]{inputenc}
\usepackage{graphicx}   
\usepackage{slashed}
\usepackage{epstopdf}
\usepackage{verbatim}   
\usepackage{color}      
\usepackage{subfigure}  
\usepackage{multirow}
\usepackage{hyperref}   
\usepackage{float}
\usepackage{epsfig,rotating}
\usepackage{amsmath,amssymb}
\usepackage{dsfont}
\restylefloat{table}
\numberwithin{equation}{section}

\newcommand{\vx}{\vec{x}}
\newcommand{\vp}{\vec{p}}

\newcommand{\vk}{\vec{k}}

\newcommand{\Om}{\Omega}

\newcommand{\be}{\begin{equation}}
\newcommand{\ee}{\end{equation}}
\newcommand{\bea}{\begin{eqnarray}}
\newcommand{\eea}{\end{eqnarray}}

\newcommand{\ket}[1]{|#1\rangle}
\newcommand{\bra}[1]{\langle#1|}

\begin{document}

\title{  Misalignment dynamics of  Scalar Condensates with Yukawa coupling: \\ Particle  and Entropy Production.}

\author{Nathan Herring}
\email{nherring@hillsdale.edu} \affiliation{Department of Physics, Hillsdale College, Hillsdale, MI 49242}
  \author{Daniel Boyanovsky}
\email{boyan@pitt.edu} \affiliation{Department of Physics, University of Pittsburgh, Pittsburgh, PA 15260}

\date{\today}

\begin{abstract}
 Misalignment dynamics, the non-equilibrium evolution of a scalar (or pseudoscalar) condensate in a potential landscape, broadly describes a solution to the strong CP problem, a mechanism for cold dark matter production and (pre) reheating post inflation. Often,   radiative corrections are included phenomenologically by replacing the potential by the effective potential which is a \emph{static quantity}, its usefulness is restricted to (near) equilibrium situations. We study the misalignment dynamics of a scalar condensate Yukawa coupled to $N_f$ fermions in a manifestly energy conserving, fully renormalized Hamiltonian framework.  A large $N_f$ limit allows us to focus on the fermion degrees of freedom, which yield a negative contribution to the effective potential, a   radiatively induced instability and ultraviolet divergent field renormalization. We introduce an adiabatic basis and an adiabatic expansion that embodies the derivative expansion in the effective action, the zeroth order is identified with
the effective potential, higher orders   account for the derivative expansion including field renormalization and  describe profuse particle production. Energy conserving dynamics leads to the conjecture of emergent asymptotic highly excited stationary states with a distribution function $n_k(\infty)\propto 1/k^6$ and an extensive entropy which is identified with an entanglement entropy.  Subtle aspects of renormalization associated with the initial value problem are analyzed and resolved. Possible  new manifestations of asymptotic spontaneous symmetry breaking (SSB) as a consequence of the dynamics  even in absence of tree level (SSB), and cosmological inferences are discussed.

\end{abstract}


\maketitle

\section{Introduction}\label{sec:intro}

Misalignment dynamics of a scalar condensate refers to the non-equilibrium evolution of a homogeneous condensate of a scalar (or pseudoscalar) field in a potential landscape from an initial value away from the equilibrium minima of such potential. Such a mechanism was originally invoked as a possible solution of the strong CP problem where the angle describing CP violation in the strong interaction was elevated to a (pseudoscalar) field evolving in a potential towards CP conserving minima\cite{pq1,pq2}.  This misalignment mechanism was recognized  as  a compelling alternative  for dark matter production\cite{preskill,abbott,dine}, whose abundance is determined by the initial value of the scalar (or pseudoscalar) condensate. Fundamentally, the dynamical evolution of the scalar condensate,    is an initial value problem, assigning initial conditions to a homogeneous condensate or mean field and evolving its equation of motion  with a potential. It is   customary phenomenologically to include  radiative corrections    by replacing the tree level potential by an effective potential.

The effective potential was originally introduced  to study  \emph{ equilibrium} aspects of spontaneous symmetry breaking in quantum field theory \cite{jona,goldstone}.  It is defined as the
generating functional of the single particle irreducible Green's functions at \emph{zero four momentum transfer} and describes how radiative corrections modify the symmetry breaking properties of the vacuum\cite{colewein}.  This generating functional can be obtained by summing an infinite series of Feynman diagrams\cite{colewein}, or alternatively by  functional methods\cite{ilio,jackiw,colemanbook,colepoli} providing a systematic and simple derivation in a consistent loop expansion, which has been extended to \emph{equilibrium} finite temperature field theory\cite{dolan,weinberg,branrmp}. In equilibrium at finite temperature   the effective potential describes how quantum and thermal fluctuations modify  the free energy landscape as a function of the order parameter, and   provides a very useful characterization of symmetry breaking and phase transitions that includes radiative corrections.

The use of the effective potential in the equations of motion to study the \emph{dynamical evolution} of a homogeneous condensate has recently been scrutinized in detail both at zero and finite temperature and shown to be inadequate\cite{nathan} as a consequence of   ubiquitous instabilities, such as spinodal (tachyonic) and or parametric,  that lead to profuse particle and \emph{entropy} production. Particle production as a consequence of the time evolution of the condensate is shown\cite{nathan} to be a very efficient mechanism of energy transfer, yielding a highly excited state and possibly novel asymptotic stationary states that are simply not captured by the phenomenological dynamics with an effective potential.

Perhaps not widely recognized, the non-equilibrium misalignment dynamics is also at the heart of post inflationary reheating\cite{rehe1,rehe2,rehe3,branreh,rehe4,rehe5,rehe6,rehe7} in which parametric instabilities resulting from the oscillatory dynamics of the condensate near a minimum of the (effective) potential, provide a mechanism of energy transfer between a ``zero mode'' (condensate) and quantum fluctuations, again leading to profuse particle production and a highly excited state which is conjectured to eventually thermalize, thereby providing a scenario for the transition from inflation to a radiation dominated cosmology.

\textbf{Motivation and objectives:}

The importance of misalignment dynamics in cosmology, either as a mechanism of dark matter production, solution to the CP problem or post-inflationary (pre) reheating,
along with the results of ref.\cite{nathan} that unequivocally show the \emph{inadequacy} of using the effective potential in the equation of motion of a scalar condensate, motivate us to study the case of the non-equilibrium evolution of a scalar condensate with Yukawa coupling to fermions.

 The time dependence of the scalar field expectation value induces time-dependent fermion masses affecting the fermion mode functions which self-consistently feed back on the dynamics of the condensate.

The influence of fermionic degrees of freedom in the dynamics of post inflationary preheating has been studied previously\cite{ferpre1,ferpre2,ferpre3,ferpre4,ferpre5},  with a general conclusion that Pauli blocking renders fermionic preheating less efficient than the bosonic case, yet it is clearly an important ingredient in the understanding of post inflationary cosmology.

Our focus in this study   is quite different, specifically   addressing   the very relevant issue of how radiative corrections from fermionic degrees of freedom self-consistently modify  the effective equations of motion and the dynamical evolution of the condensate, not just the effective potential. As such is more overarching, as it applies generically to any non-equilibrium mechanism that invokes the dynamical evolution of a scalar condensate Yukawa coupled to fermion degrees of freedom. There are at least two immediate aspects of the Yukawa coupling to fermions that are different from the bosonic case and further motivates this study:

\textbf{i:)}  the fermionic contribution to the zero temperature effective potential is \emph{negative} (the negative energy of the ``Dirac sea''), which could be a source of instabilities. Whereas these have been studied\cite{sher1,sher2,sher3} within the context of the effective potential in the Standard Model, with the conclusion that the values of masses and couplings entail that the negative contribution to the effective potential from the fermion sector is overwhelmed by the positive contribution from bosonic degrees of freedom, there is an intrinsic fundamental general interest in the \emph{dynamics} triggered by such instabilities. Furthermore, extensions beyond the Standard Model may feature enough extra fermionic degrees of freedom so as to make these potential instabilities an important dynamical feature of these extensions.

\textbf{ii:)} Field (wave function) renormalization arises at one loop and is ultraviolet logarithmically divergent. This is an important difference with the   bosonic case,   at the level of Feynman diagrams it arises from the $p^2$ contribution to the scalar one loop self energy, hence it multiplies the term $\partial_{\mu} \varphi \partial^\mu \varphi$ in the effective action, and  consequently also the $\partial_\mu \partial^\mu \varphi$ in the equation of motion of the scalar field. Furthermore, such field renormalization which in the fermion case arises at one loop level, is of  the same order as the leading radiative corrections to the effective potential and  enters in the renormalization of all masses and couplings. Therefore, it must be included properly in the
renormalized equations of motion and the effective potential. However, the effective potential being defined in the limit of $p^2 \rightarrow 0$ \emph{does not} include field renormalization which, therefore must be included  \emph{a posteriori} as an  \emph{ad-hoc} renormalization of the condensate, masses and couplings.

The correct equations of motion of the scalar condensate including radiative corrections from all the degrees of freedom to which it couples  must be obtained from the variation of the \emph{effective action}. It is a well established result\cite{colewein,colemanbook,ilio}, that the effective action has a representation in terms of a  \emph{spatio-temporal derivative} expansion of the expectation value of the scalar field, where the \emph{zeroth order} in this expansion is identified with the effective potential, consistently with the latter being defined for zero four momentum transfer. The effective action is renormalized, finite and renormalization group invariant\cite{colewein,colemanbook,ilio}, the derivative terms also receive radiative corrections in the form of a finite multiplicative function of the expectation value of the scalar field that depends on space time through this mean field and is only indirectly related to wave function renormalization. Hence the equations of motion derived from the effective action are generally very different from just replacing the tree level potential by an effective potential in the equation of motion.

 \vspace{1mm}

 \textbf{Objectives:}

Our objectives in this study are the following:

\textbf{i.)} to obtain the  manifestly energy conserving, fully renormalized   equations of motion for a dynamical scalar condensate with Yukawa couplings to $N_f$ fermion species, along with the equations of motion for the fermionic degrees of freedom. We invoke a large number of fermion flavors $N_f$ to systematically separate and unambiguously identify the contributions from fermionic degrees of freedom from the bosonic quantum fluctuations.

 \textbf{ii.)} To establish a clear relation between a derivative expansion of the renormalized effective action and  an adiabatic approximation that operationalize the usual assumption of a wide separation in time-scales between the dynamics of the condensate and those of the quantum fluctuations and study the behavior of the energy density and equations of motion under this condition.

   \textbf{iii.)} To compare and contrast these results with the phenomenologically inspired equations of motion which simply employ the effective potential in the dynamical equation of motion for the condensate.

We restrict ourselves to the study of zero-temperature fields quantized in Minkowski spacetime and focus on the self-consistent one-loop quantum corrections to the energy density and equations of motion and the proper renormalization of fields, couplings and masses. This allows an immediate and direct comparison to the one-loop effective potential, highlighting the salient differences. Restricting our study to zero temperature allows us to clearly separate the particle production contributions to the equations of motion without the complications associated with stimulated production or Pauli blocking   from finite temperature occupations of bosonic and fermionic degrees of freedom. Furthermore, we discuss the emergence of an extensive  entropy recognized as an entanglement entropy as a consequence of particle pair production, without ``contamination'' from thermal entropy.

Extending these methods to condensate dynamics in cosmological spacetimes or including higher order loop corrections and implementing a numerical integration of the equations of motion  represent avenues for future work.

\textbf{Brief summary of results:}

Using the Hamiltonian approach advocated in references\cite{veff,colemanbook,weinbergwu,nathan} we obtain the one loop effective potential in the case of a \emph{static} scalar condensate/mean field with Yukawa couplings to $N_f$ flavors of fermions. The fermion contribution to the effective potential is negative leading to a \emph{radiatively induced maximum} a la Coleman-Weinberg\cite{colewein}, indicating an instability.

 Subsequently, we obtain the energy density and the equation of motion of the expectation value of the scalar field, namely the mean field, by taking the expectation value of the Heisenberg Hamiltonian and the exact Heisenberg  field equations in a (coherent) state in which the scalar field and its conjugate momentum obtain expectations values. This equation of motion, along with the Dirac equations for the fermion fields define the dynamical evolution of an explicitly time-dependent scalar field expectation value. This formulation is manifestly energy conserving. The time dependent mean field implies a time dependent mass for the fermionic degrees of freedom which feed back into the evolution equation of the condensate.

  By considering the large $N_f$ limit, the bosonic contributions are suppressed relative to the fermionic terms allowing for careful and isolated treatment of the latter's impact on the effective action.

  We introduce   an adiabatic basis and adiabatic expansion which allows a clear and natural interpretation of the derivative expansion of the effective action as the  adiabatic expansion, within which we identify the zeroth order term with the usual fermionic contribution to the effective potential in the static case. Meanwhile, the adiabatic mode functions form a basis of ``particle" states. When viewed from this basis, the contributions beyond zeroth adiabatic order are associated with the derivative terms in the effective action. These entail profuse ``particle production" which dynamically encodes the wave function renormalization of the scalar field.

  A most important result is a fully renormalized, energy conserving set of equations of motion for the condensate and the bosonic and fermionic quantum fluctuations, that goes beyond the inclusion of the radiative corrections in the effective potential by also including the production of particles as a consequence of the non-equilibrium evolution of the misaligned condensate and \emph{does not rely on the adiabatic or derivative expansion}. These equations of motion along  with a given set of initial conditions on bosonic and fermionic mode functions defines a well posed initial value problem for misalignment dynamics amenable of direct numerical study.

   Thus the production of adiabatic particles is interpreted as an efficient energy transfer from the ``classical" non-equilibrium homogeneous condensate to the quantum fluctuations as time evolution proceeds. This mechanism invites the conjecturing of an asymptotic stationary state with a concomitant population of highly entangled, produced particle-antiparticle pairs and an associated (entanglement) entropy, both of which we obtain.

  An important and subtle aspect revealed by this study is the connection between the quantum state and initial conditions on fermionic mode functions, an aspect that is directly related to field renormalization and the dynamics of ``dressing'' (renormalization) of the field by quantum fluctuations, which leads to an initial time singularity as a consequence of a ultraviolet divergent field renormalization\cite{baacke,inising1,inising2,inising3,devega}. We show how this ``singularity'' is resolved self-consistently upon field and coupling renormalization, as it must for consistency with a finite energy density,  with  a finite remnant time dependence of ``dressing'' that vanishes on a time scale inversely proportional to the ultraviolet cutoff.

   We contrast our fully consistent energy conserving dynamical treatment with the usual phenomenologically inspired equations of motion wherein the effective potential is treated as the dynamical potential and the only modification of radiative corrections to the condensate/mean field's equation of motion. In such a phenomenological approach, not only must wave function renormalization   be added ``by hand", but we find that energy conservation is violated, and the presence of a possible highly excited asymptotic stationary state of high entropic content is completely missed. We point out the possibility   that the radiatively induced instability combined with profuse particle production may lead to novel asymptotic excited states featuring spontaneous symmetry breaking even when the tree level potential does not feature symmetry breaking minima.

    We argue that the adiabatic particle basis is a privileged, ``pointer'' basis and rapid dephasing of the off diagonal density matrix elements in this basis lead asymptotically to an emergent entropy which is recognized as the entanglement entropy obtained upon tracing over either particle or antiparticle states.

The article is organized as follows: in section (\ref{sec:largeN}) we review the Hamiltonian approach to obtaining the one loop effective potential in the case of a static mean field, showing that when the scalar is coupled to $N_f$ identical fermion fields, the bosonic contributions to the effective potential are suppressed in the large $N_f$ limit. In this section we discuss possible instabilities arising from fermionic contributions, and a Coleman-Weinberg\cite{colewein} radiatively induced \emph{maximum} of the effective potential. In section (\ref{sec:dynamics}) we extend the Hamiltonian framework  to the dynamical case obtaining the equations of motion for the condensate and quantum fluctuations introducing the adiabatic states and adiabatic expansion which underpins the derivative expansion of the effective action.   We obtain the energy density and equations of motion in terms of the exact time-dependent spinors and mode functions which satisfy the Dirac equation with a time-dependent mass. We then perform a Bogoliubov transformation to the adiabatic particle basis which furnishes a natural interpretation of the higher order terms in the derivative expansion   with the spontaneous production of adiabatic particles.  These higher order terms are identified with derivative terms in the effective action, and shown to generate a wave function renormalization in both the energy density and associated equations of motion.  We investigate the dependence of the fermionic vacuum upon initial conditions which lead to initial time singularities in the equation of motion of the condensate as a consequence of the ultraviolet divergent field renormalization, and analyze its self-consistent resolution. Section (\ref{sec:asymptotic}) discusses the possible asymptotic behavior of the system as demonstrated by our fully consistent dynamical treatment: a stationary state characterized by an abundance of adiabatic particles with a distribution function $n_k(\infty) \propto 1/k^6$,  with an associated entanglement entropy generated by decoherence via dephasing. A noteworthy result of this analysis is that particle production \emph{stabilizes} the evolution even for the case when initial conditions lead to evolution down the unstable part of the  effective potential, suggesting the possibility of asymptotic stationary states featuring spontaneous symmetry breaking.     In Section (\ref{sec:fervsbos}) we compare the fermionic contribution to the effective potential with the purely bosonic terms, highlighting how these can also be interpreted via an adiabatic approximation and therefore associated particle production. However, we argue that these terms differ from the fermionic fluctuations in that they generate a ultraviolet finite wave function renormalization and no initial time-singularity is induced by their effect on the equations of motion. Section (\ref{sec:discussion}) discusses important aspects and results from this study, including an interpretation of particle production and thermalization in terms of direct and inverse energy cascades,   and possible cosmological implications. Finally, section (\ref{sec:conclusions}) summarizes our conclusions and poses further questions. Several appendices are devoted to technical aspects, including a discussion of Majorana fermions.

\section{Large $N_f$ and the effective potential:}\label{sec:largeN}
\begin{figure}[ht!]
\begin{center}
\includegraphics[height=3in,width=4in,keepaspectratio=true]{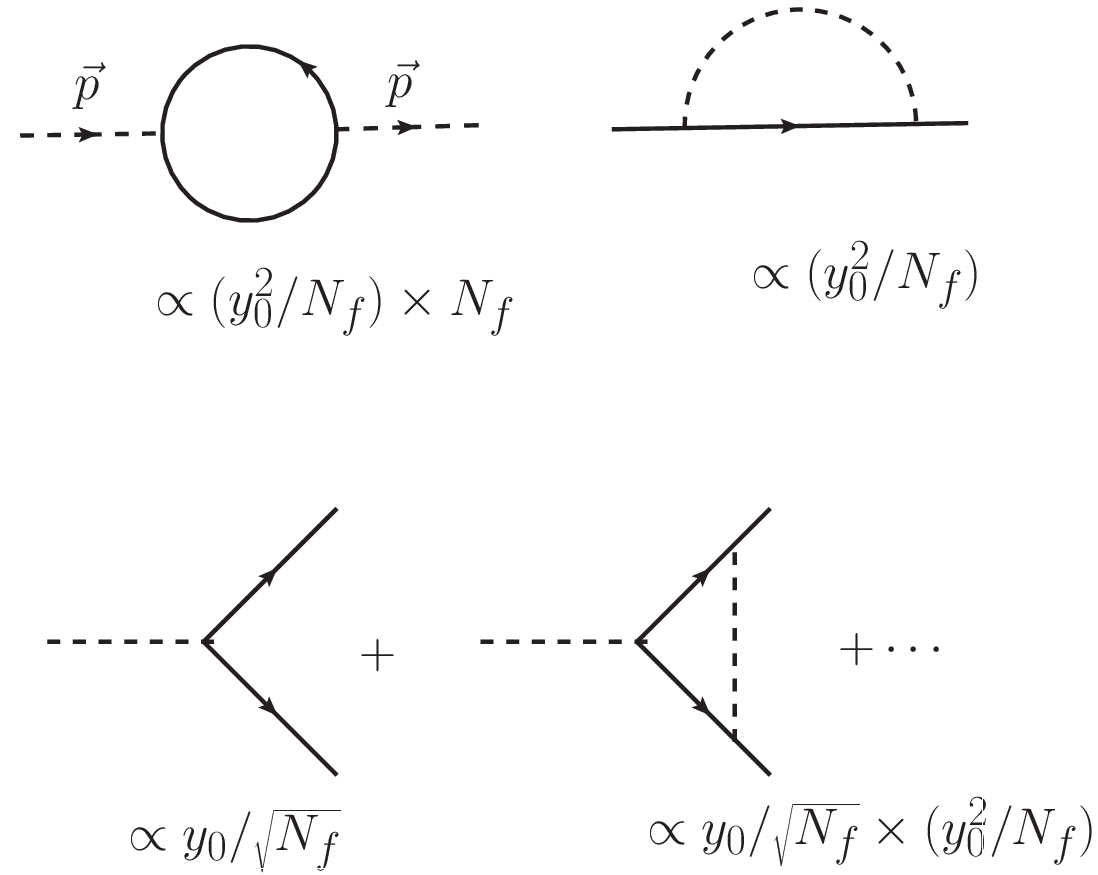}
\caption{The upper diagrams are the scalar and fermion self energies, from which the wave function renormalizations $\propto y^2_0\,\ln(\Lambda); (y^2_0/N_f)\,\ln(\Lambda)$ respectively are obtained and renormalize the Yukawa coupling. The bottom diagrams show vertex renormalization of the Yukawa coupling, which is $\propto (y^2_0/N_f)\,\ln(\Lambda)$. The fermion wave function renormalization and vertex correction are suppressed by a factor $N_f$ with respect to scalar wave-function renormalization. The loops yield the same type   $\ln(\Lambda)$ divergence with the ultraviolet cutoff $\Lambda$ as can be seen by power counting. Solid lines correspond to fermions, dashed lines to bosons. }
\label{fig:vertex}
\end{center}
\end{figure}

We consider a theory of $N_f$ flavors of fermions Yukawa coupled to a bosonic field $\phi$ with the total Lagrangian density
\be \mathcal{L} =  \frac{1}{2} \partial_\mu \phi \partial^\mu \phi- V(\phi)+ \sum_{i=1}^{N_f}  \overline{\psi}_i\Big( i{\not\!{\partial}}- \frac{y_0}{\sqrt{N_f}}\,\phi \Big)\psi_i   \label{totlag}\ee where the Fermi fields $\psi_i$ are Dirac fields with  a common Yukawa coupling $y_0/\sqrt{N_f}$ to the single scalar field $\phi$. The case of Majorana fields is discussed in appendix (\ref{app:majorana}) without any conceptual differences. The $\gamma^\mu$ matrices are chosen in the standard Dirac representation.  The formal large $N_f$ limit $N_f \gg 1$ allows us to consistently isolate and understand the contribution of the fermionic degrees of freedom to the effective potential and the effective dynamics of the expectation value of the scalar field. As in the case of the large N limit in a scalar field theory\cite{colepoli}, the large $N_f$ limit for fermions will allow us to study the dynamics  in a non-perturbative manner. An important bonus of the large $N_f$ limit is that the leading contribution to renormalization of the Yukawa coupling is from wave function renormalization of the scalar field, whereas vertex renormalization is subleading in this limit. Figure (\ref{fig:vertex}) shows that the wave function renormalization for the scalar field  from the scalar self energy is $\propto y^2_0 \times \ln(\Lambda)$ whereas the fermion self-energy and vertex renormalization is $\propto (y^2_0/N_F)\times \ln(\Lambda)$ with $\Lambda$ an ultraviolet cutoff.

Furthermore, because we consider a single scalar Yukawa coupled to the $N_f$ fermions, the scalar contribution to the effective potential and effective action are suppressed by a factor $1/N_f$ with respect to the fermionic contribution, as shown explicitly below.

\begin{figure}[ht!]
\begin{center}
\includegraphics[height=4in,width=5in,keepaspectratio=true]{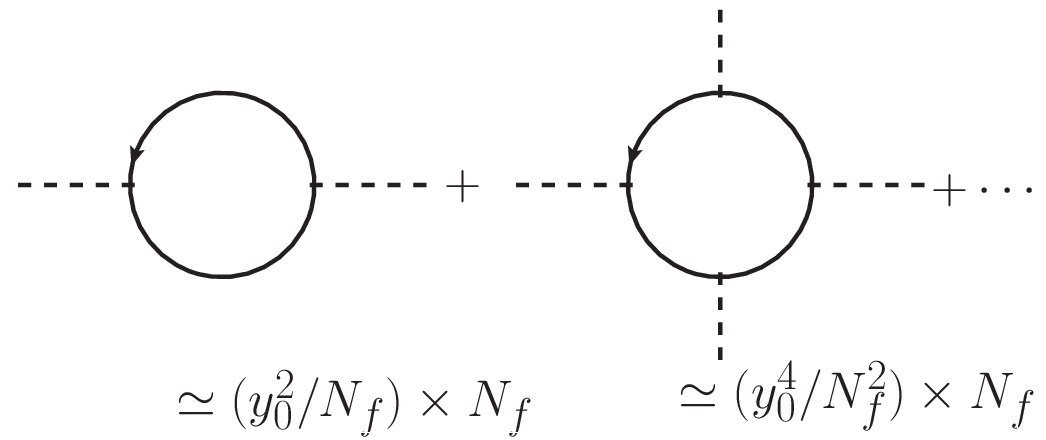}
\caption{Fermion contribution to the one loop effective potential, there are $N_f$ fermions in the loop. The first diagram is a scalar self-energy contribution, the second a quartic self-interaction. Dashed lines correspond to the scalar mean field $\varphi \sqrt{N_f}$ the solid line with arrow is the Fermion propagator. All external lines are at zero four momentum.  }
\label{fig:veffer}
\end{center}
\end{figure}

We consider the tree level potential
\be V(\hat{\phi}) = V_0+ \frac{1}{2}\,m^2_0 \hat{\phi}^2 + \frac{\lambda_0}{4N_f}\,\hat{\phi}^4 \,\label{vtree}\ee where $V_0$ is a constant, and the $1/N_f$ scaling of the quartic coupling is included so that a proper scaling with $N_f$ emerges as discussed below in more detail. This scaling can also be understood from the diagramatic expansion\cite{colewein,jackiw,ilio,colemanbook,branrmp} of the effective action shown in fig. (\ref{fig:veffer}), the contribution with four external scalar ``legs'' induce a quartic coupling proportional to $1/N_f$.

 In the large $N_f$ limit we will consider that $\lambda,y_0 \simeq \mathcal{O}(1)$, thereby allowing us to explore non-perturbative dynamics.

 Field quantization proceeds by introducing the field operator $\hat{\phi}(\vx,t)$ and its canonical conjugate momentum $\hat{\pi}(\vx,t) = \partial \mathcal{L}/\partial \dot{\phi} = \dot{\phi}$ the total Hamiltonian becomes
\be  H = \int d^3x \Bigg\{\frac{\hat{\pi}^2}{2}+\frac{(\nabla \hat{\phi})^2}{2}+V(\hat{\phi})+ \sum_{i=1}^{N_f} \psi^\dagger_i \Big(-i \vec{\alpha}\cdot \vec{\nabla} + \gamma^0\, \frac{y_0}{\sqrt{N_f}}\,\hat{\phi}\Big)\psi_i\Bigg\}\,,\label{totham} \ee and the Heisenberg field equations become
\be \ddot{\hat{\phi}}-\nabla^2\hat{\phi}+V'(\hat{\phi})+\frac{y_0}{\sqrt{N_f}} \sum_{i=1}^{N_f} \overline{\psi}_i \psi_i  = 0 \,,\label{eomfi} \ee
\be i\frac{\partial}{\partial t}\psi_i(\vx,t) = \Big(-i \vec{\alpha}\cdot \vec{\nabla} + \gamma^0\, \frac{y_0}{\sqrt{N_f}}\,\hat{\phi}\Big)\psi_i(\vx,t)\,.\label{diraceqn}\ee

  The Hamiltonian formulation of the effective potential advanced in refs.\cite{veff,weinbergwu,colemanbook,nathan}   identifies the effective potential as the \emph{minimum} of the expectation value of the Hamiltonian operator in a normalized coherent state   in which the field acquires a \emph{space-time independent expectation value},  divided by the spatial volume of quantization $\mathcal{V}$.

   Including fermions, this state is \emph{defined} to be
   \be \ket{\Phi;0_F}  \,, \label{cohstate}\ee describing a  coherent state of the scalar field and   the fermionic vacuum state, which will be clarified in the discussion below. The space-time independent expectation value of the scalar field in this state is
    \be  \bra{\Phi;0_F}\hat{\phi}(\vx,t)\ket{\Phi;0_F} \equiv \varphi\,\sqrt{N_f}~~;~~ \bra{\Phi;0_F}\hat{\pi}(\vx,t)\ket{\Phi;0_F} =0\,,\label{statexvals}\ee   namely
    \be V_{eff}(\varphi) = \frac{1}{\mathcal{V}}\,\bra{\Phi;0_F} {H}\ket{\Phi;0_F}\,.\label{veffstatic}\ee The scaling with $N_f$ of the scalar expectation value is suggested by the equation of motion (\ref{eomfi}) where the $ \sum_{i=1}^{N_f} \bra{\Phi;0_F}\overline{\psi}_i \psi_i \ket{\Phi;0_F}\propto N_f$ and will lead to a natural scaling of the total energy. We refer to $\varphi$ as a \emph{mean field}, and writing
    \be \hat{\phi}(\vx,t) = \varphi\,\sqrt{N_f}+ \hat{\delta}(\vx,t)~~;~~ \hat{\pi}(\vx,t) \equiv \hat{\pi}_\delta(\vx,t)\,, \label{fieldsplit}\ee the constraints (\ref{statexvals}) imply
    \be \bra{\Phi;0_F}\hat{\delta}(\vx,t)\ket{\Phi;0_F}=0 ~~;~~\bra{\Phi;0_F}\hat{\pi}_\delta(\vx,t)\ket{\Phi;0_F} =0\,,\label{statexvalsdel}\ee leading to
    \be   V_{eff}(\varphi)  = \,N_f\,v(\varphi)+ \frac{1}{\mathcal{V}}\, \Big[ \bra{\Phi;0_F}\hat{H}_\delta \ket{\Phi;0_F}+  \bra{\Phi;0_F}\hat{H}_F\ket{\Phi;0_F}\Big]\,,\label{veffdef} \ee where
    \be v(\varphi) = \frac{1}{2}\, m^2_0 \varphi^2 + \frac{\lambda}{4}\,\varphi^4 \,,\label{vzero}\ee and
    \be \hat{H}_\delta =
    \int d^3x \, \Big\{\frac{\hat{\pi}^2_\delta }{2}+\frac{(\nabla \hat{\delta})^2}{2}+\frac{1}{2}\,\mathcal{M}^2(\varphi)\,\hat{\delta}^2 + \cdots \Big\}\,, \label{Hamdelta} \ee
    \be \hat{H}_F =   \int d^3x \,\sum_{i=1}^{N_f} \psi^\dagger_i \Big[ -i \vec{\alpha}\cdot \vec{\nabla} + \gamma^0\, m_F(\varphi) +\cdots \Big] \psi_i \,,\label{hampsi}\ee with
     \be \mathcal{M}^2(\varphi) \equiv v''(\varphi)~~;~~ m_F(\varphi) = y_0 \, \varphi\,, \label{masses} \ee and
    where linear terms in $\hat{\delta}$ and $\hat{\pi}_\delta$ vanish by the constraints (\ref{statexvals}). The dots in (\ref{Hamdelta},\ref{hampsi}) stand for higher order terms in $1/N_f$. Each  term in the  sum in $\hat{H}_F$ in (\ref{hampsi}) yields the same contribution   to the expectation value $ \bra{\Phi;0_F}\hat{H}_F\ket{\Phi;0_F}$ because all the Fermi fields  feature the same effective mass,  therefore the sum yields  a factor $N_f$ in (\ref{veffdef}).

     \emph{Assuming} that the effective squared mass $\mathcal{M}^2(\varphi) \geq 0$\footnote{The case of spinodal (tachyonic) instabilities is discussed in detail in ref.\cite{nathan}.},  up to quadratic order the Hamiltonian in eqn. (\ref{Hamdelta}) describes a free massive field and (\ref{hampsi}) a free fermion field. Hence, we quantize as usual:
    \bea \hat{\delta}(\vx,t)  & = &  \sqrt{\frac{\hbar}{\mathcal{V}}}\,\sum_{\vk} \frac{1}{\sqrt{2\omega_k}}\,\Big[a_{\vk}\, e^{-i\omega_k t}\,e^{i\vk\cdot \vx} + a^\dagger_{\vk}\, e^{i\omega_k t}\,e^{-i\vk\cdot \vx}\Big] \,,\label{quandelta}\\ \hat{\pi}_\delta(\vx,t)  & = &  -i\sqrt{\frac{\hbar}{\mathcal{V}}}\,\sum_{\vk} \frac{\sqrt{\omega_k}}{\sqrt{2}}\,\Big[a_{\vk}\, e^{-i\omega_k t}\,e^{i\vk\cdot \vx} - a^\dagger_{\vk}\, e^{i\omega_k t}\,e^{-i\vk\cdot \vx}\Big] \,,\label{quanpidelta}\eea
    \be \psi_i(\vx,t) =  \sqrt{\frac{\hbar}{\mathcal{V}}}\,\sum_{\vk} \sum_{s=\pm 1} \Big[ {b}_{\vk,s,i} U_{\vk,s}\,  e^{-iE_k t} +  {d}^\dagger_{-\vk,s,i} V_{-\vk,s}\,e^{iE_k t}\Big]\,e^{i\vk\cdot \vx} \,,\label{psiex}\ee
     with
    \be \omega_k(\varphi) = \sqrt{k^2+\mathcal{M}^2(\varphi)} ~~;~~ E_k = \sqrt{k^2+{m}^2_F(\varphi)}\,. \label{omegak}\ee

    where $ a_{\vk},a^\dagger_{\vk}$,  $\hat{b},\hat{d}$ obey   canonical commutation and  anticommutation relations respectively,
     \be \big\{{b}_{\vk,s,i},{b}^\dagger_{\vk',s',j}\big\} = \big\{{d}_{\vk,s,i},{d}^\dagger_{\vk',s',j}\big\}= \delta_{\vk,\vk'}\delta_{s,s'}\delta_{ij} \,,\label{anticom}\ee with all other anticommutators vanishing,
     and the $U,V$ spinors are the same for all flavors because all feature the same mass $m_F(\varphi)$,  and are solutions of the Dirac equations
    \bea \Big[\vec{\alpha}\cdot\vk + \gamma^0 m_F(\varphi)\Big]U_{\vk,s} & = & E_k U_{\vk,s} \,\label{Ude} \\
    \Big[\vec{\alpha}\cdot\vk + \gamma^0 m_F(\varphi)\Big]V_{-\vk,s} & = & -E_k V_{-\vk,s} \,\label{Vde} \eea and fulfill the orthonormality relations
    \be U^\dagger_{\vk,s} U_{\vk,s'} = V^\dagger_{\vk,s} V_{\vk,s'} = \delta_{s,s'}~~;~~ U^\dagger_{\vk,s}V_{-\vk,s'} =0 \,.\label{ortofer}\ee  The state $\ket{\Phi;0_F}$ is identified with the fermionic vacuum associated with the mass $m_F = y_0\varphi$, namely
    \be \hat{b}_{\vk,s} \ket{\Phi;0_F} =0 ~;~ \hat{d}_{\vk,s} \ket{\Phi;0_F} =0 ~~~ \forall \vk,s \,.\label{fervac}\ee

    The constraints (\ref{statexvalsdel}) are implemented by requesting that
    \be a_{\vk}\ket{\Phi;0_F} =0 ~,~ \forall \vk \,, \label{anni}\ee in other words, the coherente state $\ket{\Phi;0_F}$ is the \emph{vacuum state} for the fluctuations $\hat{\delta}$ and the fermion field. We obtain
    \be V_{eff}(\varphi) = N_f \,v(\varphi)+ \frac{\hbar}{2} \int \omega_k(\varphi)\, \frac{d^3k}{(2\pi)^3} - 2 \hbar N_f\,\int E_k(\varphi)\, \frac{d^3k}{(2\pi)^3} \,.\label{finVeff}\ee the last two contributions are recognized as the zero point energies of the scalar and Fermi fields. This result clearly shows that the scalar contribution to the effective potential is subdominant in the large $N_f$ limit, therefore
    \be V_{eff}(\varphi) \equiv  N_f\,v_{eff}(\varphi)+ \mathcal{O}(1/N_f)\,,\label{boldV} \ee where setting $\hbar=1$
    \be v_{eff}(\varphi) = v(\varphi) - 2    \,\int E_k(\varphi)\, \frac{d^3k}{(2\pi)^3}  \,. \label{finbVeff}\ee

    In principle, the constraints (\ref{statexvalsdel}) are also fulfilled if  $\ket{\Phi;0_F}$ is an eigenstate of the number operator $a^\dagger_{\vk} a_{\vk}$ with eigenvalues $n_k$, however the energy is lowest for the vacuum state with $n_k=0$, it is also clear that excitations of the Fermi fields also increase the energy, therefore
    the fermi vacuum state is the lowest energy state.

Introducing an upper momentum cutoff $\Lambda$, we find in the limit $\Lambda \rightarrow \infty$
\be   v_{eff}(\varphi) = v(\varphi) - \hbar \Bigg[  \frac{\Lambda^4}{4\pi^2} + y^2_0\,\varphi^2\,\frac{\Lambda^2}{4\pi^2}-\frac{y^4_0\,\varphi^4}{16\,\pi^2}\, \ln\Big( \frac{4\Lambda^2}{\mu^2}\Big) +\frac{y^4_0\,\varphi^4}{16\,\pi^2}\,\Big[\ln\Big( \frac{y^2_0\,\varphi^2}{\mu^2}\Big)+\frac{1}{2}\Big]\Bigg]\,,  \label{vferef}\ee where we have introduced a renormalization scale $\mu$ to render dimensionless the arguments of the logarithms. The ultraviolet divergent terms are absorbed into a renormalization of the bare parameters of $v(\varphi)$, subtracting the $\varphi$ independent term $\propto \Lambda^4$, and setting $\hbar=1$
\bea m^2_1 & = &  m^2_0 -   \frac{y^2_0\,\Lambda^2}{2\pi^2}\,\label{massren} \\
\lambda_1(\mu)  & = &  \lambda_0 +  \frac{y^4_0}{4\pi^2} \ln\Bigg(\frac{4\Lambda^2}{\mu^2}\Bigg)\,, \label{lamren}\eea   yielding the effective potential in the large $N_f$ limit $V_{eff}= N_f \, v_{eff}(\varphi)$ with
\be v_{eff}(\varphi) =  \Bigg\{ \frac{1}{2}\,m^2_1 \, \varphi^2 + \frac{\lambda_1(\mu)}{4}\,\varphi^4 - \frac{y^4_0\,\varphi^4}{16\,\pi^2}\,\Big[\ln\Big( \frac{y^2_0\,\varphi^2}{\mu^2}\Big)+\frac{1}{2}\Big]\Bigg\}\,.  \label{Veffren}\ee

The renormalization conditions (\ref{massren},\ref{lamren}) do not yet define the fully renormalized mass and couplings, because to leading order in the large $N_f$, the scalar self-energy depicted by the first diagram in figure (\ref{fig:vertex}) yields a logarithmic divergent wave function renormalization. This renormalization is \emph{not} present in the effective potential, because $V_{eff}$ is obtained for \emph{space-time constant expectation values}, whereas wave function renormalization  is extracted from the coefficient of the $p^2$ term in the scalar propagator including the self-energy from the fermion loop. As we will find in the next sections, wave function renormalization \emph{emerges naturally} when considering the dynamics. Anticipating the emergence of wave function renormalization $Z_{\phi}$ we introduce the renormalized field, mass and coupling as
\be \varphi_R = \frac{\varphi}{\sqrt{Z_\phi}}~~;~~ m^2_R = Z_{\phi}\,m^2_1 ~~;~~ \lambda_R = Z^2_{\phi} \lambda_1~~;~~ y_R = \sqrt{Z_{\phi}}\,y_0 \,,\label{rentot}\ee in terms of which the renormalized $v_{eff}(\varphi_R)$ is given by

\be v_{eff}(\varphi_R) =   \Bigg\{ \frac{1}{2}\,m^2_R \, \varphi^2_R + \frac{\lambda_R(\mu)}{4}\,\varphi^4_R - \frac{y^4_R\,\varphi^4_R}{16\,\pi^2}\,\Big[\ln\Big( \frac{y^2_R\,\varphi^2_R}{\mu^2}\Big)+\frac{1}{2}\Big]\Bigg\}\,,  \label{Veffrentot}\ee note that the product $y_R  \,\varphi_R = y_0\,\varphi$ is invariant under multiplicative renormalization. A simple calculation of the one loop scalar self energy shown in the first diagram of fig. (\ref{fig:vertex})    with an ultraviolet cutoff $\Lambda$ and a renormalization scale $\mu$ yields
\be Z^{-1}_\phi = 1+ \frac{y^2_0}{4\pi^2}\, \ln\Big( \frac{2\Lambda}{\mu}\Big)\,,  \label{Z}\ee a result that will be confirmed by the consistent treatment of the dynamical evolution.

 We emphasize this point: within the framework of the effective potential, wave function renormalization must be added \emph{ad-hoc} because the effective potential corresponds to zero four momentum transfer, whereas $Z_\phi$ is obtained from the coefficient of $p^2$ in the self-energy and is of the same order as the radiative correction to the effective potential.
 As discussed in detail in the next sections, wave function renormalization emerges \emph{naturally} from the dynamics of the mean field and will be shown to agree with the Feynman calculus result (\ref{Z}).

\begin{figure}[ht!]
\begin{center}
\includegraphics[height=3.5in,width=3.5in,keepaspectratio=true]{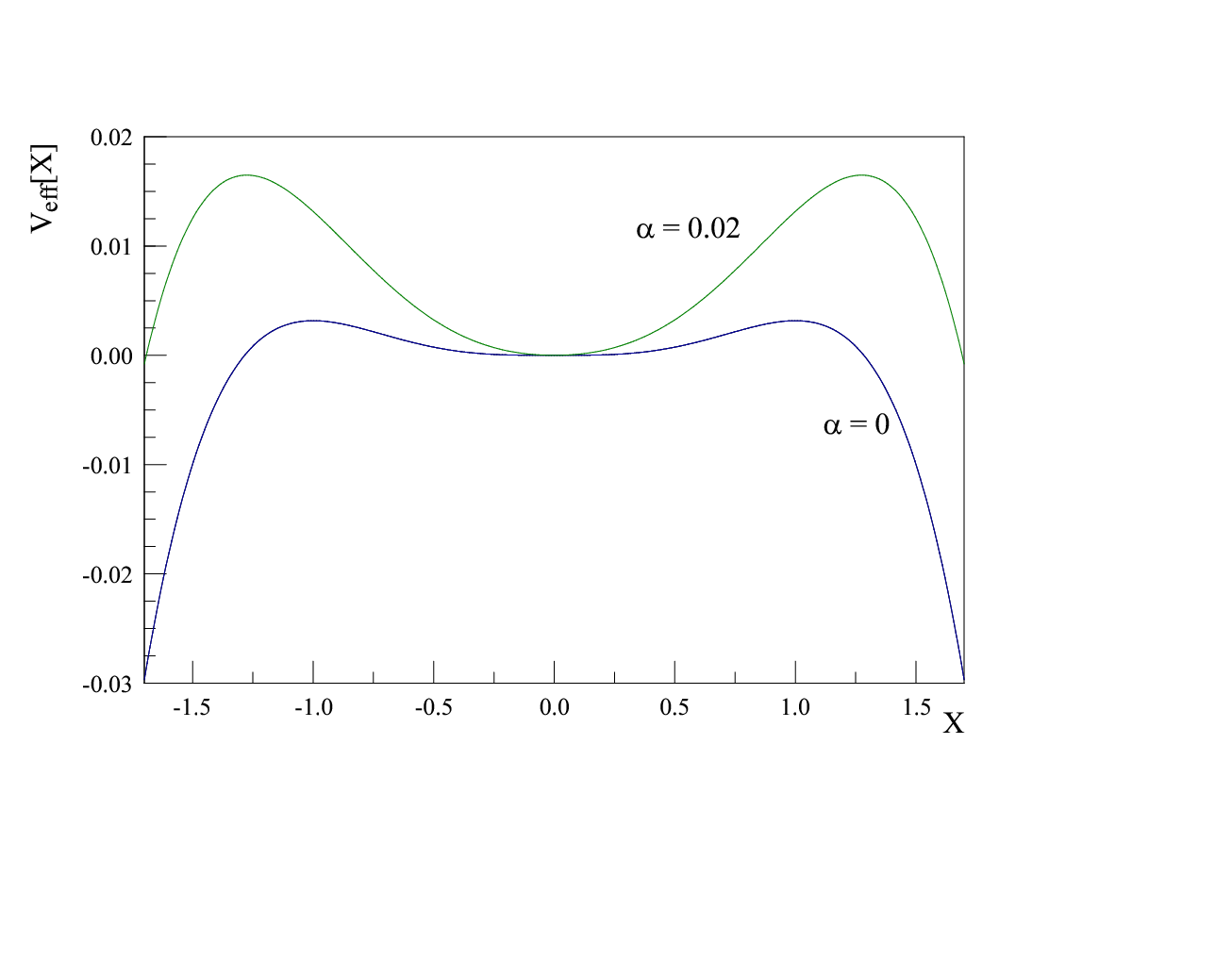}
\caption{$\frac{V_{eff}[X] }{N_f \overline{m}^{\,4}}$ vs. $X=m_f/\overline{m}$ for $\alpha =0,0.02$ }
\label{fig:veffofx}
\end{center}
\end{figure}

 The extrema of $v_{eff}$, namely $v'_{eff}(\varphi_R)=0$ correspond to the roots of the equation
 \be m^2_R \varphi_R+\frac{ y^4_R}{4\pi^2}\,\varphi^3_R\,\Bigg[4\pi^2\,\frac{\lambda_R}{y^4_R} -\ln\Big( \frac{m^2_F(\varphi_R)\,e}{\mu^2}\Big)  \Bigg]=0 ~~;~~ m^2_F(\varphi_R) = y^2_R\varphi^2_R \,,\label{rutsveff}\ee it is convenient to introduce a scale $\overline{m}$ by defining
 \be 4\pi^2\,\frac{\lambda_R}{y^4_R} \equiv \ln\Big( \frac{\overline{m}^{\,2}\,e}{\mu^2}\Big) \,,\label{mbar}\ee so that the extrema of the effective potential are the
 solutions of the equation
  \be m^2_R - \frac{y^4_R}{4\pi^2}\,\varphi^2_R \, \ln\Big( \frac{m^2_F(\varphi_R)}{\overline{m}^{\,2}}\Big) =0 \,.\label{extmbar}\ee The scale $\overline{m}$ has a clear interpretation: $m_F(\varphi_R) = \pm \, \overline{m}$   are the position of the extrema of $V_{eff}$ when $m^2_R=0$\cite{devega}, this is an example of dimensional transmutation\cite{colewein,colemanbook}: when $m_R=0$ the theory is classically scale invariant, however renormalization introduces a scale $\mu$, which can be traded by the  scale in the effective potential corresponding to a \emph{radiatively induced }extremum\cite{colewein,colemanbook}. In terms of this scale, and to leading order in $N_f$ the effective potential is given by
   \be V_{eff}(\varphi_R) = N_f\, \overline{m}^{\,4}\,\Bigg\{ \frac{\alpha}{2}\, \Big(\frac{m_F}{\overline{m}}\Big)^2 - \frac{1}{16\pi^2} \Big( \frac{m_F}{\overline{m}}\Big)^4 \,\Bigg[\ln\Big(\frac{m^2_F}{\overline{m}^{\,2}}\Big)-\frac{1}{2}\Bigg]\Bigg\}~~;~~ \alpha= \frac{m^2_R}{y^2_R \overline{m}^{\,2}}\,. \label{Veffdimless}\ee The effective potential is positive for $m^2_F < \overline{m}^{\,2}$  whereas it is negative for $m^2_F \gg \overline{m}^{\,2}$, therefore there is a minimum at $m_F =0$, namely $\varphi_R =0$, and  there are two \emph{maxima}, which for $\alpha=0$ are at
    $m_F = \pm |\overline{m}|$, namely $\varphi_R =  |\overline{m}|/ y_R$,  reflecting an instability for large $m_F$. As $\alpha$ increases the position of the maxima increases in absolute value. Figure (\ref{fig:veffofx}) displays $\frac{V_{eff}[X] }{N_f \overline{m}^{\,4}}$ vs. $X=m_f/\overline{m}$ for $\alpha =0;0.02$ clearly showing the maxima,

The scalar contribution to the effective potential, the second term in equation (\ref{finVeff})  is manifestly positive for $\mathcal{M}^2(\varphi)>0$ (see discussion in ref.\cite{nathan} for the case $\mathcal{M}^2(\varphi)<0$), and of the same form as the term in bracket in equation (\ref{vferef}) but with $y^2_0\varphi^2$ replaced by $\mathcal{M}^2(\varphi)$ given by eqn. (\ref{masses}).

Negative radiative corrections to the effective potential from large Yukawa couplings leading to potential vacuum instabilities had been previously studied\cite{sher1,sher2,sher3}, with the conclusion that within the standard model, for the values of masses,  the strength of the Yukawa couplings are not large enough to destabilize the vacuum\cite{sher1}.  However instabilities from Yukawa couplings remain a possibility in extensions beyond the standard model. The main objective of the discussion above is \emph{not} the possible vacuum instability, but to highlight the main aspects of the fermionic contribution. The large $N_f$ limit allows us to consider these systematically neglecting the contributions of the scalar fields, which will be considered in a later section.

\section{Dynamics}\label{sec:dynamics}

 The fully renormalized effective action for a background (mean)  field is in general written as a \emph{derivative expansion}\cite{colewein,colemanbook,ilio,branrmp}
 \be S_{eff}[\varphi] = \int d^4 x \Big[ - V_{eff}(\varphi)+  \mathcal{Z}[\varphi]\,\partial_\mu \varphi \partial^\mu \varphi + \cdots \Big] \,,\label{effac}\ee where the dots stand for higher space-time derivatives of $\varphi$ and  $\mathcal{Z}[\varphi]$ is   a \emph{finite} function of the background field $\varphi(\vx,t)$, it is only \emph{indirectly} related to the (divergent) wave-function renormalization. An \emph{adiabatic} expansion in the space-time derivatives of the background (mean) field   is   implicit in the derivative expansion in the  definition (\ref{effac}). In other words, the effective action defined by  equation (\ref{effac}) is valid for slowly varying background fields as a derivative expansion. In the case of spatially constant (homogeneous) mean fields, it is an expansion in the time derivatives of the field, namely an adiabatic expansion.

 The effective equations of motion are obtained from the variational derivatives of the effective action, and in principle feature higher order derivatives corresponding to the dots in (\ref{effac}). Typically in phenomenological treatments of the dynamics of misaligned scalar fields the equation of motion for spatially homogeneous but time dependent scalar condensates is simplified to
 \be \ddot{\varphi}(t) + V'_{eff}(\varphi(t)) =0 \,,\label{pheneom}\ee which would be obtained from the variation of the effective action (\ref{effac}) by ignoring both $\mathcal{Z}[\varphi]$ \emph{and} higher derivative terms which are implicit in the expansion (\ref{effac}). Since there is no explicit time dependence in the effective potential, the phenomenological equation of motion (\ref{pheneom}) features a conserved quantity:
 \be \varepsilon = \frac{1}{2} \,\dot{\varphi}^{\,2}(t) + V_{eff}(\varphi(t)) = \mathrm{constant}\,.\label{phenE}\ee

 An important question is if $\varepsilon$ is the correct energy density, namely the expectation value of the total Hamiltonian (divided by the volume) when the mean field is dynamical. As we will discuss in detail in the following analysis, the answer is negative. The quantity $\varepsilon$ does not have a physical meaning as the energy density, it is simply conserved under the \emph{assumption} of the validity of the equation of motion (\ref{pheneom}).

 Adopting uncritically equations (\ref{pheneom},\ref{phenE}) for the
 case of fermions, with the renormalized effective potential (\ref{Veffrentot}) and with  an initial condition for $\varphi_R$ beyond the maximum of the effective potential would lead to runaway solutions as a consequence of the instability, with an ever increasing ``kinetic energy'' $\dot{\varphi}^2/2$ in (\ref{phenE}) because eventually the effective potential becomes large and negative. Large derivatives suggest large radiative corrections to the gradient terms in the effective action which are not accounted for in the phenomenological equation of motion (\ref{pheneom}).

 A critical analysis\cite{nathan} of the validity of the effective potential in the dynamical evolution of condensates at zero and finite temperature with only scalar fields revealed that ubiquitous mechanisms of particle production such as spinodal and parametric instabilities render the simple phenomenological equation (\ref{pheneom}) invalid. These processes, which modify substantially the dynamical evolution of the condensate, are \emph{not} included in the effective action (\ref{effac}) as they entail a breakdown of adiabaticity and cannot be reliably described by keeping the lowest order term in a derivative expansion.

 We begin by obtaining the expectation values of the conserved energy density and equations of motion in the coherent state $\ket{\Phi;0_F}$   allowing for dynamical evolution of the expectation value (mean field), and its canonical momentum
 \be \bra{\Phi;0_F}\hat{\phi}(\vx,t)\ket{\Phi;0_F} = \sqrt{N_f}\,\varphi(t)~~;~~ \bra{\Phi;0_F}\hat{\pi}(\vx,t)\ket{\Phi;0_F} = \sqrt{N_f}\,\dot{\varphi}(t)\,,\label{exvalsfipi}\ee
and writing
\be \hat{\phi}(\vx,t) \equiv \sqrt{N_f}\,\varphi(t)+ \hat{\delta}(\vx,t)~~;~~ \hat{\pi}(\vx,t) = \sqrt{N_f}\,\dot{\varphi}(t)+ \hat{\pi}_{\delta}(\vx,t) \,\label{shifts}\ee
with
\be \hat{\delta}(\vx,t) \simeq \mathcal{O}(1)~~;~~    \hat{\pi}_{\delta}(\vx,t)\simeq \mathcal{O}(1) \,,\label{ordone}\ee in the large $N_f$ limit,  and the constraints
\be \bra{\Phi;0_F}\hat{\delta}(\vx,t)\ket{\Phi;0_F} =0 ~~;~~ \bra{\Phi;0_F}\hat{\pi}_\delta(\vx,t)\ket{\Phi;0_F} =0 \,. \label{constraints}\ee With the shifts (\ref{shifts}), the Heisenberg equations of motion (\ref{eomfi},\ref{diraceqn}) become
\be \sqrt{N_f} \Big[\ddot{\varphi}(t)+ v'(\varphi(t))+y_0\, \overline{\psi}\psi \Big] +\ddot{\hat{\delta}}(\vx,t)-\vec{\nabla}^2\,\hat{\delta}(\vx,t) +  \mathcal{M}^2(\varphi(t))\,\hat{\delta}(\vx,t)+  \frac{v^{'''}(\varphi(t))}{2\sqrt{N_f}} \,\hat{\delta}^2(\vx,t) + \cdots = 0\,, \label{eomshiftfi}\ee  where
\be  \overline{\psi}\psi  \equiv \frac{1}{N_f} \sum_{i=1}^{N_f} \overline{\psi}_i(\vx,t)\psi_i(\vx,t) \,,\label{chicon}\ee and
\be i\frac{\partial}{\partial t}\psi_i(\vx,t) = \Big(-i \vec{\alpha}\cdot \vec{\nabla} + \gamma^0\, m_F(\varphi(t))\, \Big)\psi_i(\vx,t)+ \cdots\,.\label{diraceqn2}\ee
with $ v(\varphi);\mathcal{M}^2(\varphi),m_F(\varphi)$   given by eqns. (\ref{vzero},\ref{masses}) respectively with $\varphi(t)$, and  the dots stand for terms $\propto 1/\sqrt{N_f}; 1/N_f \cdots$ that are subleading in the large $N_f$ limit.

\subsection{Quantization}\label{subsec:quant}

Quantization proceeds by expanding the quantum fields in the solutions of the linearized equations of motion
\be \ddot{\hat{\delta}}(\vx,t)-\vec{\nabla}^2\,\hat{\delta}(\vx,t) +  \mathcal{M}^2(\varphi(t))\,\hat{\delta}(\vx,t) =0 \,,\label{quanbos}\ee
\be  i\frac{\partial}{\partial t}\psi_i(\vx,t) = \Big(-i \vec{\alpha}\cdot \vec{\nabla} + \gamma^0\, m_F(\varphi(t))\, \Big)\psi_i(\vx,t)\,.\label{quanfer} \ee

The field operators  $\hat{\delta}(\vx,t)~;~\hat{\pi}_{\delta}(\vx,t)$  are expanded in Fourier modes in the   quantization volume  $\mathcal{V}$,
   \bea \hat{\delta}(\vx,t)  & =  & \frac{\sqrt{\hbar}}{\sqrt{\mathcal{V}}}\,\sum_{\vk} \Big[ a_{\vk}\,g_k(t)\,e^{i\vk\cdot\vx} + a^\dagger_{\vk}\,g^*_k(t)\,e^{-i\vk\cdot\vx} \Big]\,,\label{expadelta}\\ \hat{\pi}_\delta(\vx,t) & = & \frac{\sqrt{\hbar}}{\sqrt{\mathcal{V}}}\,\sum_{\vk} \Big[ a_{\vk}\,\dot{g}_k(t)\,e^{i\vk\cdot\vx} + a^\dagger_{\vk}\,\dot{g}^*_k(t)\,e^{-i\vk\cdot\vx} \Big]\,,\label{expapidelta}
   \eea
   and the mode functions, $g_k(t)$, obey the equation of motion
    \be \ddot{g}_k(t) +\omega^2_k(t) g_k(t) = 0\;\; ; \; \omega^2_k(t) \equiv \big[k^2+\mathcal{M}^2(\varphi(t))\big]  \label{ModeTimeEvo}\;, \ee
with the Wronskian condition
    \be \dot{g}_k(t) g^{*}_k(t) -g_k(t)\dot{g}^{*}_k(t)=-i\,. \label{wronsk} \ee The annihilation and creation operators $a_{\vk},a^\dagger_{\vk}$  are time independent because the mode functions $g_k(t)$  are solutions of the mode equations (\ref{ModeTimeEvo}), thereby the fluctuation field $\hat{\delta}(\vx,t)$ is a solution of the linearized Heisenberg field equation (\ref{quanbos}). These operators  obey standard canonical commutation relations as a consequence of the Wronskian  (\ref{wronsk}),  and the condition
    \be a_{\vk}\ket{\Phi;0_F} =0   \,,\label{coherentcond}\ee  ensures the fulfillment of the constraint (\ref{constraints}). Whereas $\ket{\Phi;0_F}$ could, in principle, be an eigenstate of the number operator $a^\dagger_{\vk} a_{\vk}$, states with non-vanishing occupation numbers increase the energy.

    The fermion operators are expanded as in eqn. (\ref{psiex}) but now with spinors that depend on time,
    \be \psi_i(\vx,t) =  \sqrt{\frac{\hbar}{\mathcal{V}}}\,\sum_{\vk} \sum_{s=\pm 1} \Big[ {b}_{\vk,s,i} U_{\vk,s}(t)  +  {d}^\dagger_{-\vk,s,i} V_{-\vk,s}(t) \Big]\,e^{i\vk\cdot \vx} \,,\label{psiext}\ee where the operators $b,d$ are time independent while the spinors obey the time dependent Dirac equations (setting $\hbar=1$)
    \be i\frac{d}{d t} U_{\vk,s}(t) = \Big[\vec{\alpha}\cdot\vk + \gamma^0 m_F(\varphi(t))\Big]\,  U_{\vk,s}(t)~~;~~   -i\frac{d}{d t} U^\dagger_{\vk,s}(t) =   U^\dagger_{\vk,s}(t)\, \Big[\vec{\alpha}\cdot\vk + \gamma^0 m_F(\varphi(t))\Big] \,,\label{DEU}\ee
\be i\frac{d}{d t} V_{-\vk,s}(t) = \Big[\vec{\alpha}\cdot\vk + \gamma^0 m_F(\varphi(t))\Big]\,  V_{-\vk,s}(t)~~;~~   -i\frac{d}{d t} V^\dagger_{-\vk,s}(t) =   V^\dagger_{-\vk,s}(t) \,\Big[\vec{\alpha}\cdot\vk + \gamma^0 m_F(\varphi(t))\Big]  \,,\label{DEV}\ee and the orthonormality relations
\be U^\dagger_{\vk,s}(t) U_{\vk,s'}(t) = \delta_{s,s'}~~;~~V^\dagger_{\vk,s}(t)V_{\vk,s'}(t) = \delta_{s,s'}~~;~~ U^\dagger_{\vk,s}(t)V_{-\vk,s'}(t) =0 \,,\label{ortot}\ee
along with  the conditions
\be b_{\vk,s,i}\ket{\Phi;0_F} =0~;~ d_{\vk,s,i}\ket{\Phi;0_F} =0, \forall \vk,s,i \,.\label{bdanni}\ee
The Dirac equations   (\ref{DEU},\ref{DEV}) lead to the constancy of the orthonormality relations, namely
\be \frac{d}{dt} \Big( U^\dagger_{\vk,s}(t) U_{\vk,s}(t)\Big) = 0 ~;~  \frac{d}{dt} \Big( V^\dagger_{\vk,s}(t)V_{\vk,s}(t) \Big) =0 ~;~ \frac{d}{dt} \Big(U^\dagger_{\vk,s}(t)V_{-\vk,s'}(t) \Big) =0 \,.\label{constortho}\ee

The expectation value of the Heisenberg equation of motion (\ref{eomshiftfi}) in the state $\ket{\Phi;0_F}$ is given by
\be   \sqrt{N_f} \Bigg[\ddot{\varphi}(t)+ v'(\varphi(t))+y_0\,\frac{1}{\mathcal{V}}\sum_{\vk,s}V^\dagger_{-\vk,s}(t)\gamma^0 V_{-\vk,s}(t) + \frac{v^{'''}(\varphi(t))}{2\,N_f}\,\frac{1}{\mathcal{V}} \sum_{\vk} |g_k(t)|^2 + \cdots \Bigg]   = 0 \,. \label{exeom} \ee where the dots stand for terms of higher order in $1/N_f$, and we used the constraints (\ref{constraints}), the quantization (\ref{expadelta},\ref{expapidelta}) and the conditions (\ref{coherentcond}).

We can now obtain the energy density
\be \mathcal{E}  = \frac{1}{\mathcal{V}}\,\bra{\Phi;0_F}H\ket{\Phi;0_F} \,,\label{enerdens}\ee with the total Hamiltonian given by (\ref{totham}), since the total Hamiltonian is time independent and the Heisenberg picture state $\ket{\Phi;0_F}$ is time independent, the energy density is a constant. Implementing the shift (\ref{shifts}) along with the quantization described above we find
\be \mathcal{E} = N_f\Bigg\{\frac{\dot{\varphi}^2}{2} + v(\varphi)+  \frac{1}{\mathcal{V}}\sum_{\vk,s}V^\dagger_{-\vk,s}(t)\Big[\vec{\alpha}\cdot\vk + \gamma^0 m_F(\varphi(t))\Big] V_{-\vk,s}(t)+ \frac{1}{2N_f\,\mathcal{V}} \sum_{\vk}\Big[ |\dot{g}_k(t)|^2+\omega^2_k(t)|g_k(t)|^2\Big]  \Bigg\}\,.\label{enerdensf} \ee Using the equations (\ref{ModeTimeEvo},\ref{DEU},\ref{DEV}) and the definitions (\ref{masses}) we find for the time derivative of the energy density
\be \dot{\mathcal{E}} = \Big(\sqrt{N_f}\,\dot{\varphi}\Big)\, \sqrt{N_f} \Bigg[\ddot{\varphi}(t)+ v'(\varphi(t))+y_0\,\frac{1}{\mathcal{V}}\sum_{\vk,s}V^\dagger_{-\vk,s}(t)\gamma^0 V_{-\vk,s}(t) + \frac{v^{'''}(\varphi(t))}{2\,N_f}\,\frac{1}{\mathcal{V}} \sum_{\vk} |g_k(t)|^2 \Bigg]  \,,\label{Edot}\ee therefore, the equation of motion (\ref{exeom}) along with the mode equations (\ref{ModeTimeEvo},\ref{DEU},\ref{DEV}) imply that
\be \dot{\mathcal{E}} = 0 \,.\label{Econst}\ee In other words, the energy density $\mathcal{E}$ \emph{is conserved} when the expectation value of the Heisenberg field equation along with the mode equations for scalar and fermion fluctuations are fulfilled.

 The scalar contributions to the energy density, equation of motion of the mean field and effective potential are suppressed in the large $N_f$ limit as is evident in the expressions (\ref{exeom},\ref{enerdensf}), these  have been studied in detail in ref.\cite{nathan}. In the analysis that follows we neglect these,  focusing solely on the fermionic contributions. To leading order in the large $N_f$ limit, the equation of motion and energy density become
 \be \ddot{\varphi}(t)+ v'(\varphi(t))+y_0\, \sum_{s}\,\int  \, V^\dagger_{-\vk,s}(t)\gamma^0 V_{-\vk,s}(t) \,\frac{d^3k}{(2\pi)^3} =0  \,\label{eomNf}\ee
 \be  \mathcal{E} = N_f\Bigg\{\frac{\dot{\varphi}^2}{2} + v(\varphi)+   \sum_{s}\,\int \,V^\dagger_{-\vk,s}(t)\Big[\vec{\alpha}\cdot\vk + \gamma^0 m_F(\varphi(t))\Big] V_{-\vk,s}(t) \,\frac{d^3k}{(2\pi)^3}\Bigg\} \,\label{edensNf}\ee along with the Dirac equations (\ref{DEV}), from which it follows that
 \be \dot{\mathcal{E}}(t) = N_f\,\dot{\varphi} \Bigg[\ddot{\varphi}(t)+ v'(\varphi(t))+y_0\, \sum_{s}\,\int  \, V^\dagger_{-\vk,s}(t)\gamma^0 V_{-\vk,s}(t) \,\frac{d^3k}{(2\pi)^3} \Bigg] \,,\label{dotEfer}\ee confirming again, that the equation of motion (\ref{eomNf}) is obtained from energy conservation and the Dirac equation.

 \subsection{Fermion mode functions:}\label{subsec:fermodes}
 We now obtain the spinors $U,V$ in terms of fermion mode functions so as to provide a compact formulation that will be amenable to a numerical study. It is convenient to rewrite the Dirac equations (\ref{DEU},\ref{DEV}) in the form

 \bea
&& \Bigg[i \; \gamma^0 \;  \partial_t- \vec{\gamma}\cdot \vec{k}
-m_F(t) \Bigg]U_{\vk,s}(t)   =  0 \label{Uspinor} \\
&& \Bigg[i \; \gamma^0 \;  \partial_t- \vec{\gamma}\cdot \vec{k} -m_F(t)
\Bigg]V_{-\vk,s}(t)   =   0 \,. \label{Vspinor}
\eea

 It proves
convenient to write
\bea
U_{\vk,s}(t) & = & \Bigg[i \; \gamma^0 \;  \partial_t -
\vec{\gamma}\cdot \vec{k} +m_F(t)
\Bigg]f_k(t)\, u_s \label{Us}\\
V_{-\vk,s}(t) & = & \Bigg[i \; \gamma^0 \;  \partial_t -
\vec{\gamma}\cdot \vec{k} +m_F(t)
\Bigg]h_k(t)\,v_s \label{Vs}
\eea
\noindent with $u_s\,,\,v_s$ being
constant spinors  obeying
\be
\gamma^0 \; u_s  =  u_s
\label{Up} \qquad , \qquad
\gamma^0 \;  v_s  =  -v_s \,.
\ee

We choose the spinors $u_s \,;\,v_s$  as
\be u_s = \left(\begin{array}{c}
                             \xi_s\\
                            0
                          \end{array}\right)~~;~~ v_s = \left(\begin{array}{c}
                            0\\
                            \xi_s
                          \end{array}\right) \label{helispinors}\,, \ee  where the Weyl spinors $\xi_s$ are chosen to be helicity eigenstates, namely
                                                   \be \vec{\sigma}\cdot\vec{k} = s \, k \, \xi_s ~~;~~ s = \pm 1\,,  \label{helicity} \ee and fulfill the orthonormality condition
                                                   \be \xi^\dagger_s \xi_{s'} = \delta_{s,s'} \,.\label{xiorto}\ee

Inserting the ansatz (\ref{Us},\ref{Vs}) into the Dirac equations (\ref{Uspinor},\ref{Vspinor}) we find that the mode functions $f_k(t);h_k(t)$ obey the following
equations
\bea \left[\frac{d^2}{dt^2} +
E^2_k(t)-i \; \dot{m}_F(t)\right]f_k(t) & = & 0 \,, \label{feq}\\
\left[\frac{d^2}{dt^2} + E^2_k(t)+i \; \dot{m}_F(t)\right]h_k(t)
& = & 0 \,,\label{heq}
\eea   where
\be E_k(t) = \sqrt{k^2+m^2_F(t)} \,,\label{instener}\ee is  the ``instantaneous'' energy.

We will consider ``antiparticle'' initial conditions for the mode functions $h_k(t)$, namely
\be h_k(0) = f^*_k(0)  ~;~\dot{h}_k(0) = \dot{f}^*_k(0) \,,\label{inih}\ee these are the conditions corresponding to $e^{\mp iE_k t}$ for the case of constant $m_F$ as in the expansion (\ref{psiex}), we will adopt these ``particle-antiparticle'' initial conditions even for the time dependent case. The mode equation (\ref{heq}) with these initial conditions leads to the relation
\be h_k(t) = f^*_k(t) \,,\label{hfrel}\ee which will be seen below to yield the correct orthonormality conditions for the $U,V$ spinors. We choose
\be   f_k(0) =1 \,,\label{fk0}\ee and leave open the initial condition $\dot{f}_k(0)$ to study the dependence of the dynamics on initial conditions. In the next section we will establish a relation between the initial condition $\dot{f}_k(0)$ and the nature of the fermionic ground state.

The spinor solutions  are

\be U_{\vk,s}(t) = N_k\,\left( \begin{array}{c}
                                     \mathcal{F}_k(t)\, \xi_s\\
                                    k\,f_k(t) \, s \, \xi_s
                                  \end{array}\right)\,,  \label{Uspin}\ee

 \be V_{-\vk,s}(t) = N_k\,\left( \begin{array}{c}
                                      -k\,f^*_k(t) \, s\,\xi_s\\
                                    \mathcal{F}^*_k(t)  \,   \xi_s
                                  \end{array}\right)\,,  \label{Vspin}\ee  where we introduced

 \be \mathcal{F}_k(t) =    i\dot{f}_k(t)+m_F(t) f_k(t)\,, \label{capF}\ee and
                                  $N_k$ is a (constant) normalization factor chosen to be real, it is determined by the orthonormality conditions (\ref{ortot})
                                    yielding
 \be N^2_k\Big[\mathcal{F}^*_k(t)\,\mathcal{F}_k(t)+ k^2 f^*_k(t)\, f_k(t) \Big] = 1\,, \label{normaN} \ee in fact, using the mode equation (\ref{feq}) it is straightforward
 to find that
 \be \frac{d}{dt} \Big[\mathcal{F}^*_k(t)\,\mathcal{F}_k(t)+ k^2 f^*_k(t)\, f_k(t) \Big] =0 \,,\label{const}\ee therefore the bracket in (\ref{normaN}) is time independent and is fixed at $t=0$, yielding
 \be N_k = \Big[|\dot{f}_k(0)|^2 + E^2_k(0)|f_k(0)|^2+i\,m_F(0)\big(\dot{f}_k(0)\,f^*_k(0)-\dot{f}^*_k(0)\,f_k(0) \big)  \Big]^{-1/2} \,.\label{finN}\ee

 Using the Dirac equations (\ref{DEU},\ref{DEV}) leading to the constancy of the orthonormality relations (\ref{constortho}), one finds that the time derivative of the bracket in (\ref{normaN}) vanishes, confirming that $N_k$ is indeed a constant fixed by the normalization condition above. The orthogonality condition
 \be  U^\dagger_{\vk,s}(t)  V_{-\vk,s'}(t)=0 ~;~ s,s' = \pm 1 \,,\label{orto2}\ee is automatically fulfilled at all times.

 \subsection{Adiabatic expansion:}\label{subsec:wkb}

 As discussed in the previous section, the general form of the effective action (\ref{effac}) implies an expansion in derivatives\cite{colewein,colemanbook,ilio,branrmp} of the mean field, which  in the case of
 spatially homogeneous mean fields correspond to time derivatives. Such expansion is identified  with an adiabatic expansion, for which the Wentzel-Kramers-Brillouin (WKB) ansatz for the fermion mode function $f_k(t)$
 \be f_k(t) = e^{-i\int^t_0 \Omega_k(t') dt'}\,,\label{wkb}\ee
 is ideally suited. Inserting this ansatz into the mode equation (\ref{feq}) we find that $\Om_k(t)$ obeys the non-linear equation
 \be \Om^2_k(t)+ i\,\dot{\Om}_k(t)+ i\,\dot{m}_F(t) - E^2_k(t) = 0 \,,\label{Omeq} \ee writing $\Om_k(t) = \Om_{Rk}(t) + i \, \Om_{Ik}(t)$ with real $\Om_R,\Om_I$ it follows that
 \bea \Omega^2_{Rk}(t) & = & E^2_k(t) \Big[ 1 +  \frac{\Om^2_{Ik}(t)}{E^2_k(t)}+ \frac{\dot{\Om}_{Ik}(t)}{E^2_k(t)}\Big]\,\label{realOm}\\
 \Omega_{Ik}(t) & = & -\frac{1}{2} \Bigg( \frac{\dot{\Om}_{Rk}(t)}{\Om_{Rk}(t)} + \frac{\dot{m}_F(t)}{\Om_{Rk}(t)} \Bigg) \,,\label{imagOm}\eea
 inserting (\ref{imagOm}) into (\ref{realOm}) we find
 \be \Omega^2_{Rk}(t) = E^2_k(t) - \frac{1}{2} \Bigg(\frac{\ddot{\Om}_{Rk}(t)}{\Omega_{Rk}(t)}- \frac{3}{2}\,\frac{\dot{\Om}^{\,2}_{Rk}(t)}{\Om^2_{Rk}(t)}  \Bigg)+ \Bigg( \frac{\dot{m}^{\,2}_F(t)}{4\Om^2_{Rk}(t)} + \frac{\dot{m}_F(t)\,\dot{\Om}_{Rk}(t)}{\Om^2_{Rk}(t)} \Bigg)-\frac{\ddot{m}_F(t)}{2\Om_{Rk}(t)}\,.\label{omeR} \ee The first three terms in (\ref{omeR}) are similar to the scalar case\cite{nathan}. The last three terms originate in the second term in (\ref{imagOm}), which   can be traced back to the
 term $-i\dot{m}_F(t)$ in the mode equation (\ref{feq}),   a consequence of the fact that the Dirac equation is first order in time derivatives and an important difference between the bosonic (\ref{ModeTimeEvo}) and fermionic (\ref{feq}) mode equations, with profound implications.

The $\Om_{Rk}(t);\Om_{Ik}(t)$  can be expanded in  even and odd powers of derivatives   of $m_F(t)$ respectively:
\bea \Om_{Rk}(t)  & = & E_k(t)  - \frac{1}{4} \Bigg(\frac{\ddot{E_k}(t)}{E^2_k(t)}- \frac{3}{2}\,\frac{\dot{E}^{\,2}_{k}(t)}{E^2_k(t)}  \Bigg)+ \frac{1}{2}\Bigg( \frac{\dot{m}^{\,2}_F(t)}{4E^3_k(t)} + \frac{\dot{m}^{\,2}_F(t)\,m_F(t)}{E^4_{k}(t)} \Bigg)-\frac{\ddot{m}_F(t)}{4E^2_{k}(t)} \cdots \,,\label{omrex}\\ \Om_{Ik}(t) & = & -\frac{\dot{m}_F(t)}{2E_k(t)}\,\Bigg[1+\frac{m_F(t)}{E_k(t)}\Bigg]+\cdots \,,\label{omimex} \eea
and refer to terms with n-derivatives as n-th order adiabatic. Because $m_F(t) = y_0 \varphi(t)$ the adiabatic expansion is, in fact, an expansion in time derivatives of the
mean field. In other words,  the derivative terms in the effective action (\ref{effac}) correspond to an adiabatic expansion, the zeroth order is (minus) the effective potential and the higher order terms are associated with the derivatives in the effective action (\ref{effac}).
With $f_k(t)$ given by (\ref{wkb}) the normalization condition (\ref{normaN}) becomes
\be N^2_k |f_k(t)|^2\Big[|\Om_k(t)+m_F(t)|^2+ k^2  \Big] = 1 \,,   \label{newnorm}\ee and the   normalized $U,V$ spinors are
\be U_{\vk,s}(t) = \frac{e^{-i\int^t_0 \Om_{Rk}(t')dt'}}{\Big[E^2_k(t)+|\Om_k(t)|^2+2\Om_{Rk}(t)m_F(t)\Big]^{1/2}}\,\left( \begin{array}{c}
                                     \Big(\Om_k(t)+m_F(t)\Big)\, \xi_s\\
                                    k\, s \, \xi_s
                                  \end{array}\right)\,,  \label{Uspin2}\ee

 \be V_{-\vk,s}(t) = \frac{e^{i\int^t_0 \Om_{Rk}(t')dt'}}{\Big[E^2_k(t)+|\Om_k(t)|^2+2\Om_{Rk}(t)m_F(t)\Big]^{1/2}}\,\left( \begin{array}{c}
                                      -k\  s\,\xi_s\\
                                   \Big(\Om^*_k(t)+m_F(t)\Big)  \,   \xi_s
                                  \end{array}\right)\,,  \label{Vspin2}\ee

The zeroth adiabatic order fermion mode function is
\be f^{(0)}_k(t) = e^{-i \int^t_0 E_k(t')dt'} \,\label{zerof}\ee with which we construct the normalized zeroth adiabatic order spinors
\be \mathcal{U}_{\vk,s}(t) = \frac{e^{-i\int^t_0 E_k(t')dt'}}{\Big[2E_k(t)\Big(E_k(t)+m_F(t)\Big)\Big]^{1/2}}\,\left( \begin{array}{c}
                                     \Big(E_k(t)+m_F(t)\Big)\, \xi_s\\
                                    k\, s \, \xi_s
                                  \end{array}\right)\,,  \label{Uspinzero}\ee

 \be \mathcal{V}_{-\vk,s}(t) = \frac{e^{i\int^t_0 E_k(t')dt'}}{\Big[2E_k(t)\Big(E_k(t)+m_F(t)\Big)\Big]^{1/2}}\,\left( \begin{array}{c}
                                      -k\  s\,\xi_s\\
                                   \Big(E_k(t)+m_F(t)\Big)  \,   \xi_s
                                  \end{array}\right)\,,  \label{Vspinzero}\ee  which obey the \emph{time independent} orthonormality relations

\be \mathcal{U}^\dagger_{\vk,s}(t)\mathcal{U}_{\vk,s'}(t) = \delta_{s,s'}~;~ \mathcal{V}^\dagger_{-\vk,s}(t)\mathcal{V}_{-\vk,s'}(t)=\delta_{s,s'}~;~ \mathcal{U}^\dagger_{\vk,s}(t)\mathcal{V}_{-\vk,s'}(t)=0\,.\label{orthozero}\ee Furthermore, we note that although these spinors are \emph{not} solutions of the time dependent Dirac equations (\ref{DEU},\ref{DEV}),  they feature the  important property of being \emph{ instantaneous} eigenstates of the  momentum space Dirac Hamiltonian operator, namely
\bea \Big[\vec{\alpha}\cdot\vk + \gamma^0 m_F(\varphi(t))\Big]\,  \mathcal{U}_{\vk,s}(t) & = &  E_k(t)\,\mathcal{U}_{\vk,s}(t) \,,\label{eigenUzero} \\
\Big[\vec{\alpha}\cdot\vk + \gamma^0 m_F(\varphi(t))\Big]\,  \mathcal{V}_{-\vk,s}(t) & = &  -E_k(t)\,\mathcal{V}_{-\vk,s}(t) \,.\label{eigenVzero}   \eea

The orthonormality properties (\ref{orthozero}) entail that these spinors form a ``basis'', and we can expand the exact spinors $U,V$ in this basis,
\bea  U_{\vk,s}(t) & = & A_{k,s}(t) \,\mathcal{U}_{\vk,s}(t) + B_{k,s}(t) \,\mathcal{V}_{-\vk,s}(t) \,,\label{Uexp}\\
V_{-\vk,s}(t) & = & C_{k,s}(t) \, \mathcal{V}_{-\vk,s}(t)+ D_{k,s}(t)\, \mathcal{U}_{\vk,s}(t)   \,.\label{Vexp}\eea  This is a Bogoliubov transformation and the coefficients $A,B,C,D$ are Bogoliubov coefficients. These are obtained by projections and are given by
\bea A_{k,s}(t) & = & \mathcal{N}^*_k(t)\,N_k\,\Big[\mathcal{F}_k(t)\,\big(E_k(t)+m_F(t)\big)+k^2f_k(t)\Big] = C^*_{k,s}(t) \,,\label{ACbogen}\\
B_{k,s}(t) & = & \mathcal{N}_k(t)\,N_k\, \big(-s\,k\big)\Big[i\,\dot{f}_k(t)-E_k(t)\,f_k(t) \Big] = -D^*_{k,s}(t)\,. \label{BDbogen}\eea

The orthonormality of both sets of spinors $U,V$ and $\mathcal{U},\mathcal{V}$ imply that
\be |A_{k,s}(t)|^2+|B_{k,s}(t)|^2 = |C_{k,s}(t)|^2+|D_{k,s}(t)|^2 =1 \,,\label{unitary}\ee which is the statement of a unitary transformation and is straightforwardly confirmed from the expressions above.

Using the Bogoliubov transformation (\ref{Uexp},\ref{Vexp}) in the field expansion (\ref{psiext}) we expand the fermi field operator in the zeroth order adiabatic basis
\be \psi_i(\vx,t) =  \sqrt{\frac{\hbar}{\mathcal{V}}}\,\sum_{\vk} \sum_{s=\pm 1} \Big[ \widetilde{b}_{\vk,s,i}(t) \, \mathcal{U}_{\vk,s}(t)  +  \widetilde{d}^{\,\dagger}_{-\vk,s,i}(t) \, \mathcal{V}_{-\vk,s}(t) \Big]\,e^{i\vk\cdot \vx} \,,\label{psiextad}\ee where the annihilation and creation operators now depend on time and are given by
\bea  \widetilde{b}_{\vk,s,i}(t) & = &  {b}_{\vk,s,i}\,A_{k,s}(t) - {d}^\dagger_{-\vk,s,i}\,B^*_{k,s}(t)\,,\label{btil} \\
\widetilde{d}^{\,\dagger}_{-\vk,s,i}(t)  & = & {d}^\dagger_{-\vk,s,i}\,A^*_{k,s}(t) +   {b}_{\vk,s,i}\, B_{k,s}(t) \,,\label{ddtil}\eea the relations (\ref{unitary}) imply that
the operators $\widetilde{b},\widetilde{d}$ obey the standard anticommutation relations. We identify $\widetilde{b},\widetilde{d}$ as the annihilation operators of \emph{adiabatic particles and antiparticles}, since the adiabatic spinors $\mathcal{U},\mathcal{V}$ are orthonormal instantaneous eigenfunctions of the Dirac Hamiltonian (see equations (\ref{eigenUzero},\ref{eigenVzero})).   With the field expansion in the adiabatic basis (\ref{psiextad}), the fermion Hamiltonian  (\ref{hampsi}) for the dynamical case  becomes to leading order in the large $N_f$ limit
\be H_F =  \int d^3x \,\sum_{i=1}^{N_f} \psi^\dagger_i \Big[ -i \vec{\alpha}\cdot \vec{\nabla} + \gamma^0\, m_F(\varphi(t)) \Big] \psi_i = \sum_{i=1}^{N_f}\sum_{\vk,s} E_k(t) \Big[\widetilde{b}^{\,\dagger}_{\vk,s,i}(t)\widetilde{b}_{\vk,s,i}(t)+ \widetilde{d}^{\,\dagger}_{\vk,s,i}(t)\widetilde{d}_{\vk,s,i}(t) -1\Big] \,.\label{adhami} \ee
The \emph{adiabatic vacuum state},  $\ket{\Phi;\widetilde{0}(t)}$ is defined such that
\be \widetilde{b}_{\vk,s,i}(t)\ket{\Phi;\widetilde{0}(t)}=0 ~;~\widetilde{d}_{\vk,s,i}(t)\ket{\Phi;\widetilde{0}(t)}=0  ~~\forall t \,.\label{adiavac}\ee

 The conditions (\ref{bdanni}) yield the number of adiabatic particles and antiparticles in the vacuum state $\ket{\Phi;0_F}$ as
\be \bra{\Phi;0_F}\widetilde{b}^{\,\dagger}_{\vk,s,i}(t)\widetilde{b}_{\vk,s,i}(t)\ket{\Phi;0_F} = n_{k}(t)= \bra{\Phi;0_F}\widetilde{d}^{\,\dagger}_{\vk,s,i}(t)\widetilde{d}_{\vk,s,i}(t)\ket{\Phi;0_F} = \overline{n}_{k}(t)= |B_{k,s}(t)|^2 ~~\forall i\,.\label{adnumber}\ee   The distribution function of produced adiabatic particles and antiparticles is given by
\be n_k(t)= \overline{n}_k(t) = \frac{N^2_k\,k^2\, \Big|i\dot{f}_k(t)-E_k(t)f_k(t)\Big|^2}{\Big[2E_k(t)\Big(E_k(t)+m_F(t)\Big)\Big]}\,,\label{nkoft}\ee where $f_k(t)$ is an exact solution of the mode equations (\ref{feq}). We note that the distribution functions do not depend on helicity or flavor, therefore yielding the same contribution to the energy density for each flavor.

Writing the spinor $V_{-\vk,s}(t)$ as in equation (\ref{Vexp}), using the relations (\ref{ACbogen},\ref{BDbogen},\ref{adnumber}), and the unitarity relation (\ref{unitary}) we find the fermionic contribution to the energy density in equation (\ref{enerdensf})
\be \frac{1}{\mathcal{V}}\sum_{\vk,s}V^\dagger_{-\vk,s}(t)\Big[\vec{\alpha}\cdot\vk + \gamma^0 m_F(\varphi(t))\Big] V_{-\vk,s}(t) = -2 \int E_k(t) \frac{d^3k}{(2\pi)^3} + 2 \int E_k(t) \Big[n_k(t)+\overline{n}_k(t)\Big] \frac{d^3k}{(2\pi)^3} \,,\label{ferenerfin}   \ee therefore neglecting the scalar contribution to the energy density, which is suppressed by $1/N_f$ in the large $N_f$ limit, we find
\be  \mathcal{E} = N_f\Bigg\{\frac{\dot{\varphi}^2}{2} +\underbrace{ v(\varphi)-2 \int E_k(t) \frac{d^3k}{(2\pi)^3}}_{v_{eff}(\varphi)} + 2 \int E_k(t) \underbrace{\Big[n_k(t)+\overline{n}_k(t)\Big]}_{2\,|B_{k,s}(t)|^2} \frac{d^3k}{(2\pi)^3}\Bigg\}\,,\label{enedensfin}  \ee   The second and third terms correspond to the effective potential with
the negative fermion vacuum energy (\ref{finbVeff}), the last two terms are manifestly \emph{positive}, describe adiabatic particle production and are of second adiabatic order or higher, corresponding to derivatives of $\varphi$, in direct agreement with the form of the effective action (\ref{effac}).

In order to obtain the equation of motion from the conservation of energy $\dot{\mathcal{E}}=0$, we need the time derivative of the particle production contribution. The result (\ref{const}) and the normalization condition (\ref{normaN}) allows us to write
\be  E_k(t)\,n_k(t) =\frac{ k^2}{2} \Bigg[\frac{1}{\Big(E_k(t)+m_F(t)\Big)}- i\,N^2_k\,\Big(\dot{f}_k(t)\,f^*_k(t)- f_k(t)\,\dot{f}^*_k(t) \Big) \Bigg]\,,\label{enen}   \ee using the mode equations (\ref{feq}) we find
\be \frac{d}{dt} \Big[ E_k(t)\,n_k(t)  \Big]  =  \dot{\varphi}(t) \, y_0\, k^2 \Bigg[N^2_k \Big|f_k(t)\Big|^2 -\frac{1}{2E_k(t)\Big(E_k(t)+m_F(t)\Big)}  \Bigg]\,,\label{dtenen}  \ee  yielding
\be \dot{\mathcal{E}} = N_F\,\dot{\varphi}\,\Bigg\{\ddot{\varphi} + \frac{d}{d\varphi}v_{eff}(\varphi) + 4\,y_0  \int   k^2 \Bigg[N^2_k \Big|f_k(t)\Big|^2 -\frac{1}{2E_k(t)\Big(E_k(t)+m_F(t)\Big)}   \Bigg]\,      \frac{d^3k}{(2\pi)^3}  \Bigg\} \,,\label{edott}\ee energy conservation yields the equation of motion
\be   \ddot{\varphi} + \frac{d}{d\varphi}v_{eff}(\varphi) + 4\,y_0  \int   k^2 \Bigg[N^2_k \Big|f_k(t)\Big|^2 -\frac{1}{2E_k(t)\Big(E_k(t)+m_F(t)\Big)}   \Bigg]\,      \frac{d^3k}{(2\pi)^3}    =0  \,,\label{fineommf}\ee and $f_k(t)$ is the exact solution of the mode equations  (\ref{feq}).

The last term in equation (\ref{fineommf}) yields a manifestly energy conserving dynamics, in contrast to the  equation of motion (\ref{pheneom}) often used in the literature which does not.

This form of the  equation of motion is of course equivalent to (\ref{eomNf}),  its main advantage is that it explicitly separates the zeroth adiabatic order radiative corrections into the effective potential, and the higher adiabatic orders describing adiabatic particle production embodied in the last term. In other words, the equation of motion (\ref{fineommf}) is the explicit manifestation of  the derivative expansion, i.e.,  the adiabatic expansion of the effective action (\ref{effac}).

This analysis emphasizes the importance of the zeroth-order adiabatic basis: it directly exhibits the main features of the effective action (\ref{effac}) separating the contribution
of the effective potential (zeroth adiabatic order) and derivative terms (second and higher adiabatic orders) which correspond to terms describing the production of adiabatic particles in eqns. (\ref{enedensfin},\ref{fineommf}), which are of second and higher adiabatic order. The constant energy density (\ref{enedensfin}) is strikingly different from the phenomenological constant of motion (\ref{phenE}): the contributions to $\mathcal{E}$ (\ref{enedensfin}) from adiabatic particle production are \emph{manifestly positive}. In particular let us consider the case when the mean field is ``rolling down'' the unstable part of the effective potential, as the time derivative of $\varphi$ increases, so does the number of particles, contributing positively to the energy density, therefore, \emph{slowing down} the dynamical evolution as the potential energy is drained by particle production. This mechanism
of energy transfer between the mean field and quantum fluctuations is definitely \emph{not} described by the phenomenological equation (\ref{pheneom}) and its conserved quantity (\ref{phenE}), which evidently is \emph{not}  the total energy density. Therefore, using the equation of motion (\ref{pheneom}) explicitly violates total energy conservation, and completely misses the energy transfer to adiabatic modes.

\subsection{Results for WKB solutions:}\label{subsec:wkbcoefs}

The results (\ref{nkoft},\ref{enedensfin},\ref{fineommf}) are general for any mode function $f_k(t)$ which is an exact solution of the mode equations (\ref{feq}). With the purpose of extracting the high energy behavior of the various quantities introduced above, we summarize their expressions for the case when the exact solution  of the mode equation $f_k(t)$ is given by the WKB ansatz (\ref{wkb}) and $\Omega_k(t)$ is the exact solution of equation (\ref{Omeq}).

Using equation (\ref{newnorm}) the Bogoliubov coefficients become
\bea A_{k,s}(t) & = & \mathcal{N}^*_k(t)\,\widetilde{\mathcal{N}}_k(t)\,\Big[ \big(E_k(t)+m_F(t)\big)\,\big(\Om_{k}(t)+m_F(t) \big)+ k^2\Big] = C^*_{k,s}(t) \,,\label{ACbog}\\
B_{k,s}(t) & = & \mathcal{N}_k(t)\,\widetilde{\mathcal{N}}_k(t)\, \big(-s\,k\big)\Big[\Om_k(t)-E_k(t) \Big] = -D^*_{k,s}(t)\,, \label{BDbog}\eea
where
\be \mathcal{N}_k(t) = \frac{e^{-i\int^t_0 E_k(t')dt'}}{\Big[2E_k(t)\Big(E_k(t)+m_F(t)\Big)\Big]^{1/2}}~~;~~ \widetilde{\mathcal{N}}_k(t) = \frac{e^{-i\int^t_0 \Om_{Rk}(t')dt'}}{\Big[E^2_k(t)+|\Om_k(t)|^2+2\Om_{Rk}(t)m_F(t)\Big]^{1/2}}\ \,,\label{normfacs}\ee
yielding
\be n_k(t) = \overline{n}_k(t) = |B_{k,s}(t)|^2 =  \frac{k^2\,\Big[\Big(\Om_{Rk}(t)-E_k(t)\Big)^2+\Om^2_{Ik}(t)\Big]}{2E_k(t)\Big(E_k(t)+m_F(t)\Big)\,\Big[E^2_k(t)+|\Om_k(t)|^2+2\Om_{Rk}(t)m_F(t)\Big]}  \,,\label{bsquare}\ee therefore the distribution functions are of second and higher adiabatic order. This result confirms that  the last term in the energy density (\ref{enedensfin}) contains the radiative corrections to the derivative terms, leading us to one of the important conclusions of this study:  the radiative corrections to the derivative terms in the equation of motion for the condensate arise from the production of adiabatic particle-antiparticle production.

The denominator in (\ref{bsquare}) can be written as
\be  E^2_k(t)+|\Om_k(t)|^2+2\Om_{Rk}(t)m_F(t)    =  2E_k(t)(E_k(t)+ m_F(t))+ \mathcal{D}(t) \,\label{denomi}\ee where
\be \mathcal{D}(t)   =  \Big[ \Om^2_{Rk}(t)-E^2_k(t) + \Om^2_{Ik}(t)+ 2 m_F(t)\,\big(\Om_{Rk}(t)-E_k(t)\big)\Big] \,,\label{Ddef}\ee is of second and higher adiabatic order.

The leading order term in the large $k$ limit for $n_k(t)$ arises from the $\Om^2_{Ik}(t)$ contribution, with the leading adiabatic order given by eqn. (\ref{omimex}) we find the large $k$ behaviour of the distribution function
\be n_k(t)~ {}_{ \overrightarrow{k\rightarrow \infty}} ~  \frac{y^2_0 \dot{\varphi}^2(t)\,k^2 }{16\,E^6_k(t)} + \mathcal{O}(1/k^6)+\cdots \label{nkasyk}\ee Therefore for large wavevector  $n_k(t) \propto \dot{\varphi}^2(t)/k^4$, hence the total number of particles (and antiparticles) produced is finite, however
\be E_k(t)\, n_k(t)  ~ {}_{ \overrightarrow{k\rightarrow \infty}} ~  \frac{y^2_0 \dot{\varphi}^2(t)\,k^2 }{16\,E^5_k(t)} + \mathcal{O}(1/k^5)+\cdots \,,\label{enasyk}\ee yielding an ultraviolet logarithmic divergence in the total energy density (\ref{enedensfin}) proportional to $\dot{\varphi}^2(t)$, which leads to wave function renormalization  as discussed below .

Using the normalization condition  (\ref{newnorm}) along with the relation (\ref{realOm}), we find the particle production contribution to the equation of motion (\ref{fineommf})

\be k^2\,\Bigg[N^2_k \Big|f_k(t)\Big|^2 -\frac{1}{2E_k(t)\Big(E_k(t)+m_F(t)\Big)}   \Bigg]  =   - k^2\,\frac{\Bigg[ \Om^2_{Rk}(t)-E^2_k(t)  +\Om^2_{Ik}(t)  +2m_F(t)\,\big(\Om_{Rk}(t)-E_k(t) \big) \Bigg]}{2E_k(t)\Big(E_k(t)+m_F(t)\Big)\,\Big[2E_k(t)(E_k(t)+ m_F(t))+ \mathcal{D}(t)\Big]}   \,. \label{dtennwkb}\ee  The large $k$ behaviour is determined by the $-\ddot{m}_F(t)/2E_k(t)$ term in $\Om^2_{Rk}(t)-E^2_k(t)$ in eqn. (\ref{omeR}), yielding
\be k^2\,\Bigg[N^2_k \Big|f_k(t)\Big|^2 -\frac{1}{2E_k(t)\Big(E_k(t)+m_F(t)\Big)}   \Bigg]~ {}_{ \overrightarrow{k\rightarrow \infty}} ~  \frac{y_0 \ddot{\varphi}(t)\,k^2}{8 E^5_k(t)}+ \mathcal{O}(1/k^5)+\cdots\,, \label{denasyk}\ee which of course is consistent with the result (\ref{enasyk}) as it must. This asymptotic behavior gives rise to an ultraviolet logarithmic divergence in the momentum integral for the equations of motion (\ref{edott}), as discussed below.

\subsection{On wave function renormalization and the origin of $\mathcal{Z}[\varphi]$:}\label{subsec:Z}
Since both $\mathcal{Z}[\varphi]$ and wave function renormalization emerge as coefficients of $\dot{\varphi}^2(t)$ in the effective action and the coefficient of $\ddot{\varphi}$ in the equation of motion (the term $\propto p^2$ in the fermion propagator), their origin must be in the particle production terms (last terms)
in (\ref{enedensfin}, \ref{edott}), since both are of second adiabatic order. The ultraviolet divergent terms contributing these coefficients are precisely those identified in equations (\ref{enasyk},\ref{denasyk}) in the large $k$ limit.
With the result (\ref{enasyk}) we find the ultraviolet divergent contribution from the second adiabatic order

\be 4 \int E_k(t)\,n_k(t) \,\frac{d^3k}{(2\pi)^3}\Bigg|_{\mathrm{UV-div-2nd~ order}} = \frac{\dot{m}^2_F(t)}{8\pi^2}\, \int^{\Lambda}_0 \frac{k^4}{E^5_k(t)}\, dk =\frac{1}{2}\,\dot{\varphi}^2(t)  \,\frac{y^2_0}{4\pi^2}\, \Big[ \ln\Big( \frac{2\Lambda}{|m_f(t)|}\Big)-\frac{4}{3}\Big]   \,,\label{zetaad} \ee where we have introduced an ultraviolet cutoff $\Lambda$ and neglected terms that do not yield ultraviolet divergences, in particular, higher adiabatic orders yield ultraviolet finite contributions. Neglecting the constant term in the $\Lambda \rightarrow \infty$ limit, we find up to \emph{second adiabatic order} the energy density (\ref{enedensfin})
\be   \mathcal{E}_{2nd} = N_f\Bigg\{\frac{\dot{\varphi}^2}{2}\,\Big[1 + \frac{y^2_0}{4\pi^2}\,\ln\Big( \frac{2\Lambda}{|m_f(t)|}\Big) \Big]  + v_{eff}(\varphi) \Bigg\}  \,,\label{Esecord} \ee where $v_{eff}(\varphi)$ is given by equation (\ref{Veffrentot}).
The kinetic term can be written as
\be \frac{1}{2}\,y^2_0\, \dot{\varphi}^2(t) \Bigg[\frac{1}{y^2_R(\mu)}+ \frac{1}{4\pi^2} \,\ln\Big(\frac{\mu}{|m_F(t)|}\Big)\Bigg] \,,\label{kinter}\ee where
\be \frac{1}{y^2_R(\mu)} = \Bigg[\frac{1}{y^2_0}+ \frac{1}{4\pi^2} \,\ln\Big(\frac{2\Lambda}{\mu}\Big) \Bigg]\equiv \frac{Z^{-1}_\phi}{y^2_0} \,,\label{zitfi}\ee where we recognize that $Z^{-1}_\phi$ coincides with the wave function renormalization (\ref{Z}) obtained from the one loop scalar self energy. Therefore, with the renormalization conditions (see eqn. \ref{rentot})
\be \varphi_R = \frac{\varphi}{\sqrt{Z_\phi}}~~;~~   y_R = \sqrt{Z_{\phi}}\,y_0  \Rightarrow y^2_0 \dot{\varphi}^2(t) =y^2_R \dot{\varphi}^2_R(t) \,,\label{rentotfin}\ee we can write the conserved energy density up to second adiabatic order, consistently with the expansion of the effective action (\ref{effac}) as
\be  \mathcal{E}_{2nd} = N_f\Bigg\{\frac{\dot{\varphi}^2_R}{2}\,\mathcal{Z}[\varphi_R]  + v_{eff}(\varphi_R) \Bigg\}    \,,\label{Esecordfin} \ee where we identify
\be \mathcal{Z}[\varphi_R] = \Bigg[1- \frac{y^2_R(\mu)}{4\pi^2} \,\ln\Big(\frac{|m_F(t)|}{\mu}\Big)\Bigg]\,.\label{matZ}\ee Energy conservation yields
\be \dot{\mathcal{E}} = N_f \dot{\varphi}_R \Bigg[ \ddot{\varphi}_R \mathcal{Z}[\varphi_R] + \frac{d}{d\varphi_R} v_{eff}(\varphi_R) + \cdots \Bigg] =0 \,,\label{consi}\ee hence the terms in the bracket are the equation of motion for $\varphi_R$. The dots stand for derivatives of $\mathcal{Z}$ and higher order derivative terms.  These results are worth emphasizing: field renormalization arises from the derivative terms in the adiabatic expansion, which are associated with particle production.

This analysis is confirmed directly from the equation of motion for $\varphi$, (\ref{fineommf}), from the result (\ref{denasyk}) we obtain up to second adiabatic order and neglecting terms that are finite in the $\Lambda \rightarrow \infty$ limit,
\be  \ddot{\varphi}(t)\,\Bigg[1 + \frac{y^2_0}{4\pi^2}\,\int_0^{\Lambda}   \frac{k^4}{E^5_k(t)}\, dk   \Bigg]  + \frac{d}{d\varphi}\,v_{eff}(\varphi)=0 \,.\label{eomfiZ}\ee Following the renormalization steps discussed above, yields the renormalized equation of motion up to second adiabatic order
\be  \frac{1}{\sqrt{Z_\phi}}\Big[ \ddot{\varphi}_R(t)\,\mathcal{Z}[\varphi_R]+ \frac{d}{d\varphi_R}\,v_{eff}(\varphi_R(t)) + \cdots \Big] =0 \,,\label{eomZ} \ee  consistently with (\ref{consi}). This result is also confirmed in appendix (\ref{app:eom})   starting directly from the equation of motion in the form given by equation
(\ref{eomNf})  up to second adiabatic order and extracting the ultraviolet divergence.

The function $\mathcal{Z}[\varphi_R]$ has an important meaning: the definition of $y^2_R(\mu)$ in equation (\ref{zitfi}) implies the ``running'' of the coupling with the scale $\mu$:
\be \frac{1}{y^2_R(\mu')}= \frac{1}{y^2_R(\mu)}+  \frac{1}{4\pi^2} \,\ln\Big(\frac{\mu}{\mu'}\Big)\,,\label{runy}\ee therefore the bracket in equation (\ref{kinter}) is identified as
\be \Bigg[\frac{1}{y^2_R(\mu)}+ \frac{1}{4\pi^2} \,\ln\Big(\frac{\mu}{|m_F(t)|}\Big)\Bigg]  =  \frac{1}{y^2_R(|m_F(t)|)} \Rightarrow \mathcal{Z}[\varphi_R] = \frac{y^2_R(\mu)}{y^2_R(|m_F(t)|)} \,,\label{runmf}\ee
 where the last result follows from equation (\ref{matZ}).
Therefore, the coupling ``runs'' with $\varphi_R(t)$ and $\mathcal{Z}$ in the effective action describes the renormalization group running of the effective coupling with $\varphi_R(t)$
\be y^2_R(\varphi_R(t)) = \frac{y^2_R(\mu) }{\mathcal{Z}[\varphi_R(t)]}\,.\label{runyt}\ee

Although the result (\ref{Esecordfin}) is consistent with the form of the effective action (\ref{effac}) which as we argued   is of second   adiabatic order, the function $\mathcal{Z}[\varphi_R(t)]$ features some pathologies:  Firstly for large $|m_F(t)|$, for example in the region of instability of $v_{eff}$ both terms in (\ref{Esecordfin}) become large and negative,   counter to the conservation of $\mathcal{E}$. In this limit the effective coupling $y^2_R(\varphi(t))$ given by (\ref{runyt}) blows up when $\mathcal{Z}$ vanishes, this is the Landau pole ubiquitous in non-asymptotically free theories, such as Quantum Electrodynamics and $y^2_R(\varphi)$ becomes \emph{negative} as this pole is passed. Secondly  as $|m_F| \rightarrow 0$, $v_{eff} \rightarrow 0$ but $\mathcal{Z} \rightarrow \infty$ entailing
$\dot{\varphi}$ must go rapidly to zero to maintain energy conservation.  Both these limits entail a \emph{breakdown} of the adiabatic expansion and result in unphysical behavior that
contradicts the exact conservation of $\mathcal{E}$: In the large $|m_F(t)|$ limit (in which the Landau pole is present) the derivatives  of $\varphi$ are large as per the equations of motion, entailing a breakdown of adiabaticity. In the opposite limit as $|m_F(t)|\rightarrow 0$ there is an infrared divergence in the integrals in (\ref{zetaad}), which is
a consequence of having neglected the contribution from $\mathcal{D}(t)$, in the denominators in eqns. (\ref{bsquare},\ref{dtennwkb}) which are of second order. As a result, the numerator and denominator are of equal order   again a breakdown of adiabaticity. Therefore the pathologies of $\mathcal{Z}[\varphi(t)]$ in these limits are a consequence of the loss of validity of the adiabatic expansion and are not physical. In turn, this translates into  a breakdown of the derivative expansion of the effective action for long wavelength fluctuations when the effective time dependent fermion mass becomes very small. Furthermore, energy conservation entails that the dynamical evolution does not feature any pathology  at the Landau pole.

The analysis above clarifies that the radiative corrections to the derivative terms in the energy and the equation of motion for the condensate originate in the particle production contribution yielding both the ultraviolet divergent field renormalization and the finite field dependent function $\mathcal{Z}[\varphi]$ that multiplies the derivative terms in the effective action (\ref{effac}).

  For a thorough discussion of the breakdown of the adiabatic approximation within the context of the effective action see the recent set of lectures in ref.\cite{dunnecern}.

We emphasize that the derivative expansion is discussed above solely within the context of establishing direct contact with the form and content of the effective action (\ref{effac}). The \emph{only} aspect of this expansion that is used in this study is the emergence of field renormalization at the second order in this expansion, and for this
only the large $k$ behavior is relevant, for which the WKB expansion is certainly reliable. As is discussed in detail below the final set of fully renormalized and energy conserving equations of motion \emph{do not} involve the adiabatic approximation in any form beyond extracting the ultraviolet divergent wave function renormalization.

\subsection{Initial conditions: initial time singularity}\label{subsec:inicon}

In order to study the dynamical evolution as an initial value problem, we must append initial conditions on the homogeneous condensate $\varphi_R(t)$ \emph{and} the mode functions $f_k(t)$. For the homogeneous mode we consider the   initial conditions
\be \varphi_R(t=0) \equiv \varphi_R(0)~~;~~ \dot{\varphi}_R(t=0) =0 \,,\label{iniconfi}\ee generically there is always some particular value of time, $t_*$  at which $\dot{\varphi}_R(t=t_*)=0$, therefore without loss of generality, we define $t_*$ as the origin of time.  It remains to initialize the mode functions $f_k(t)$, since their evolution equation (\ref{feq}) is second order in time derivative, we must specify $f_k(0)$ and $\dot{f}_k(0)$.

  With  $\Om_k(t)$   a solution of equations (\ref{Omeq},\ref{realOm},\ref{imagOm}), the WKB mode functions (\ref{wkb}) are solutions of the mode equation (\ref{feq}) with initial condition
  \be f_k(0)=1~;~\dot{f}_k(0) = -i\,\Om_k(0)\,,\label{inif}\ee where $\Om_k(0)= \Om_{Rk}(0)+ i\, \Om_{Ik}(0)$ is the solution of equations (\ref{omeR},\ref{imagOm}) at time $t=0$, which require $\ddot{\varphi}_R(0)$ (and higher derivatives),  and which, in turn, is a solution of the equation of motion (\ref{eomNf}) at the initial time  for consistency with energy conservation. Therefore, initialization of the mode functions with initial conditions (\ref{inif}) entails \emph{solving self-consistently} the non-linear equation (\ref{Omeq}) and  the renormalized equations of motion for $\varphi_R$ along with higher derivatives   at the initial time\cite{inising1}.

   This can be seen from  the expression (\ref{dtennwkb}) (factoring out the $\dot{\varphi}$ that multiplies the full expression),  which when evaluated at $t=0$ depends on $\ddot{\varphi}(0)$ and higher derivatives even when $\dot{\varphi}(0) =0$ and as shown in eqn. (\ref{denasyk}) the coefficient of $\ddot{\varphi}(0)$ is ultraviolet divergent when integrated in momentum. This implies that when evolving the set of equations, one must solve initially  a highly non-linear self-consistent problem including the ultraviolet divergences that must be absorbed into a consistent renormalization. Furthermore,  in the adiabatic approximation with $\Om_{Rk}(t),\Om_{Ik}(t)$ given by equations (\ref{omrex},\ref{omimex}), even when $\dot{m}_F(0) =0$ it follows that $\big(\Om_k(0)-E_k(0)\big) = - y_R\ddot{\varphi}_R(0)/4E^2_k(0) $,   generically $\ddot{\varphi}_R(0) \neq 0 $ and must be solved self-consistently from  the equation of motion (\ref{eomNf}). As a result the adiabatic particle number $n_k(0) = \overline{n}_k(0) \neq 0$ and the vacuum state does not coincide with the adiabatic vacuum for the initial conditions (\ref{inif}).

  From the expression for the adiabatic particle distributions (\ref{nkoft}) it follows that $n_k(0)= \overline{n}_k(0)=0$ if the exact mode functions are solution of the mode equations (\ref{feq}) with initial conditions
  \be \dot{f}_k(0) = -i\,E_k(0)f_k(0) ~~;~~ E_k(0) = \sqrt{k^2 + y^2_R\,\varphi^2_R(0)}\,.\label{iniE}\ee Therefore, we adopt the following initial conditions for   $f_k(t)$
  \be f_k(0)=1 ~~;~~ \dot{f}_k(0) = -i\,E_k(0) \,,\label{fininif}\ee which yield $n_k(0)=\overline{n}_k(0)=0$ and do not involve solving a non-linear self consistent system of equations to initialize the dynamics. For these initial conditions, the normalization factor $N_k$ of the exact Dirac spinors (\ref{Uspin},\ref{Vspin}) is simply (see equation (\ref{finN}))
  \be N_k = \Big[2E_k(0)\,\big(E_k(0)+m_F(0)\big)\Big]^{-1/2}\,. \label{nofkini}\ee Therefore, expressions (\ref{nkoft},\ref{dtenen}) with (\ref{nofkini})  entail that   these initial conditions yield
  \be n_k(0) =0 ~;~ \Bigg[N^2_k \Big|f_k(t)\Big|^2 -\frac{1}{2E_k(t)\Big(E_k(t)+m_F(t)\Big)}  \Bigg]_{t=0} =0 \,.\label{zeroini}\ee  The second identity implies that the particle production contribution to the equation of motion for $\varphi$ (\ref{fineommf}) vanishes at $t=0$.

 These initial conditions imply that the vacuum $\ket{\Phi;0_F}$ coincides with the adiabatic vacuum state (\ref{adiavac}) at $t=0$, namely $\ket{\Phi;0} = \ket{\Phi;\widetilde{0}(0)}$ and that with the initial conditions (\ref{iniconfi}) on the condensate the energy density is given by

  \be \mathcal{E}= N_f\, v_{eff}(\varphi_R(0))\,. \label{enerinicon}\ee

  Although eventually the full dynamics must be studied numerically and setting initial conditions on the condensate and mode functions suffices, as we learned in the previous
  sections, consistent renormalization requires understanding the large momentum behavior of the mode functions, for which the WKB solution is uniquely suited. Therefore,
  we seek to obtain the mode functions with the initial conditions (\ref{fininif}) in a WKB approach to identify if and how these initial conditions modify the high energy behavior
  and the renormalization program.

  The mode equations (\ref{feq}) feature two linearly independent solutions, let us call them $f_{1k}(t),f_{2k}(t)$  for which their Wronskian
  \be \mathcal{W}_{12}= \dot{f}_{1k}(t)\,f_{2k}(t) - \dot{f}_{2k}(t)\,f_{1k}(t) = \mathrm{constant} \neq 0 \Rightarrow \frac{d}{dt} \Big(\frac{f_{2k}(t)}{f_{1k}(t)} \Big)= -\frac{\mathcal{W}_{12}}{f^2_{1k}(t)} \,,\label{wronskf}\ee allowing to obtain $f_{2k}(t)$ once we know $f_{1k}(t)$, namely
  \be f_{2k}(t) = -\mathcal{W}_{12} \,f_{1k}(t)\,\int^t_0 \frac{dt'}{f^2_{1k}(t')} \,.\label{f2oft}\ee We choose $f_{1k}(t)$ to be the WKB solution (\ref{wkb}), namely
  \be f_{1k}(t)= e^{-i\int^t_0 \Om_k(t')\,dt'}\,,\label{ef1}\ee with $\Om_k(t)$ is the solution of the equation (\ref{Omeq}) and
  \be f_k(t) = \alpha_k \,f_{1k}(t)+ \beta_k\,f_{2k}(t)~~;~~ f_k(0)=1~~;~~ \dot{f}_k(0) = -i\,E_k(0) \,,\label{totfsol}\ee yielding
  \be \alpha_k=1 ~~;~~ \beta_k = -i\frac{\big(\Om_k(0)-E_k(0) \big)}{\mathcal{W}_{12}}\,. \label{ABcoefs}\ee With the initial condition (\ref{iniconfi}) and the expansions (\ref{omrex},\ref{omimex}) it follows that up to second adiabatic order
  \be \big(\Om_k(0)-E_k(0) \big) = -\frac{1}{4} \frac{\ddot{m}_F(0)}{E^2_k(0)}\,\Big[1+ \frac{m_F(0)}{E_k(0)} \Big] + \cdots \label{difex}\ee Gathering the results of appendix (\ref{app:Icalc}) we find up to second adiabatic order
  \be f_k(t) = f_{1k}(t)  \,\Bigg[ (1-\eta_k)+ \eta_k\, e^{2i \int^{t}_0 \Om_{Rk}(t')dt'}\, e^{\int^{t}_0 \frac{\dot{m}_F(t')}{\Om_{Rk}(t')}dt'} + \cdots \Bigg] \,,\label{ftot}\ee where
  \be \eta_k \equiv \Big(\frac{\Om_k(0) - E_k(0)}{2\,\Om_{Rk}(0)}\Big) = -\frac{\ddot{m}_F(0)}{8\,E^3_k(0)}\,\Big[1+\frac{m_F(0)}{E_k(0)} \Big]+ \cdots \label{eta2nd}\ee and the dots in the expressions above correspond to terms of third and higher adiabatic order. Up to second adiabatic  order we find
     \be i\frac{\dot{f}_k(t)}{f_k(t)} \equiv \widetilde{\Om}_k(t) = \Om_k(t) - 2\Om_{Rk}(t)\,\eta_k\,\mathcal{H}_k(t) \,. \label{tilome}\ee where we defined
  \be \mathcal{H}_k(t) \equiv \mathcal{H}_{Rk}(t) + i\,\mathcal{H}_{Ik}(t) =  e^{2i \int^{t}_0 \Om_{Rk}(t')dt'}\, e^{\int^{t}_0 \frac{\dot{m}_F(t')}{\Om_{Rk}(t')}dt'} \,.\label{hofk}\ee
  Note that
  \be \widetilde{\Om}_k(0) = E_k(0)~;~ \Rightarrow \widetilde{\Om}_{Rk}(0)= E_k(0)~~;~~ \widetilde{\Om}_{Ik}(0)=0 \,,\label{reimtilo}\ee are \emph{exact} relations to all orders in the adiabatic expansion as a consequence of the initial conditions (\ref{totfsol}).

  The modification, if any, of the ultraviolet divergences by the initial conditions are obtained by resorting to a WKB expansion just as in the treatment above.  From equation (\ref{bsquare}) with the replacement $\Om_k(t) \rightarrow \widetilde{\Om}_k(t)$ we obtain the expression for $n_k(t)$ modified by the initial conditions (\ref{totfsol})

\be n_k(t) = \overline{n}_k(t)=  \frac{k^2\,\Bigg[\Big(\widetilde{\Om}_{Rk}(t)-E_k(t)\Big)^2+\widetilde{\Om}^2_{Ik}(t)\Bigg]}{\Big[2E_k(t)\Big(E_k(t)+m_F(t)\Big)\Big]\,\Big[E^2_k(t)+|\widetilde{\Om}_k(t)|^2+2\widetilde{\Om}_{Rk}(t)m_F(t)\Big]}  \,,\label{bsquaremod}\ee the change to the distribution functions from initial conditions to leading adiabatic order becomes
\be \delta n_k(t) = \delta \overline{n}_k(t) =    -4\,k^2\,\Om_{Rk}(t)\,\eta_k\,\,\frac{ \Bigg[\Big(\Om_{Rk}(t)-E_k(t)\Big)\,\mathcal{H}_{Rk}(t)+ \Om_{Ik}(t)\, \mathcal{H}_{Ik}(t)\Bigg]}{\Big[2E_k(t)\Big(E_k(t)+m_F(t)\Big)\Big]^2} + \cdots \,,\label{b2adini}\ee  The leading contribution in the large $k$ limit arises from the $\Om_{Ik}(t) \propto \dot{m}_F(t)/E_k(t)$ yielding $\delta n_k(t) \propto 1/E^5_k(t)$ for large $k$, therefore $\delta n_k(t)$ gives  a finite contribution to the total number of produced adiabatic particles. Similarly, the change  to the energy density from the extra terms from the different initial conditions is
\be \delta \mathcal{E} = 4 \int E_k(t)\,\delta n_k(t) \,\frac{d^3k}{(2\pi)^3}\,,\label{enerchange}\ee  is also finite since the integrand  $\propto 1/E^4_k(t)$ for large $k$. Therefore the total energy density only features the ultraviolet divergence associated with wave function renormalization which is not affected by the change of initial conditions. However, the \emph{time derivative} of $ \delta \mathcal{E}$ features a \emph{new} ultraviolet divergence because the time derivative of $\mathcal{H}_{Ik}(t)$ brings in an extra factor $E_k(t)$ in the leading adiabatic order $\Om_{Rk}(t) \simeq E_k(t)$. Therefore, we expect a new divergence in the  equation of motion of $\varphi$. We analyze this new divergence directly from the equation of motion (\ref{fineommf}).

To simplify notation, let us define the particle production contribution to the equation of motion (\ref{fineommf}) as
\be \mathcal{P}(\varphi) \equiv   \int   k^2 \Bigg[N^2_k \Big|f_k(t)\Big|^2 -\frac{1}{2E_k(t)\Big(E_k(t)+m_F(t)\Big)}   \Bigg]\,      \frac{d^3k}{(2\pi)^3}~~;~~ \mathcal{P}(\varphi)\Big|_{t=0} =0 \,\label{ppeom}\ee where $f_k(t)$ are the exact solutions of the mode equations (\ref{feq}) with initial conditions (\ref{fininif}).

\begin{figure}[ht!]
\begin{center}
\includegraphics[height=3.5in,width=3.5in,keepaspectratio=true]{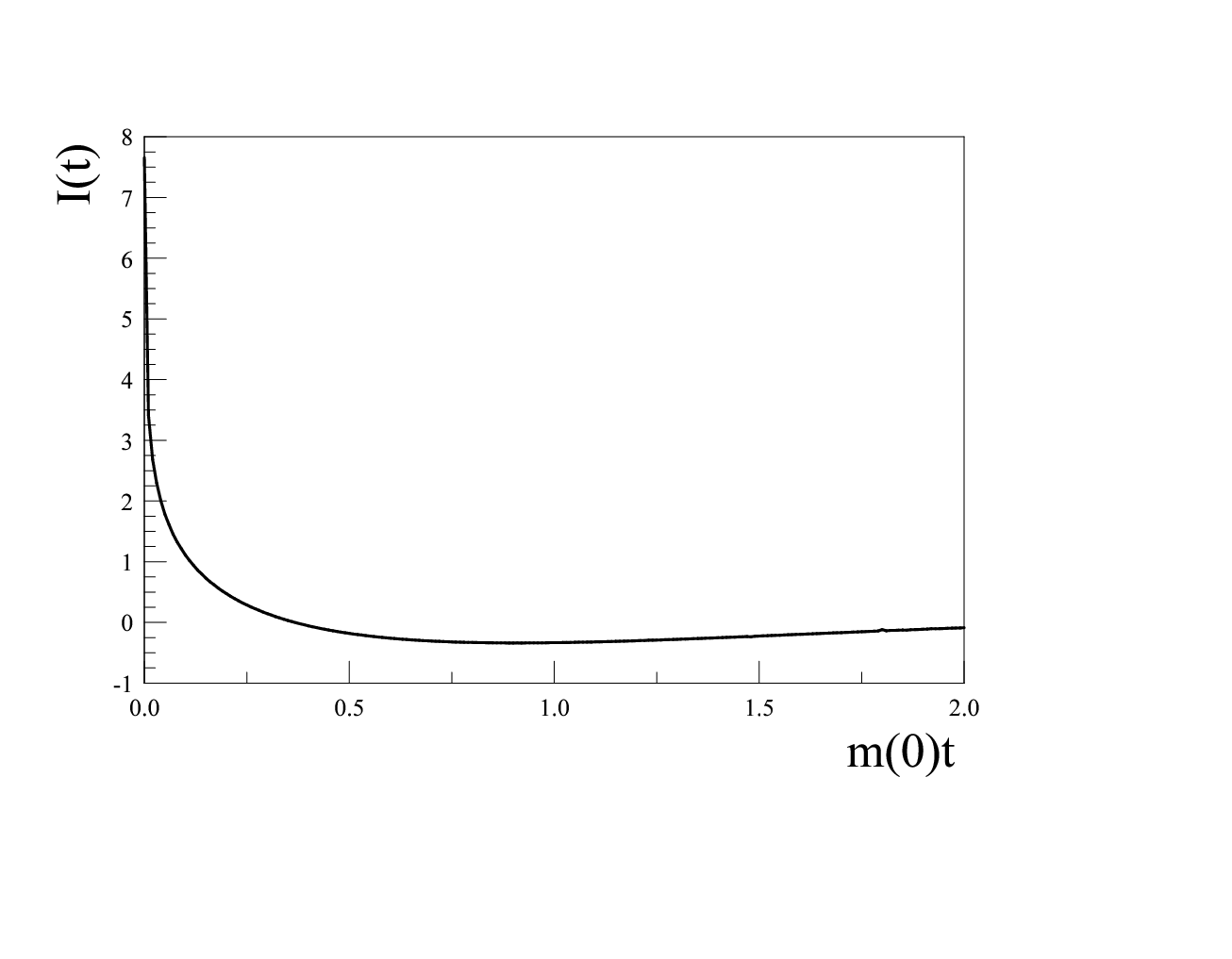}
\caption{$I(t) = \int^{{\Lambda}/{m_F(0)}}_0 \frac{x^4\,\cos\big(2\,\sqrt{x^2+1}\,m_F(0)t\big)}{(x^2+1)^{5/2}}\,dx $ vs. $m_F(0)\,t$ for $\Lambda/m_F(0) = 4000$.}
\label{fig:inising}
\end{center}
\end{figure}

The change   in the particle production contribution to the equation of motion for $\varphi$,   $\mathcal{P}(\varphi)$ given by (\ref{ppeom}),    is obtained from eqn. (\ref{dtennwkb}) by replacing $\Om_{k}(t) \rightarrow \widetilde{\Om}_k(t)$, the leading contribution
arises from the term $\widetilde{\Om}^2_{Rk}(t)-E^2_k(t) \simeq \Om^2_{Rk}(t)-E^2_k(t) -4\Om^2_{Rk}(t)\,\eta_k \,\mathcal{H}_{Rk}(t)$ as analyzed previously for wave function renormalization, yielding for the ultraviolet divergent  contribution to the integral
\be     4\,y_0\,\mathcal{P}(\varphi)\,\Bigg|_{UV-div} =    \frac{y^2_0}{4\pi^2} \int^{\Lambda}_0   k^4\,\Bigg[\frac{\ddot{\varphi}(t)}{E^5_k(t)}- \frac{\ddot{\varphi}(0)}{E^2_k(t)\,E^3_k(0)}\,\cos\Big(2\,\int^t_0 E_k(t')\,dt' \Big) \Bigg] dk\,,\label{intddfi} \ee
 note that this contribution vanishes at $t=0$, as it must since as noted above the total particle production contribution to the equation of motion vanishes at $t=0$ with the initial conditions (\ref{iniE}) for the mode functions $f_k(t)$. The first term on the
right hand side  of eqn. (\ref{intddfi}) is recognized from equation (\ref{eomfiZ})  with the wave function renormalization,  the second term featuring the  cosine, arises from the change in initial conditions and \emph{exactly cancels} the ultraviolet divergence associated with wave function renormalization at $t=0$, thereby introducing a new singularity at the initial time.  However,  the integral of the oscillatory   term $\propto \ddot{\varphi}(0)$,  vanishes rapidly for $t \gg 1/\Lambda$ by dephasing. Figure (\ref{fig:inising}) displays  this integral replacing $E_k(t) \rightarrow E_k(0)$, showing clearly the singularity as $t\rightarrow 0$ and the rapid fall off on a time scale $\propto 1/\Lambda$. The integral is finite for $t \gg 1/\Lambda$  and   vanishes  for $m_F(0) t \gtrsim 1$.

The interpretation of the dynamics described by the competition between the two terms   in equation (\ref{intddfi}), which is discussed further in section (\ref{sec:discussion}), is the \emph{dressing} of the bare fields into the fully renormalized field. The  initial time singularity in the oscillatory term cancels exactly   the ultraviolet divergent wave function renormalization (the first term   in (\ref{intddfi}))  at $t=0$, and vanishes  for $t \gg 1/\Lambda$. The combination of the two terms describes  the dynamics of \emph{dressing} on short time scales $t \simeq 1/\Lambda$, since the
total integral  in (\ref{intddfi}) rises from zero up to the ultraviolet divergent wave function renormalization on this short time scale. In other words, the bare field \emph{fully dresses} into the renormalized field by fermionic quantum fluctuations on a short time scale $\simeq 1/\Lambda$.

\subsection{Renormalized energy and equations of motion.}\label{subsec:reneom}
The full dynamics of the scalar condensate and the fermion fluctuations is completely determined by the equation of motion (\ref{fineommf}) along with the fermion   mode equations (\ref{feq}) with initial conditions (\ref{fininif}), which together yield the conservation of the energy density given by equation (\ref{enedensfin}).
The particle production contribution to the energy density features the ultraviolet divergence (\ref{zetaad}), namely
 \be 4 \int E_k(t)\,n_k(t) \,\frac{d^3k}{(2\pi)^3}\Bigg|_{\mathrm{UV-div}} =  \frac{1}{2}\,\dot{\varphi}^2(t)\, \frac{y^2_0}{4\pi^2}\, \, \int^{\Lambda}_0 \frac{k^4}{E^5_k(t)}\, dk \,. \label{uvdivEdens} \ee  This is a consequence of the asymptotic   behavior of the distribution function $n_k(t) \simeq \dot{\varphi}^2(t)/k^4$ as $k\rightarrow \infty$ which is responsible for field renormalization.

 As discussed above   the particle production contribution  $\mathcal{P}(\varphi)$ to the equation of motion of the condensate, namely the last term in (\ref{edott}), features two ultraviolet divergences given by equation (\ref{intddfi}), where the first term is associated with wave function renormalization and the second with an initial time singularity.  The momentum integrals in  equations (\ref{uvdivEdens}, \ref{intddfi})
   are logarithmically divergent in the limit $\Lambda \rightarrow \infty$, let us introduce the following function
\be \mathcal{I}\big[\frac{\Lambda}{\mu};\mu t\big] \equiv \int^{\Lambda}_0 \frac{k^4}{(k^2+\mu^2)^{\frac{5}{2}}}\,\cos\big(2\,(k^2+\mu^2)\,t\big)\,dk\,,\label{Ifun}\ee    which for $t=0$ and  $\Lambda/\mu \gg 1$ becomes
\be \mathcal{I}\big[\frac{\Lambda}{\mu};0\big] = \ln\Big(\frac{2\Lambda}{\mu}\Big)  \,,\label{Ilarge}\ee where $\mu$ is a renormalization scale which we have chosen consistently with the renormalization program for the effective potential and neglected a finite constant of $\mathcal{O}(1)$ in the $\Lambda/\mu \rightarrow \infty$ limit.
Let us \emph{define} the \emph{renormalized} (R)  particle production contribution to the energy density as
\be 4 \int E_k(t)\,n_k(t) \,\frac{d^3k}{(2\pi)^3}\Bigg|_{R} = 4 \int E_k(t)\,n_k(t) \,\frac{d^3k}{(2\pi)^3} -  \frac{1}{2}\,\dot{\varphi}^2(t)\, \frac{y^2_0}{4\pi^2}\, \mathcal{I}\big[\frac{\Lambda}{\mu};0\big] \,,\label{renEdens} \ee which is manifestly ultraviolet finite in the limit $\Lambda \rightarrow \infty$ and it also vanishes at $t=0$ because of the initial condition $\dot{\varphi}(0) =0$ which has been used all throughout the analysis. The energy density (\ref{enedensfin}) is now written as
\be   \mathcal{E}  = N_f\Bigg\{\frac{\dot{\varphi}^2}{2}\,\Bigg[1 + \frac{y^2_0}{4\pi^2}\,\mathcal{I}\big[\frac{\Lambda}{\mu};0\big] \Bigg]  + v_{eff}(\varphi) + 4 \int E_k(t)\,n_k(t) \,\frac{d^3k}{(2\pi)^3}\Bigg|_{R} \Bigg\}  \,, \label{renEg1}\ee the bracket multiplying $\dot{\varphi}^2/2$ is precisely $Z^{-1}_{\phi}$ given by eqn. (\ref{Z}), therefore, using the renormalization conditions (\ref{rentotfin}) the energy density in terms of fully renormalized fields, masses and couplings is given by
\be \mathcal{E}  = N_f\Bigg\{\frac{\dot{\varphi_R}^2}{2}\,  + v_{eff}(\varphi_R) + 4 \int E_k(t)\,n_k(t) \,\frac{d^3k}{(2\pi)^3}\Bigg|_{R} \Bigg\}\,,\label{renEgfin}\ee since $v_{eff}$ is solely a function of the renormalized field, couplings and masses and the distribution function $n_k(t)$ is solely a function of $m_F(t) = y_0 \varphi(t) = y_R \varphi_R(t)$. However, unlike the expression (\ref{Esecordfin}) the particle production integral  in (\ref{renEgfin}) still contains contributions of second adiabatic order ($\propto \dot{\varphi_R}^2$) and higher which are ultraviolet and infrared finite. The renormalization prescription adopted in (\ref{renEgfin}) lumps the contribution from the finite $\mathcal{Z}[\varphi_R]$ in (\ref{Esecordfin}) into the renormalized  particle production contribution which contains all orders in the adiabatic expansion thereby making explicit that  the unphysical singularity (Landau pole)  discussed above, is a result of the breakdown of the adiabatic approximation.

The final form of the energy density (\ref{renEgfin}) is conserved and solely a function of the renormalized fields, masses and couplings, it is ultraviolet and infrared finite, and  the particle production contribution vanishes at $t=0$.

We proceed in a similar fashion to obtain a fully renormalized equation of motion (\ref{fineommf}) which with the definition (\ref{ppeom}) we write as
\be  \ddot{\varphi}(t) + \frac{d}{d\varphi}v_{eff}(\varphi(t)) + 4\,y_0\, \mathcal{P}(\varphi(t))   =0 \,,\label{eomcondi} \ee

 The result given by equation (\ref{intddfi}) leads us  now to \emph{define}
\be  4\,y_0\,\mathcal{P}_R(\varphi(t)) \equiv  4\,y_0\,\mathcal{P}(\varphi(t)) - \frac{y^2_0}{4\pi^2} \Big[\ddot{\varphi}(t)\,\mathcal{I}\big[\frac{\Lambda}{\mu};0\big]- \ddot{\varphi}(0)\,\mathcal{I}\big[\frac{\Lambda}{\mu};\mu t\big]\Big] ~~;~~  \mathcal{P}_R(\varphi(t))\Big|_{t=0}=0\,,\label{Pren}\ee from this definition, it is clear that $P_R(\varphi(t))$ is ultraviolet \emph{finite}, vanishes at $t=0$  and does not feature an initial time singularity. Writing $\mathcal{P}(\varphi(t))$ in terms of $\mathcal{P}_R(\varphi(t))$, the equation of motion (\ref{eomcondi}) becomes
\be   \ddot{\varphi}(t)\,\Big[1+\frac{y^2_0}{4\pi^2} \,\mathcal{I}\big[\frac{\Lambda}{\mu};0\big] \Big]  + \frac{d}{d\varphi}v_{eff}(\varphi(t)) + 4\,y_0\, \mathcal{P}_R(\varphi(t))   = \frac{y^2_0}{4\pi^2}\,\ddot{\varphi}(0) \,\mathcal{I}\big[\frac{\Lambda}{\mu};\mu \,t\big]  \,.\label{eomPP} \ee With the renormalized coupling given by (\ref{zitfi}) and the renormalization conditions (\ref{rentotfin}) the equation of motion becomes
\be \frac{1}{\sqrt{Z_\phi}}\Bigg\{  \ddot{\varphi}_R(t)+\frac{d}{d\varphi_R}v_{eff}(\varphi_R(t))+ 4\,y_R \,\mathcal{P}_R(\varphi_R(t))   \Bigg\} =  \frac{y^2_R}{4\pi^2\,\sqrt{Z_{\phi}}}\,\ddot{\varphi}_R(0) \,\mathcal{I}\big[\frac{\Lambda}{\mu};\mu \,t\big]\,.\label{inteom}  \ee The value of $\ddot{\varphi}_R(0)$ is found self-consistently by evaluating this equation of motion at $t=0$, since the equation of motion is fulfilled at all times, yielding
\be \ddot{\varphi}_R(0) =\frac{ \frac{d}{d\varphi_R}v_{eff}(\varphi_R(t))\Big|_{t=0}  }{  \frac{y^2_R}{4\pi^2}\,\mathcal{I}\big[\frac{\Lambda}{\mu};0\big]-1}\,,\label{ddotphiR}  \ee where we used that $\mathcal{P}_R(t=0)=0$. Finally we obtain the fully renormalized equation of motion
\be   \ddot{\varphi}_R(t)+\frac{d}{d\varphi_R}v_{eff}(\varphi_R(t))+ 4\,y_R \,\mathcal{P}_R(\varphi_R(t)) =  \frac{d}{d\varphi_R}v_{eff}(\varphi_R(t))\Big|_{t=0} \,\, \Bigg[\frac{\mathcal{I}\big[\frac{\Lambda}{\mu};\mu \,t\big]}{\mathcal{I}\big[\frac{\Lambda}{\overline{\mu}};0\big]}\Bigg]~~;~~ \overline{\mu} = \mu\,e^{\frac{4\pi^2}{y_R}}\,,  \label{reneomfin}\ee with initial conditions
\be \varphi_R(t=0) = \varphi_R(0)~;~\dot{\varphi}_R(t)\Big|_{t=0} =0 \,.\label{reninicons}\ee This is the final form of the equation of motion for the scalar condensate, solely in terms of renormalized fields, couplings and masses. The last term on the left hand side is the contribution from particle production, it vanishes at the initial time and is manifestly ultraviolet finite.

\begin{figure}[ht!]
\begin{center}
\includegraphics[height=3.5in,width=3.5in,keepaspectratio=true]{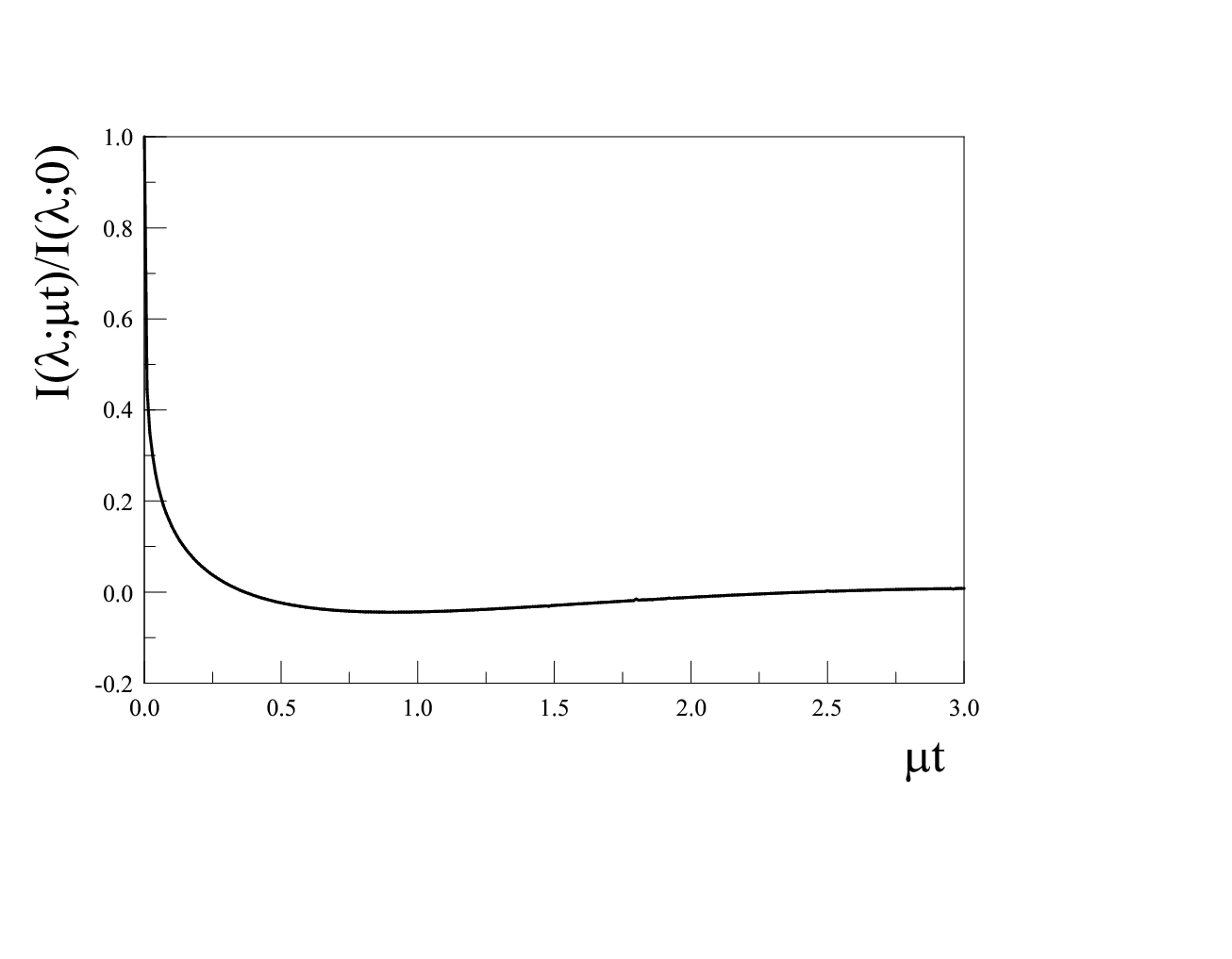}
\caption{$I(\lambda;\mu t)/I(\lambda;0)$ vs $\mu t$ for $\lambda=4000$.}
\label{fig:source}
\end{center}
\end{figure}

The source term on the right hand side of (\ref{reneomfin}) is a remnant of the dressing dynamics. As discussed above, the dressing from the bare into the renormalized field is a dynamical process that occurs on a time scale $\simeq 1/\Lambda$, however, we have renormalized the field with the \emph{time independent} and ultraviolet divergent renormalization constant $Z_\phi$, the time dependence of the source term accounts for the remaining dynamics of dressing. Furthermore, whereas in terms of the bare fields there is an initial time singularity,  the source term does not feature any initial time singularity after field and coupling renormalization,  it is finite and independent of $\Lambda$ in the limit $\Lambda/\mu \rightarrow \infty$ consistently with a finite and conserved energy. In this limit the bracket in the source term is a function starting at one at $t=0$ and sharply falling off to zero on a time scale $t \simeq 1/\Lambda$. To highlight this aspect numerically, figure (\ref{fig:source}) displays the ratio $\mathcal{I}\big[\frac{\Lambda}{\mu};\mu \,t\big]/\mathcal{I}\big[\frac{\Lambda}{\mu};0\big]$ as a function of $\mu t$ for $\Lambda/\mu =4000$.

With the purpose of providing a well defined and consistent dynamical framework for future numerical study of misalignment dynamics, we gather together the full set of renormalized equations and initial conditions that must be solved self-consistently:

\textbf{Fermion mode equations:}
\be   \left[\frac{d^2}{dt^2} +
E^2_k(t)-i \; \dot{m}_F(t)\right]f_k(t)   =   0 ~~;~~ f_k(0)=1~;~\dot{f}_k(0) = -i E_k(0) \,\label{fmeq}\ee
\be E_k(t) = \sqrt{k^2+m^2_F(t)}~~;~~ m_f(t) = y_R\,\varphi_R(t) \,\label{enere}\ee

\textbf{Scalar condensate equation of motion:}
\be   \ddot{\varphi}_R(t)+\frac{d}{d\varphi_R}v_{eff}(\varphi_R(t))+ 4\,y_R \,\mathcal{P}_R(\varphi_R(t)) =  \frac{d}{d\varphi_R}v_{eff}(\varphi_R(t))\Big|_{t=0} \,\,\Bigg[\frac{\mathcal{I}\big[\frac{\Lambda}{\mu};\mu \,t\big]}{\mathcal{I}\big[\frac{\Lambda}{\overline{\mu}};0\big]}\Bigg]\,,\label{scaleom}  \ee with initial conditions
\be \varphi_R(t=0) = \varphi_R(0)~~;~~\dot{\varphi_R}(t)\Big|_{t=0}=0 \,.\label{iniconre}\ee The self-consistent evolution of (\ref{fmeq},\ref{scaleom}) guarantees the conservation of the energy density
\be  \frac{ \mathcal{E}}{N_f}  = \Bigg\{\frac{\dot{\varphi_R}^2}{2}\,  + v_{eff}(\varphi_R) + 4 \int E_k(t)\,n_k(t) \,\frac{d^3k}{(2\pi)^3}\Bigg|_{R} \Bigg\} =  v_{eff}(\varphi_R(0))\,,\label{conseER} \ee

We emphasize that   to obtain the fully renormalized equations of motion and energy density,  all we did was to add and subtract the ``counterterms'' in the definitions (\ref{renEdens},\ref{Pren}). This subtraction rendered the renormalized definitions finite in the limit $\Lambda \rightarrow \infty$, and the addition was absorbed into the (time independent) field renormalization and the source term in the equation of motion. Nothing depends on the renormalization scale $\mu$, namely all the equations and conserved energy are manifestly renormalization group invariant and independent of the arbitrary scale $\mu$.

 This set of fully renormalized, self-consistent, energy conserving equations are some of the main results of this study, the last term on the left hand side of equation (\ref{scaleom}) accounts for the contribution from particle production.

 While as we have stated above, it is not our intention in this article to pursue the numerical study of the dynamics,   some practical aspects merit discussion: in the strict $\Lambda \rightarrow \infty$  the bracket on the right hand side of the final equation of motion (\ref{scaleom}) equals $1$ at $t=0$ and vanishes for $t>0$, for a finite but large $\Lambda$ it vanishes  very rapidly on the scale $t\simeq 1/\Lambda$, therefore  for all practical purposes with a time step of  integration   much larger than this scale, the right hand side of (\ref{scaleom})  \emph{can be safely set to zero}. Secondly, even when the set of equations is renormalization group invariant and does not depend on the scale $\mu$, it is clear that in order to numerically implement the integrations even with a large cutoff, this scale   must be specified. A particularly natural choice of scale is $\mu=m_F(0) = y_R\,\varphi_R(0)$ this scale  obviously determines the energy density (\ref{conseER}),  is one of the important physical scales in the problem, and is numerically appealing since $m_f(t)$ is  the only scale that appears in the integrals. Since this is the renormalization group scale, all renormalized couplings are referred to it.

\section{Asymptotics: excited stationary fixed points of the dynamics:}\label{sec:asymptotic}
The form of the energy density, equation (\ref{conseER}) suggests the \emph{possibility} of asymptotic stationary fixed points of the dynamics: with the initial conditions (\ref{iniconfi}) and (\ref{iniE}) the latter ones yielding $n_k(0)=\overline{n}_k(0) =0$ and the energy density given by (\ref{enerinicon}), there is the possibility of an asymptotic stationary fixed point of the dynamics with $\varphi(t \rightarrow \infty) \equiv \varphi(\infty); \dot{\varphi}(\infty)=0;\ddot{\varphi}(\infty) =0; n_k(\infty) = \overline{n}_k(\infty)  \neq 0$ so that conservation of energy implies
\be \frac{\mathcal{E}}{N_f} =   v_{eff}(\varphi_R(\infty))+ 4 \int E_k(\infty) \, n_k(\infty) \,\frac{d^3k}{(2\pi)^3}   =  \,v_{eff}(\varphi_R(0)) \,.\label{eneasy}\ee Note that the particle production contribution agrees in this limit with its subtracted, renormalized counterpart (\ref{renEdens}) because $\dot{\varphi_R}(\infty) =0$.  As analyzed above the ultraviolet divergence in the contribution to the energy density from particle production is $\propto \dot{\varphi}(t)$    since this is the origin of wave function renormalization, which vanishes in an asymptotic stationary state with $\dot{\varphi}(t)~{}_{\overrightarrow{t\rightarrow \infty}} ~~ 0 $.  Therefore in this limit the contribution from particle production is finite and given by
\be \int E_k(\infty) \, n_k(\infty)\,\frac{d^3k}{(2\pi)^3} = \frac{1}{4} \Big(v_{eff}(\varphi_R(0))- v_{eff}(\varphi_R(\infty))  \Big)  = \mathrm{finite}\,>0\,.\label{asynumb}\ee

The physical meaning of this
result is that particle production has drained energy from the condensate, which eventually settles   with vanishing time derivatives at an asymptotic value of the effective potential below the initial one. In particular for the case when the initial value of the field is on the unstable part of the effective potential, the field evolves towards
lower values of the effective potential all the while creating particles which eventually stabilize the dynamics at the stationary fixed point.

The emergence of a stationary  asymptotic limit as a consequence of particle production must be contrasted with the phenomenological equation of motion (\ref{pheneom}) and its conserved ``energy'' (\ref{phenE})  that neglect the important contribution from particle production: considering the same initial conditions $\varphi_R(t=0) = \varphi_R(0);\dot{\varphi_R}(0)=0$, the only stationary point predicted by these equations is the minimum of the effective potential and only if the initial value of the condensate is at the minimum, hence no evolution whatsoever. Since there is no mechanism of energy transfer to a continuum   in (\ref{pheneom}), if the initial value of the condensate is not at the minimum of the effective potential it would oscillate forever. Particle production is an efficient mechanism of energy transfer from one single mode, the homogeneous scalar condensate, into a continuum of modes even with  very large $k$, thereby providing an effective dissipative mechanism of relaxation with manifest energy conservation and unitary time evolution since the evolution equations follow from the exact Heisenberg equations of motion.

The asymptotic limit of the mode functions can be understood from the \emph{exact} solution of the mode equations (\ref{f2oft}-\ref{ABcoefs})
\be f_k(t) = e^{-i\int^t_0 \Om_k(t')\,dt'}\,\Big[1 + i  \big(\Om_k(0)-E_k(0) \big)   \,  \int^t_0  e^{2i\int^{t'}_0 \Om_k(t'')\,dt''} {dt'} \Big]  \,.\label{exafk}\ee \emph{If} an asymptotic stationary limit emerges with $\varphi(t) \rightarrow \varphi(\infty)$ and $\dot{\varphi},\ddot{\varphi} \cdots {}_{\overrightarrow{t\rightarrow \infty}} ~~  0$, it follows from equations (\ref{Omeq}-\ref{imagOm}) that $\Om_k(t)~ {}_{\overrightarrow{t\rightarrow \infty}} ~  E_k(\infty)$, consequently the analysis of the integral in (\ref{exafk}) in appendix (\ref{app:Icalc}), and equation (\ref{ftot})  yield the asymptotic behavior
\be f_k(t)~ {}_{\overrightarrow{t\rightarrow \infty}}~\widetilde{\alpha}_k\, e^{-iE_k(\infty) t}\,+ \widetilde{\beta}_k\,e^{iE_k(\infty) t} \,,\label{fkasy}\ee where $\widetilde{\alpha}_k,\widetilde{\beta}_k$ depend on the initial condition and the \emph{history} of the dynamics, and
\be \widetilde{\beta}_k =\eta_k \, \Big(1 + \cdots\Big) \label{widbet}\ee where the $ \eta_k $ is given by equation (\ref{eta2nd}) and  dots stand for terms of higher adiabatic order that arise from the time integrals, namely the dynamical history as discussed in appendix (\ref{app:Icalc}). That $f_k(t)$ given by  (\ref{fkasy}) is an asymptotic solution if $\varphi(t) \rightarrow \varphi(\infty); \dot{\varphi} \rightarrow 0$ can be directly seen from the mode equation (\ref{feq}).

Introducing the asymptotic solution (\ref{fkasy}), and the normalization factor (\ref{nofkini}) into the adiabatic distribution function (\ref{nkoft}) we find the asymptotic result
\be n_k(\infty) = |\widetilde{\beta}_k|^2 \,  \frac{ \Big(1-\frac{m_F(0)}{E_k(0)}\Big)}{  \Big(1+\frac{m_F(\infty)}{E_k(\infty)}\Big)}\,.\label{nkasy}\ee Although $\widetilde{\beta}_k$ must be proportional to $\eta_k$ as shown in equation (\ref{widbet}) the proportionality function within the brackets in equation (\ref{widbet}) must be obtained numerically because it is a result of the history of the dynamical evolution. However the leading dependence of $\widetilde{\beta}_k$ on $k$ in the large $k$ limit is determined by $\eta_k \propto y_R \ddot{\varphi}_R(0)/k^3$, (see equation (\ref{eta2nd})), therefore in the asymptotic long time limit, $n_k \propto 1/k^6$ and its contribution to the energy density is ultraviolet finite, as it must be to satisfy the asymptotic condition (\ref{asynumb}).

This \emph{must} be the case: we have previously shown that the ultraviolet divergence to the energy density from particle production is solely proportional to $\dot{\varphi}^2$, yielding wave function renormalization and the initial time singularity which is irrelevant in the long time limit, therefore in the asymptotic stationary limit with $\dot{\varphi}\rightarrow 0$ the coefficient of the ultraviolet divergence \emph{vanishes}.

For consistency, such an asymptotic limit must also emerge from the equations of motion  (\ref{fineommf}) in the long time limit. Assuming that a stationary solution emerges with $\ddot{\varphi}(t)\rightarrow 0;\dot{\varphi}(t) \rightarrow 0; \varphi(t)\rightarrow \varphi(\infty)$, with $f_k(t)$ asymptotically given by equation (\ref{fkasy}), the stationary solution is such that
\be v'_{eff}(\varphi(\infty)) + 4\,y_0  \int   k^2 \Bigg[\frac{|\widetilde{\alpha}|^2+|\widetilde{\beta}|^2}{2E_k(0)\Big(E_k(0)+m_F(0)\Big)} -\frac{1}{2E_k(\infty)\Big(E_k(\infty)+m_F(\infty)\Big)}   \Bigg]\,      \frac{d^3k}{(2\pi)^3}    = 0\,. \label{staeom}\ee where interference terms from $|f_k(t)|^2$ vanish by dephasing (Riemann-Lebesgue) upon integration in the long time limit in a manner similar to figure (\ref{fig:inising}). While the coefficients $\widetilde{\alpha},\widetilde{\beta}$ must be found numerically as they depend on the full history of the dynamics, the discussion above makes clear that the integral in (\ref{staeom}) is finite: the ultraviolet divergences of the    integral in the equation of motion (\ref{fineommf}) were identified with wave function renormalization and the initial time singularity, the former being proportional to $\ddot{\varphi}(t)$   vanishes in the long time limit if there is an emergent stationary state, and the latter only has support on the short time scale $\simeq 1/\Lambda$ and is irrelevant in the long time limit.

\vspace{1mm}

\subsection{Spontaneous symmetry breaking?:}\label{subsec:ssb}

The emergence of an asymptotic highly excited stationary state brings an intriguing and hitherto unexplored possibility, that of \emph{spontaneous symmetry breaking} even with the tree level potential (\ref{vzero}) with $m_0 > 0$ which does not feature spontaneous symmetry breaking minima. Consider the
case where the initial condition $\varphi_R(0)$ is such that $|\varphi_R(0)|$ is beyond the \emph{radiatively induced maxima}, on the unstable side of the effective potential (\ref{Veffdimless}) (see figure (\ref{fig:veffofx})). The dynamics of misalignment in this case leads to profuse particle production because the time derivatives of the scalar condensate are ever increasing, which in turn would shut-off the instability because the potential energy is drained by particle production, leading to the asymptotic stationary state with $|\varphi_R(\infty)|\neq 0$ satisfying the condition (\ref{asynumb}). This would be a novel mechanism of spontaneous symmetry breaking, in the case of the tree level potential (\ref{vzero}) the discrete symmetry being $Z_2$,  as a consequence of the combination of  \emph{static and dynamical } aspects, \textbf{i:)} the radiatively induced maxima of the static effective potential a l'a Coleman-Weinberg which leads to the instability, and \textbf{ii:)} misalignment dynamics with profuse particle production with conserved energy which leads to
the asymptotic highly excited stationary state. Unlike the case of finite temperature wherein highly excited (thermal) states lead to symmetry restoration, in this case  a larger density of particle-antiparticle pairs would imply a  larger value of the order parameter $|\varphi_R(\infty)|$ because the larger values of $|\varphi_R(\infty)|$ imply a larger energy transfer from the condensate to pairs as a consequence of the result (\ref{asynumb}) from energy conservation with $v_{eff}(\varphi_R(\infty))$ large and negative $|v_{eff}(\varphi_R(\infty))|\gg |v_{eff}(\varphi_R(0))|$. This mechanism is distinctly associated with fermionic degrees of freedom because of their negative contribution to
the effective potential, but is different from spontaneous symmetry breaking a l\'a Coleman-Weinberg\cite{colewein} in that it relies on the dynamics of misalignment with profuse particle production. This dynamical mechanism  of spontaneous symmetry breaking may play an important role in cosmology or perhaps  in extensions beyond the Standard Model with a large number of fermionic degrees of freedom.

\subsection{Entropy of the asymptotic state}\label{subsec:entropy}
If the asymptotic long time limit is described by the excited stationary stationary state discussed above, part of the energy originally stored in the homogeneous condensate is in the form of excited adiabatic   particle - antiparticle pairs produced with  all momenta $k$. The dynamical evolution has ``spread the information'' initially stored in the zero momentum condensate to all $k\neq 0$ momentum modes in phase space. This \emph{suggests} a growth in entropy, however time evolution is unitary and energy conserving, and in the Heisenberg picture, as discussed all throughout this study, states do not evolve in time.

The Bogoliubov transformation to the adiabatic basis (\ref{btil},\ref{ddtil}) determines that the coherent vacuum state $\ket{\Phi;0_F}$ is a squeezed state of adiabatic particle- antiparticle pairs as discussed in detail in appendix (\ref{app:bogofer}). Let us define the asymptotic values of the Bogoliubov coefficients in equations (\ref{btil},\ref{ddtil})
as
\be A_{k,s}(\infty) \equiv \cos(\theta_k) \,e^{i(\varphi_+ + \varphi_-)}~~;~~ B_{k,s}(\infty) \equiv \sin(\theta_k) \,e^{i(\varphi_+ - \varphi_-)}  \label{asybogos}\ee where
\be \cos(\theta_k) = \big|A_{k,s}(\infty)\big|~;~ \sin(\theta_k) = \big|B_{k,s}(\infty)\big| \Rightarrow 0 \leq \theta_k \leq \frac{\pi}{2} \,,\label{tetak}\ee and
$\theta_k$ does not depend on helicity (s). The results of appendix (\ref{app:bogofer}) lead to
\be \ket{\Phi;0_F} = \Pi_{\vk,s}\Bigg\{\Big[\cos(\theta_k)\Big]~ \,\sum_{n_{\vk,s}=0}^{1} \Bigg(-e^{2i\varphi_-(k)}\,\tan(\theta_k) \Bigg)^{n_{\vk,s}} \ket{n_{\vk,s};\overline{n}_{-\vk,s}}\Bigg\} \,, \label{vacext}\ee where the adiabatic fermionic   particle-antiparticle pair states are given by
\be  \ket{n_{\vk,s};\overline{n}_{-\vk,s}}  = \frac{\Big(\widetilde{b}^\dagger_{\vec{k},s}(\infty)\Big)^{n_{\vk,s}}}{\sqrt{n_{\vk,s} !}}~
\frac{\Big(\widetilde{d}^\dagger_{-\vec{k},s}(\infty)\Big)^{n_{\vk,s}}}{\sqrt{n_{\vk,s} !}}\, \ket{{\Phi;\widetilde{0}(\infty)}} ~~;~~ n_{\vk,s} = 0, 1\,, \label{excist}\ee
and  $\ket{{\Phi;\widetilde{0}(\infty)}}$ is the asymptotic adiabatic vacuum state (\ref{adiavac}).

The coherent vacuum state $\ket{\Phi;0_F}$ is a pure squeezed entangled state of adiabatic particle-antiparticle pairs of zero total momentum. The pure state density matrix in the Heisenberg picture
\be \rho =  \ket{\Phi;0_F}\bra{\Phi;0_F}  \ee does not evolve in time. Expectation values (or correlation functions) of Heisenberg field operators in this density matrix are obtained as
\be \bra{\Phi;0_F} \mathcal{O}(t)  \ket{\Phi;0_F}  = \bra{\Phi;0_F} U^{-1}(t;0)\mathcal{O}(0)U(t;0)  \ket{\Phi;0_F} = \mathrm{Tr} \rho_S(t)\,\mathcal{O}(0) \,,\label{exheis}\ee where
\be \rho_S(t) = U(t;0)\, \rho\,  U^{-1}(t;0) \,\label{rhosh}\ee is the density matrix in the Shroedinger picture wherein it depends on time but operators do not.

Let us consider a time $t \gg t^*$ where $t^*$ is the time scale at which the stationary state is reached, and the fermion Hamiltonian $H_F$ (\ref{adhami}) is constant, the Schroedinger picture density matrix becomes
\be \rho_S(t) =  U(t;t^*)) \ket{\Phi;0_F}\bra{\Phi;0_F}U^{-1}(t;t^*) \,,\label{rhos}  \ee  where $U(t;t^*)$ is the unitary time evolution operator. In the asymptotic long time limit for $t \gg t^*$
\be U(t;t^*) = e^{-i H_F(\infty) (t-t^*)} \,\label{uoft}\ee with
\be H_F(\infty) =   \sum_{i=1}^{N_f}\sum_{\vk,s} E_k(\infty) \Big[\widetilde{b}^{\,\dagger}_{\vk,s,i}(\infty)\widetilde{b}_{\vk,s,i}(\infty)+ \widetilde{d}^{\,\dagger}_{\vk,s,i}(\infty)\widetilde{d}_{\vk,s,i}(\infty) \Big] \,.\label{Hinfty}\ee where we have neglected the zero point energy which is absorbed into the effective potential.

In the asymptotic long time limit after the stationary state has been reached by the dynamics, the Schroedinger picture density matrix $\rho_S(t)$ is   given by
\be \rho_S(t) = \Pi_{\vk,s} \Pi_{\vp,s'}  \sum_{ {n}_{\vk,s}=0}^1  \sum_{ {m}_{\vp,s'}=0}^1 \mathcal{C}^*_{{m}_{\vp,s'}}(p)~ \mathcal{C}_{{n}_{\vk,s}}(k)~\ket{n_{\vk,s};\overline{n}_{-\vk,s}}\bra{m_{\vp,s'};\overline{m}_{-\vp,s'}}~
e^{2i  \big({m}_{\vp,s'}\,E_p(\infty)-{n}_{\vk,s}\,E_k(\infty) \big)(t-t^*)} \,, \label{rhosoutfer}\ee where  (see appendix (\ref{app:bogofer}))
\be \mathcal{C}_{ {n}_{\vk,s}}(k)  = \cos(\theta_k)\,\Big(-e^{2i\varphi_-(k)}\,\tan(\theta_k) \Bigg)^{ {n}_{\vk,s}} ~~;~~  {n}_{\vk,s}=0,1 \,.  \label{coefsfer}\ee The off diagonal density matrix elements, namely the coherences for $n_{k,s} \neq m_{p,s};p\neq k$ feature rapidly oscillating phases that average out in the long time limit. Formally one can perform a long time average as $\frac{1}{T}\int_{t^*}^{T-t^*} (\cdots ) dt$ as $T \rightarrow \infty$, the main point is that the rapidly varying phases with energies in  a continuum of momenta lead to decoherence of the density matrix via dephasing yielding the diagonal form
\be \rho^{(d)}_S =  \Pi_{\vk,s} \big[\cos^2(\theta_k) \big]  \sum_{n_{\vk,s}=0}^1 \Big(\tan^2(\theta_k)\Big)^{n_{\vk,s}}
\ket{n_{\vk,s};\overline{n}_{-\vk,s}}\bra{n_{\vk,s};\overline{n}_{-\vk,s}}\,. \label{diagrhofer}\ee

This can also be seen by considering the time average of expectation values (\ref{exheis}) of Heisenberg operators for $t \gg t^*$, with $\mathcal{O}(t) = U^{-1}(t;t^*) \mathcal{O}(t^*) U(t;t^*)$, namely

 \be   \frac{1}{T} \int^T_{t_0}  \mathrm{Tr}\mathcal{O}_{\delta}(t^*){\rho}_s(t) \, dt_{~~\overrightarrow{T\rightarrow \infty}~~}\mathrm{Tr}\mathcal{O}_{\delta}(t^*)\hat{\rho}^{(d)}_S \,.  \ee

The diagonal density matrix (\ref{diagrhofer}) describes a \emph{mixed state} and can be written in a familiar quantum statistical mechanics form by introducing a fiducial Hamiltonian
\be \widehat{\mathcal{H}} = \sum_{\vk,s}  \mathcal{E}_k \,\widehat{\mathbb{ N}}_{k,s}
 \,,\label{fiduHfer} \ee where
 \be \mathcal{E}_k = -\ln[\tan^2(\theta_k)]\,,\label{fiden}\ee and
 \be \widehat{\mathbb{ N}}_{k,s} \equiv  \sum_{n_{k,s}=0}^1   \,n_{ks}\,\ket{n_{\vk,s};\overline{n}_{-\vk,s}}\bra{n_{\vk,s};\overline{n}_{-\vk,s}} \,.\label{nuf}\ee
  $\widehat{\mathbb{ N}}_{k,s} $ is the operator number of particle and antiparticle pairs,
  \be \widehat{\mathbb{ N}}_{k,s} \ket{m_{\vk,s};\overline{m}_{-\vk,s}} = m_{k,s}\, \ket{m_{\vk,s};\overline{m}_{-\vk,s}}\,,\label{pairop} \ee
  with the property
   $\widehat{\mathbb{ N}}^2_{k,s} =\widehat{\mathbb{ N}}_{k,s}$,    therefore for   fixed $\vk,s$ its eigenvalues are $0,1$, and represents the extrapolation of the usual number operator of single particle Fock states to the correlated pair states.   $\widehat{\mathcal{H}}$ is a familiar free fermion Hamiltonian in the occupation basis representation, it is just that this basis is not a single particle basis but a correlated pair basis,   so that
\be \rho^{(d)}_S = \frac{e^{-\widehat{\mathcal{H}}}}{\mathcal{Z}}~~;~~ \mathcal{Z}= \mathrm{Tr}\,e^{-\widehat{\mathcal{H}}} \equiv e^{-\mathbb{F}} \,,\label{rhofidufer}\ee with $\mathbb{F}$ the fiducial free energy  and the  partition function is given by
\be \mathcal{Z} = \Pi_{\vk,s}[\cos^2(\theta_k)]^{-1} = \Pi_{\vk,s}[1+\tan^2(\theta_k)] \,. \label{zetfer}\ee

  The equivalence between (\ref{diagrhofer}) and (\ref{rhofidufer}) can be straightforwardly confirmed by comparing their matrix elements in the basis of particle-antiparticle pairs.    From the relations (\ref{asybogos}, \ref{adnumber}) it follows that
\be \tan^2(\theta_k) = \frac{n_k(\infty)}{1-n_k(\infty)} \,. \label{tan2fer}\ee Therefore, $\rho^{(d)}_S$ simply describes  a  quantum statistical canonical ensemble of free fermions (the ``temperature'' has been absorbed into the eigenvalues $\mathcal{E}_k$). The free energy is as usual
\be \mathbb{F} = U - S ~~;~~ U =  \mathrm{Tr}\rho^{d}_S \widehat{\mathcal{H}} \,,\label{frin}\ee where $U,S$ are the internal energy and entropy respectively, and as in a free
fermion case
\be U = \sum_{k,s} \frac{\mathcal{E}_k}{e^{\mathcal{E}_k}+1} = 2 \sum_{k} n_k(\infty) \,\ln\Bigg[ \frac{1-n_k(\infty)}{n_k(\infty)}\Bigg]\,,\label{Uff}\ee consequently the entropy is given by
\be S= -2\,\sum_{\vk} \Bigg\{\Big(1-n_k(\infty)\Big)\,\ln\Big(1-n_k(\infty)\Big) + n_k(\infty)\,\ln n_k(\infty)  \Bigg\} \,,\label{entro}\ee which can be confirmed from
the more general form of the Von Neumann entropy
\be S= - \mathrm{Tr} \rho^{(d)}_S \,\ln \rho^{(d)}_S  = - \sum_{\vk,s}\sum_{n_{k,s}=0}^1 p_{n_{k,s}}\, \ln p_{n_{k,s}} \,,\label{sdef}\ee with
\be  p_{n_{k,s}} = \frac{\Big(\tan^2(\theta_k)\Big)^{n_{k,s}}}{1+\tan^2(\theta_k)}\,.  \label{provs}\ee

 This argument, based on \emph{decoherence by dephasing} at long time yielding a density matrix diagonal in the ``energy'' basis underpins the  \emph{eigenstate thermalization hypothesis}\cite{sred,rigol,deutsch} and is at the heart of the arguments on thermalization in closed quantum systems, a subject of much current theoretical and experimental interest.

While decoherence by dephasing leads to  the diagonal density matrix (\ref{diagrhofer}), from which the Boltzmann entropy (\ref{entro}) follows directly, it is also
identified with the \emph{entanglement entropy} obtained by tracing over one member of the correlated pairs in the full pure state density matrix\cite{nathan,beilok}. To see this clearly, consider for example a detector that only detects particles (with for example, negative charge) but is impervious to antiparticles, tracing the pure state density matrix (\ref{rhosoutfer}) over the antiparticle states yields the \emph{reduced} mixed state density matrix
\be \rho^{(r)} =  \Pi_{\vk,s} \big[\cos^2(\theta_k) \big]  \sum_{n_{\vk,s}=0}^1 \Big(\tan^2(\theta_k)\Big)^{n_{\vk,s}}
\ket{n_{\vk,s}}\bra{n_{\vk,s}}\,,\label{rhored}\ee yielding the same entropy (\ref{entro}) as can be straightforwardly confirmed. In particular the probabilities $ p_{n_{k,s}}$ for the reduced density matrix (\ref{rhored}) are the same as for $\rho^{(d)}_S $ (\ref{diagrhofer}). In fact, any correlation function in the pure state $\ket{\Phi;0_F}$ of operators that ``measure'' only one member of the correlated pairs is equivalent to the correlation function of the operator in the mixed state reduced density matrix $\rho^{(r)}$ (if the operator only measures particles, otherwise replace $\ket{n_{k,s}} \rightarrow \ket{\overline{n}_{k,s}}$).   Therefore the entropy (\ref{entro}) is identified with the \emph{entanglement entropy} a result of tracing over one member of a \emph{correlated} pair state. With the translation $\sum_{\vk} \rightarrow \mathcal{V} \int d^3k/(2\pi)^3 $ it becomes clear that S is extensive, as is the total energy.

The emergence of an extensive Boltzmann entropy from a pure quantum state is a consequence of one single zero mode transferring energy to a continuum of $k$ modes even to large wavevectors, this continuum is also the reason for decoherence by dephasing in the long time limit.

  \section{Fermionic vs. Bosonic contributions: a comparison}\label{sec:fervsbos}

The bosonic contribution to the energy density (\ref{enerdensf}) and equations of motion for the mean field, the bracket in equation (\ref{Edot}) have been studied in detail in ref.\cite{nathan} where it is also obtained in terms of the bosonic adiabatic particle number. It is illuminating to compare the results for the fermionic contributions to the energy density and equations of motion (which dominate in the large $N_f$ limit) to those from the bosonic fluctuations as this comparison highlights fundamental differences between bosonic and fermionic degrees of freedom for the dynamics of misaligned condensates. We will only focus on the case when the tree level potential does not feature symmetry breaking, thus avoiding the important issue of spinodal instabilities, which are not germane to the main aspects of the comparison, the reader is referred to (\cite{nathan}) for an in depth discussion of that case.

Gathering the results in this reference, for the case $\mathcal{M}^2(\varphi) = v''(\varphi) > 0$ (see equations (\ref{Hamdelta}),\ref{masses}) the  contribution to the energy density from bosonic fluctuations is (see eqn. (\ref{enerdensf}))
\be \mathcal{E}_b = \frac{1}{2} \int \Big[ |\dot{g}_k(t)|^2+\omega^2_k(t)|g_k(t)|^2\Big]\,\frac{d^3k}{(2\pi)^3}  =   \frac{1}{2} \int \,\omega_k(t) \Big[1 + 2 \widetilde{\mathcal{N}}_k(t) \Big]\, \frac{d^3k}{(2\pi)^3}  \,,\label{ebos}\ee
where the mode functions $g_k(t)$ are solutions of equation (\ref{ModeTimeEvo}) and
\be \widetilde{\mathcal{N}}_k(t) = \frac{1}{2\omega_k(t)}\Big[|\dot{g}_k(t)|^2+ \omega^2_k(t) |g_k(t)|^2\Big]- \frac{1}{2}  \,\label{Nbos}\ee
 is the adiabatic particle number\cite{nathan}.

    A WKB ansatz for the bosonic mode functions $g_k(t)$ obeying the Wronskian condition (\ref{wronsk}) was analyzed in ref.\cite{nathan}
\be g_{k}(t) = \frac{e^{-i\int_0^t\,W_k(t')dt'}}{\sqrt{2W_k(t)}}\,,\label{adiadef} \ee where $W_k(t)$ is   real and the  solution of equation
  \be W^2_k(t) = \omega^2_k(t)-\frac{1}{2}\Bigg[ \frac{\ddot{W}_k}{W_k} -\frac{3}{2}\frac{\dot{W}_k^2}{W_k^2}\Bigg]\,. \label{wkbos}\ee whose solution in an adiabatic expansion is
   \be W_k^2(t) = \omega_k^2(t) \Bigg[1-\frac{1}{2}\frac{\ddot{\omega}_k}{\omega_k^3}+\frac{3}{4}\Big(\frac{\dot{\omega_k}}{\omega_k^2}\Big)^2+\cdots \Bigg]\,. \label{adiaexp1}\ee It is straightforward to confirm that this WKB solution is up to a normalization constant (fixed by the Wronskian condition (\ref{wronsk}) the same as that obtained from equation (\ref{Omeq}) by setting $\Om_k(t) \rightarrow W_k(t)$, $E_k(t) \rightarrow \omega_k(t)$ and $\dot{m}_F(t) =0$, which results in setting to zero the second term in the bracket in (\ref{imagOm}) and the last three terms in (\ref{omeR}), whereby equation (\ref{wkbos}) is precisely the same as the first three terms in (\ref{omeR}). The difference can be traced back to the term $-i\dot{m}_F(t)$ in the fermion mode equation (\ref{feq}) as compared to the bosonic mode equation (\ref{ModeTimeEvo}), which in turn is a consequence of the fact that Dirac spinors obey an equation \emph{first order in time derivatives}. These important differences yield up to  second adiabatic order\cite{nathan}
   \be  \mathcal{E}_b = \frac{1}{2} \int \omega_k(t) \,\frac{d^3k}{(2\pi)^3}  +\frac{\dot{\varphi}^2(t)}{64}\,\big(v^{'''}(\varphi(t))\big)^2\,\int\frac{d^3k}{(2\pi)^3}\frac{1}{\omega^5_k(t)} + \cdots \,, \label{ebos1} \ee unlike the fermionic case the second momentum integral is ultraviolet finite, yielding a finite wave function renormalization, with the result (assuming $  v^{''}(\varphi(t))>0\ $, see ref.\cite{nathan})
   \be \mathcal{E}_b = \frac{1}{2} \int \omega_k(t) \,\frac{d^3k}{(2\pi)^3} +   \frac{\dot{\varphi}^2}{384\pi^2}\frac{(v^{'''}(\varphi(t)))^2}{v^{''}(\varphi(t))} +\cdots  \,.\label{ebos2}\ee Comparing to the second adiabatic contribution from fermions, equation (\ref{Esecord}) we see that the bosonic contribution is ultraviolet \emph{finite}, and yields a bosonic contribution to $\mathcal{Z}[\varphi]$
   \be \mathcal{Z}_b[\varphi] =  1 + \frac{1}{192\,\pi^2}\frac{(v^{'''}(\varphi))^2}{v^{''}(\varphi)}\,,\label{Zb} \ee which coincides with the result found by functional methods in ref.\cite{ilio}\footnote{See eqn.(4.23) in this reference.}. Taking the time derivative of (\ref{ebos2}) we obtain the bosonic contribution to the equations of motion, which again reveal that the particle production contribution is finite consistently with the  finite wave function renormalization $\mathcal{Z}_b[\varphi]$.

   \vspace{1mm}

   \textbf{On initial conditions:} The WKB mode functions (\ref{adiadef}) satisfy the initial conditions
   \be g_k(0) = \frac{1}{\sqrt{2W_k(0)}}~~;~~ \dot{g}_k(0) = -i  \, \sqrt{\frac{W_k(0)}{2}}\,, \label{gini}\ee where $W_k(0)$ is the solution of eqn. (\ref{wkbos}) at $t=0$ which requires  $\ddot{\varphi}(0)$,   which, in turn, is the solution of the full equation of motion at the initial time. Therefore, just as in the fermionic case, the initialization of the dynamics even with the initial condition  $\dot{\varphi}(0)=0$, entails solving a highly non-linear self-consistent equation at the initial time, before the dynamical evolution can be studied. Instead, following reference\cite{nathan} we consider mode functions solutions of equation (\ref{ModeTimeEvo}) with the following initial conditions
   \be g_k(0) = \frac{1}{\sqrt{2\omega_k(0)}}~~;~~ \dot{g}_k(0) = -i  \, \sqrt{\frac{\omega_k(0)}{2}}\,,\label{newgini}\ee  yielding $\widetilde{\mathcal{N}}_k(0)=0$,
     hence describing the adiabatic vacuum. These mode functions are linear combinations of the WKB solution (\ref{wkbos}) and a second linearly independent solution of (\ref{ModeTimeEvo}). It is straightforward to show that the complex conjugate of the WKB solution (\ref{wkbos}) is such function, a result that can also be obtained by following the procedure of appendix (\ref{app:Icalc}). The combination that fulfills the initial conditions (\ref{newgini}) are
   \be g_k(t) = \alpha_k \,\frac{e^{-i\int_0^t\,W_k(t')dt'}}{\sqrt{2W_k(t)}}+ \beta_k \,\frac{e^{i\int_0^t\,W_k(t')dt'}}{\sqrt{2W_k(t)}}\,,\label{newgis}\ee with
   \be \alpha_k = \frac{W_k(0)+ \omega_k(0)}{2\sqrt{W_k(0)\omega_k(0)}}~~;~~ \beta_k = \frac{W_k(0)- \omega_k(0)}{2\sqrt{W_k(0)\omega_k(0)}}\,.\label{alfbet}\ee With these modes we find
   \be   |\dot{g}_k(t)|^2+\omega^2_k(t)|g_k(t)|^2 = \frac{W^2_k(t)+\omega^2_k(t)}{2W_k(t)}\,\Big( \alpha^2_k +\beta^2_k\Big) - 2\alpha_k\beta_k \, \frac{W^2_k(t)-\omega^2_k(t)}{2W_k(t)}\,\cos\Big(\int^t_0 W_k(t') dt' \Big)\,,\label{enik} \ee with the expansion (\ref{adiaexp1}) it follows that the interference (cosine) term is $\propto 1/k^7$ for large $k$ therefore \emph{there are no initial time singularities }in its time derivative, unlike the fermionic case. The origin of this difference can be traced back to the extra $-i\,\dot{m}_F$ term in the fermionic mode equation (\ref{feq}) itself a consequence of the Dirac equation being first order in time.

\subsection{Fermions and Bosons, $N_f=1$}\label{subsec:Nfeq1}

  Although in our analysis  we considered  the large $N_f$ limit to isolate and focus on the fermionic contributions, for completeness of presentation we now  extrapolate the results to $N_f=1$ by simply setting $N_f=1$ in all the quantities that feature $N_f$ in section (\ref{sec:largeN}),  and complement with  the results of reference\cite{nathan} for the bosonic contribution to the effective potential and the renormalization of couplings and masses from the bosonic contribution. Here will only consider the case $\mathcal{M}^2(\varphi)>0 $ (see equation (\ref{masses})) and refer the reader to reference\cite{nathan} for a thorough discussion of the case with $\mathcal{M}^2(\varphi)<0$ which features subtle aspects associated with spinodal (tachyonic) instabilities.
   Similarly, setting $N_f=1$ in equations  (\ref{enerdensf},\ref{Edot}), and  setting to zero the bracket in the latter one yields the complete equation of motion including the bosonic contributions.  Therefore the results of the previous sections in conjunction with those of reference\cite{nathan} define a  complete set of equations of motion for the condensate and bosonic and fermionic mode functions yielding  a self-consistent, fully renormalized and energy conserving framework to study the dynamics of misalignment with bosonic and fermionic degrees of freedom.

  Gathering the results for fermions and bosons for the case $\mathcal{M}^2(\varphi(t)) >0$  and setting $N_f=1$, the total energy density in terms of fully renormalized quantities is
  \be    \mathcal{E}   =  \frac{\dot{\varphi_R}^2}{2}\,  + v_{eff}(\varphi_R) + 4 \int E_k(t)\,n_k(t) \,\frac{d^3k}{(2\pi)^3}\Bigg|_{R}   + \int \,\omega_k(t) \, \widetilde{\mathcal{N}}_k(t)  \, \frac{d^3k}{(2\pi)^3}   =  v_{eff}(\varphi_R(0)) \,,\label{totefb} \ee the full equation of motion for the condensate
  \bea &&   \ddot{\varphi}_R(t)+\frac{d}{d\varphi_R}v_{eff}(\varphi_R(t))+ 4\,y_R \,\mathcal{P}_R(\varphi_R(t)) + v^{'''}_R(\varphi_R)\,\frac{1}{2} \int \Big[ |g_k(t)|^2- \frac{1}{2\omega_k(t)} \Big] \,\frac{d^3k}{(2\pi)^3}= \nonumber \\ &&  \frac{d}{d\varphi_R}v_{eff}(\varphi_R(t))\Big|_{t=0} \,\, \Bigg[\frac{\mathcal{I}\big[\frac{\Lambda}{\mu};\mu \,t\big]}{\mathcal{I}\big[\frac{\Lambda}{\overline{\mu}};0\big]}\Bigg]\,,\label{eomfb}  \eea where $v_{eff}$ is the fully renormalized effective potential including both the fermionic contribution of section (\ref{sec:largeN}) and the bosonic contribution obtained in reference\cite{nathan}, and $v_R(\varphi_R)$ is the tree level potential in terms of renormalized field, coupling and mass. The mode functions $g_k(t)$ obey the equation (\ref{ModeTimeEvo}) with the initial conditions (\ref{newgini}). This is a self-consistent, fully renormalized and energy conserving framework to study the misalignment dynamics of the condensate including both fermionic and bosonic fluctuations. Whether an instability of the effective potential ensues, is now a matter of whether the bosonic contribution to the effective potential is large enough to compensate for the negative fermionic one, of course this depends on the values of couplings.

   The main caveat is that for $N_f=1$, the vertex correction, the last Feynman diagram in figure  (\ref{fig:vertex}) is no longer suppressed and of the same order (one loop) as field, mass and coupling renormalizations. Vertex renormalization    is not accounted for by either field,  mass and coupling renormalizations as described in the previous sections. This   shortcoming notwithstanding, the set of equations ((\ref{totefb},\ref{eomfb}) is self-consistent, without ultraviolet divergences and energy conserving and can be implemented numerically.

   While we recognize this caveat in the case $N_f=1$, (this is one of the reasons why we focused on the large $N_f$ limit), it may be relevant beyond the one-particle irreducible framework analyzed here, perhaps at the two-particle irreducible level where vertex corrections would be necessary, and we postpone its treatment to a future study.

\section{Discussion:}\label{sec:discussion}

\textbf{The adiabatic basis is a priviledged ``pointer'' basis:} As discussed above, the derivative expansion of the effective action (\ref{effac}) is at heart an adiabatic expansion, where the effective potential is the zeroth adiabatic order. The adiabatic modes (\ref{zerof},\ref{Uspinzero},\ref{Vspinzero}) along with the field expansion (\ref{psiextad}) provides a natural basis which when combined with the adiabatic expansion  allows a clear separation of the zeroth order, yielding the effective potential.  The higher orders are directly associated with time derivatives of the mean field, namely the gradient terms in the effective action and the production of particles as defined by this adiabatic basis. Furthermore, this basis, in turn corresponds to the \emph{asymptotic eigenstates of the fermion Hamiltonian} when the dynamics settles into stationary states, as explicitly shown by equation (\ref{Hinfty}). In this sense, this adiabatic basis is a preferred, in other words, a ``pointer'' basis\cite{zurek,zurek2}. Decoherence via dephasing in this basis immediately identifies the emergent entropy, a result of particle production into a continuum of modes.

\textbf{Beyond reheating: efficient mechanism of energy transfer, entropy production and instability taming.} As mentioned in the introduction, the misalignment dynamics studied here is more general than the original mechanism proposed to solve the strong CP problem and dark matter production as it naturally also encompasses the non-equilibrium dynamics of reheating. As discussed in the previous sections, our study   reveals that the dynamics of the condensate leads to profuse particle production in a continuum of modes even for large wavevectors, which in turn leads to the emergent entropy. As such, we advocate the self-consistent non-equilibrium evolution of the condensate and the fermionic and bosonic degrees of freedom
to which it couples as an efficient mechanism of energy transfer and entropy production. While this aspect is well understood within the context of post-inflationary reheating as a consequence of the oscillatory behavior of the condensate near the minimum of the (effective?) potential, energy transfer to a wide continuum of modes is a feature that persists \emph{even in non-oscillatory dynamics}. In particular, the discussion on the possibility of an emergent  asymptotic stationary regime suggests that even in the case when the effective potential leads to (radiatively induced) instabilities, profuse particle production ``tames'' this instability leading to a highly excited but stationary state, in which the
contribution from particle production restores the stability of the asymptotic state and possibly leading to spontaneous symmetry breaking - a consequence of the instability combined with misalignment dynamics yielding profuse particle production as discussed in section (\ref{subsec:ssb}).

\textbf{Pauli Blocking:} Although Pauli blocking, which is manifest in the relation (\ref{unitary}) between Bogoliubov coefficients results in a less efficient mechanism of energy transfer for fermions as compared to bosons,  the above analysis clearly shows that production of fermion particle-antiparticle pairs from the dynamical evolution of the condensate leads to a substantial transfer of energy for all values of the momenta.   Perhaps this should not be surprising, even in a thermal bath for temperatures much larger than the mass, the total number of fermionic particles differs from bosonic ones by a factor of order one, even when there is a Bose-enhancement at low momenta and fermionic states are Pauli blocked. Either species features a total number of particles and entropy $\propto T^3$ and an energy density $\propto T^4$ differing in factors of order one only. As the relation (\ref{asynumb}) clearly demonstrates, if a stationary state emerges, there is a substantial energy \emph{density} transferred into particle-antiparticle production, namely the total number of produced pairs is proportional to the volume. The occupation of all momentum states also leads to an extensive entropy, which is interpreted as the entanglement entropy associated with tracing over one member of the correlated pairs.

\textbf{Initial time singularity: dynamics of dressing in an initial value problem:} The initial time singularity discussed in section (\ref{subsec:inicon}) is a noteworthy aspect of setting initial conditions such that the adiabatic vacuum coincides with the vacuum state $\ket{\Phi;0_F}$ at the initial time, namely there are no adiabatic particles in this state at the initial time. It is a consequence of the ultraviolet divergence associated with wave function renormalization and as discussed above is a dynamical consequence of ``dressing'' of the renormalized physical fields out of the bare fields by the quantum fluctuations. In references\cite{baacke,inising1,inising2,inising3} this initial time singularity was recognized and argued to be cancelled by redefining the vacuum state at the initial time, and in ref.\cite{baacke} a Bogoliubov transformation was proposed to implement this redefinition. Sorting out the technical details of these references, it becomes clear that such redefinition in fact amounts to imposing initial conditions on the fermionic mode functions given by equation (\ref{inif}), with $\Omega_k(t)$ being the exact solution of equation (\ref{Omeq}). In fact the analysis in the previous sections confirms that indeed there is no initial time singularity with these initial conditions, however as argued in section  (\ref{subsec:inicon})   the initial condition (\ref{inif}) is tantamount to solving the fully renormalized self-consistent initial value problem at $t=0$, which obviously solves for the ``fully dressed' fields. By renormalizing the fields with the time independent wave function renormalization the remnant of this initial time singularity is a \emph{finite} time dependence in the equation of motion that describes the dynamics of dressing on a time scale $\propto 1/\Lambda$.

The reason that an initial time singularity does not appear in the bosonic case is that for a renormalizable bosonic theory, wave function renormalization arises at two loop level, whereas in the fermionic theory it is at one loop. Therefore, we expect that in a two-particle irreducible formulation\cite{cornwall} of non-equilibrium dynamics\cite{ferpre5,berges,serreau} with initial conditions equivalent to (\ref{iniE}) the initial time singularities will emerge, and while the ultraviolet singularity is resolved by field renormalization there remains a time dependent finite remnant associated with the dynamics of ``dressing''.

This initial time behavior is unavoidable in an initial value problem in a renormalizable quantum field theory with an ultraviolet divergent wave function renormalization. In formal quantum field theory the adiabatic switching-on procedure by ``turning on'' the interaction from $t \rightarrow -\infty$ is precisely designed to avoid the problem of preparation of an initial state in a theory with infinite number of degrees of freedom. Since misalignment dynamics is fundamentally an initial value problem this procedure is not compatible with the dynamical evolution, and obviously much less in a cosmological setting.

\textbf{Particle production: dynamical ``cascades''} The analysis of the previous section shows that the time evolution of the condensate leads to the production of particle-antiparticle pairs (defined in the adiabatic basis) for all values of the momenta $k$, with a distribution function $n_k(t) = \overline{n}_k(t)$ given by equation (\ref{adnumber}). For $\dot{\varphi} \neq 0$ the large $k$ behavior of distribution  function is $n_k(t) \propto \dot{\varphi}^2(t)/k^4$, yielding a finite number of produced particles, but an ultraviolet  logarithmically divergent energy density. This is the origin of wave function (field) renormalization.  As the stationary state is approached $\dot{\varphi}\rightarrow 0$ the leading momentum behavior of the distribution function becomes $n_k(t)\propto 1/k^6$. This transfer of energy from a ``zero mode'' condensate to all momentum modes with vanishing occupation initially,  is similar to a direct energy cascade. If higher order processes such as collisions  lead to thermalization with
a distribution function $\propto e^{-\beta k}$ for large $k$ this means a smaller occupation as compared to that in the self-consistent one loop $\propto 1/k^6$. This entails a redistribution of energy from large to smaller momenta by the thermalization process, which may be interpreted as an inverse energy cascade. An analysis of similar processes within the context of post-inflationary reheating was carried out in ref.\cite{ferpre5}. However, one of our main points is that a ``direct cascade'' with power law distributions at large $k$ is a general result of misalignment dynamics irrespective of whether the scalar condensate oscillates or ``rolls down'' a potential (stable or unstable) without oscillations.

\textbf{Possible cosmological consequences:} Although our study focused on the dynamics of the scalar field in Minkowski space time, several aspects reveal potential impact in cosmology: \textbf{i:)} at the level of the effective potential, the detailed calculation leading to the effective potential (\ref{finVeff}) highlights the role of the quantization of fermions and bosons, in particular the mode functions featuring the exponentials $e^{\pm i \omega_k t};e^{\pm i E_k t}$ which are solutions of the wave equations in Minkowski space time, lead to the $\omega_k;E_k$ integrands in the effective potential (\ref{finVeff}).  In a cosmological setting the solutions of the wave equations even in a spatially flat Robertson-Walker cosmology, are \emph{not} these simple exponentials. For example during the inflationary stage the mode functions are Hankel functions, and during radiation domination they are parabolic cylinder functions\cite{natboy}, hence even the ``static'' effective potential will feature an explicit time dependence through the correct mode functions.    \textbf{ii:)} The equation of motion for the condensate features a ``cosmological friction term'' $3 H(t) \dot{\varphi}(t)$ consistent with covariant energy conservation, however as per the previous comment, the effective potential contribution features an \emph{explicit} time dependence from the exact mode functions. \textbf{iii:)} Typically it is argued that when the scalar condensate begins oscillating the equation of state (upon some time averaging) corresponds to a pressureless  fluid consistently with a cold dark matter component, but this ignores the contribution to the equation of state from particle production discussed in the previous sections. This contribution may yield large isocurvature perturbations since particle production is proportional to the volume and is directly related to entropy production. \textbf{iv:)} the ``two-step'' mechanism of spontaneous symmetry breaking induced by radiative corrections of the static effective potential a l\'a Coleman-Weinberg combined with misalignment dynamics on the unstable side of the potential may provide alternative intriguing possibilities for novel phases associated with highly excited states. All of these aspects merit further, and deeper study in the cosmological context.

\textbf{On Finite Temperature:} As stated in our objectives, we restricted our study to zero temperature to clearly understand particle and entropy production as a consequence of the non-equilibrium dynamical evolution of the condensate unencumbered by thermal effects. In ref. (\cite{nathan}), it was argued that rapid evolution of the condensate during misalignment dynamics may lead to a breakdown of local thermodynamic equilibrium, in which case the initial value problem with an initial thermal density matrix can be studied by modifying the initial state, rather than the vacuum for the fluctuations, one must consider a thermal equilibrium density matrix \emph{at the initial time}. This translates into stimulated production of bosonic particles, and Pauli blocking of fermionic ones. Furthermore the initial thermal equilibrium density matrix yields a thermal entropy. An unanswered question in ref.\cite{nathan} is how the emerging entropy from particle production combines with the thermal entropy associated with the initial ensemble. This is an important question that will be addressed, along with the consequences of an initial thermal state, in a future study.

\section{Conclusions and Further Questions:}\label{sec:conclusions}

Misalignment dynamics, the non-equilibrium evolution of a scalar (or pseudoscalar) condensate in a potential landscape, broadly encompasses a solution to the strong CP problem, provides a compelling mechanism of cold dark matter production and (pre) reheating the post inflationary Universe. As such it is of consequential and ubiquitous importance in cosmology. It is usually studied as an initial value problem, establishing initial conditions on the scalar condensate and evolving its equation of motion. The effect  of other degrees of freedom to which the scalar or pseudoscalar field couples is typically included phenomenologically by replacing the tree level potential by an effective potential that includes radiative (loop) corrections. By definition and construction the effective potential is a \emph{static} quantity, at zero or finite temperature its usefulness is restricted to equilibrium aspects such as the determination of the lowest (free) energy of a theory to identify possible phases and symmetry breaking.  However, recent studies have revealed the limitations of the familiar effective potential to capture the \textit{dynamics} of a (misaligned) scalar condensate \cite{nathan}.  These studies have focused on the bosonic fluctuations. Motivated by both the importance of dynamical condensates in cosmology and the fact that many models involving such condensates feature couplings to fermion fields, in this article we have investigated the real-time dynamics of a misaligned scalar condensate with Yukawa couplings to $N_f$ identical fermion fields in Minkowski space time as a prelude towards its extension to cosmology.

Yukawa interactions bring at least two important subtleties: \textit{i:)} their contribution to the effective potential is negative implying the possibility of instabilities, \textit{ii:)}   ultraviolet divergent field (wave function) renormalization at the one-loop level, in contrast to bosonic degrees of freedom. In the phenomenological approach to misalignment dynamics, which only considers radiative corrections to the effective potential, field renormalization must be introduced   ``by hand''.

We implemented a Hamiltonian framework leading to  manifestly energy conserving, self-consistent and fully renormalized equations of motion.  By considering the large $N_f$ limit, we suppress the contributions from purely scalar fluctuations and isolate the impact of the fermion degrees of freedom, however, the lessons learned are more general and apply broadly to  the cases wherein both fermions and bosons couple to the homogeneous condensate.

We introduce the adiabatic ``particle" basis  of instantaneous eigenstates of the fermion Hamiltonian, which naturally leads to an adiabatic expansion which is at the heart of the gradient expansion of the effective action. The zeroth order in this expansion yields the effective potential and the higher order terms are associated with the profuse production of ``adiabatic particles'' sourced by the dynamical evolution of the condensate and correspond to the gradient terms in the effective action.  Radiative corrections induce \emph{maxima} and instabilities in the effective potential in a manner similar to a Coleman-Weinberg mechanism.  We show that field renormalization, namely the ``dressing'' of the bare fields by quantum fluctuations, emerges naturally as a consequence of particle production.

  The dynamical evolution   requires setting initial conditions on the condensate $\varphi(t)$ and fermion mode functions  and their first time-derivatives. This turns out to be highly non-trivial:  generic initial conditions on the fermionic mode functions lead to an \textit{initial misalignment between the vacuum state and the adiabatic vacuum} and
  imply the self-consistent solution of a highly non-linear initial value problem.   We resolved this conundrum by stipulating initial conditions such that these two vacuua are initially aligned, and the condensate energy is given entirely by the effective potential at $t=0$. This leads to a well-defined initial value problem and a self-consistent, energy conserving set of equations directly amenable to numerical study. We demonstrate that the trade-off for adopting these more natural initial conditions is the emergence of an initial time singularity which we contend is associated with the physics of field dressing on a time scale $\propto 1/\Lambda$ where $\Lambda$ is an ultraviolet cutoff, which, however,  is self-consistently resolved by field, coupling and mass renormalizations. We argue that dynamical evolution in all renormalizable quantum field theories will feature these subtle aspects.

 Energy conservation entails that during the  non-equilibrium evolution of the condensate  down the potential hill energy is drained from the ``classical'' homogeneous condensate $\varphi(t)$  and transferred to quantum fluctuations in the form of particle production,    leading us to conjecture on the emergence of highly excited asymptotic stationary states characterized by $\dot{\varphi}(\infty) =0~;~\ddot{\varphi}(\infty) =0$.  The dynamical transfer of energy occurs as in a cascade: during the evolution the large $k$ behavior of the distribution function $n_k(t) \propto (\dot{\varphi}(t))^2/k^4$,  yielding a finite total number of produced particles but is responsible for the ultraviolet divergent field renormalization. At asymptotically long time when the stationary state is reached the large $k$ behavior becomes $n_k(\infty) \propto (\ddot{\varphi}(0))^2/k^6$. Therefore, regardless of Pauli blocking, particle-antiparticle pairs are created with all momenta. This transfer of energy from one ``zero mode'' to all $k$-modes   is suggestive of a spread of information across phase space and the production of entropy. Yet the dynamics of the system are carried out via the explicitly \textit{unitary}, Heisenberg picture, seemingly prohibiting entropy growth. Careful analysis of the density matrix of the asymptotic stationary state in the Schrodinger picture reveals the effects of decoherence by dephasing and an attendant \textit{entanglement entropy} in the  adiabatic basis, which is argued to be a ``privileged'' or pointer basis.

 In the case when the initial value of the  condensate leads to evolution in the unstable branch of the effective potential, we argue that this instability along with particle production leads to a stationary highly excited state featuring spontaneous symmetry breaking, an indirect consequence of both the radiatively induced \emph{maxima} of the effective potential \emph{and} profuse particle production with energy conserving dynamics even when the tree level potential does not feature symmetry breaking minima.

While our study was conducted in Minkowski spacetime, we anticipate ramifications for cosmology. In the regime where the expansion rate is small compared to the frequencies of the fermion mode functions, the adiabatic approximation can again be leveraged, and we anticipate the higher order adiabatic terms will induce profuse particle production across all momenta. However, Friedmann-Robertson-Walker spacetime does not feature a global time-like Killing vector, and so energy is not conserved. This makes identifying the emergence of an asymptotic dynamical fixed point as a consequence of energy transfer from condensate to fluctuations more subtle. However, the main results obtained here would point out that the misalignment mechanism does not yield ``cold dark matter'' but a highly excited state from particle production which  contributes to the energy momentum tensor along with the condensate component. Furthermore, the growth of entropy associated with particle production suggests the emergence of  isocurvature perturbations. Thus, the full consequences for cosmological spacetimes warrants further study which we intend to report on in future work. Additionally, extending this Minkowksi treatment to include the effects of gauge theories will mandate careful analysis of gauge invariance in both the dynamics and renormalization aspects. This is clearly beyond the scope of our present study and is a subject for future work.

While we are not aware of alternative approaches to obtaining the correct equations of motion, perhaps the avenue advocated in ref.\cite{dunne2} may prove to be such an alternative if it can be shown to lead to a manifestly energy conserving and fully renormalized set of equations of motion for the condensate and bosonic and fermionic fluctuations.

 Finally, it is worth noting that similar considerations arise within the Higgs sector of the empirically vindicated Standard Model. In the Standard Model, dynamical scalar condensates are central to particle physics: the Higgs mechanism relies on a spatially homogeneous scalar vacuum expectation value that generates fermion masses through Yukawa couplings and spontaneously breaks the electroweak symmetry. Although the Higgs vacuum expectation value is static after the electroweak phase transition, the formation of the condensate itself could have involved a period during which the Higgs field evolved toward the minimum of its symmetry-breaking potential. In principle, the particle-production mechanisms analyzed in this work apply equally to such transient Higgs dynamics and would generically lead to the nonperturbative production of Standard Model fermion–antifermion pairs during the relaxation of a displaced Higgs condensate. Important caveats, however, limit the direct applicability of these results: our analysis assumes $N_f$ fermion species with identical Yukawa couplings and neglects interactions with gauge bosons, whereas in the Standard Model the Yukawa couplings span many orders of magnitude and gauged vector interactions can strongly influence the dynamics. A dedicated treatment of Higgs-sector relaxation that includes these effects lies beyond the scope of the present work but represents a natural extension of this framework.

\acknowledgements
 D.B.  gratefully acknowledges  support from the U.S. National Science Foundation (NSF) through grants   NSF 2111743 and NSF 2412374.

\appendix

\section{Majorana fermions.}\label{app:majorana}

In this appendix we gather the main features for the quantization of Majorana fermions. With the solutions of the Dirac equation obtained in section (\ref{subsec:fermodes}), we construct self-conjugate Majorana fermions as follows.

\be U_\lambda(\vk,t) = N_k  \, \left(
                            \begin{array}{c}
                              \mathcal{F}_k(t)\, \chi_\lambda \\
                              \vec{\sigma}\cdot \vec{k}\, f_k(t)\, \chi_\lambda \\
                            \end{array}
                          \right) ~~;~~ \chi_1 = \left(
                                                   \begin{array}{c}
                                                     1 \\
                                                     0 \\
                                                   \end{array}
                                                 \right) \;; \; \chi_2 = \left(
                                                                       \begin{array}{c}
                                                                         0 \\
                                                                         1 \\
                                                                       \end{array}
                                                                     \right) \,,
 \label{Uspinorsolm} \ee  and

 \be V_\lambda(\vk,t) = N_k   \, \left(
                            \begin{array}{c}
                               \vec{\sigma}\cdot \vec{k}\,f_k(t)\, \varphi_\lambda \\
                               \mathcal{F}^*_k(t)\, \varphi_\lambda  \\
                            \end{array}
                          \right) ~~;~~ \varphi_1 = \left(
                                                   \begin{array}{c}
                                                    0 \\
                                                     1 \\
                                                   \end{array}
                                                 \right) \;; \; \varphi_2 = -\left(
                                                                       \begin{array}{c}
                                                                         1 \\
                                                                         0 \\
                                                                       \end{array}
                                                                     \right) \,,
 \label{Vspinorsolm} \ee
 where $\lambda = 1,2$.  These spinors are orthonormalized as in equation (\ref{ortot}).

    It is straightforward to confirm that  the $U$ and $V$ spinors (\ref{Uspinorsolm}, \ref{Vspinorsolm}) obey the charge conjugation relation
 \be i\gamma^2 U^*_\lambda(\vk,t) = V_\lambda(\vk,t) ~~:~~ i\gamma^2 V^*_\lambda(\vk,t) = U_\lambda(\vk,t) ~~;~~ \lambda = 1,2 \,. \label{chargeconj}\ee In terms of these spinor solutions we can construct Majorana (charge self-conjugate) fields obeying\footnote{We set the Majorana phase to zero as it is not relevant for the discussion.}
 \be \psi^c_{M,i}(\vx,t) = C (\overline{\psi}_{M,i}(\vx,t))^T  = \psi_{M,i}(\vx,t) ~~;~~ C = i\gamma^2\gamma^0 \label{majorana} \ee and given by
\be
\psi_{M,i}(\vec{x},t) =    \frac{1}{\sqrt{V}}
\sum_{\vec{k},\lambda}\,   \left[b_{\vec{k},\lambda,i}\, U_{\lambda}(\vec{k},t)\,e^{i \vec{k}\cdot
\vec{x}}+
b^{\dagger}_{\vec{k},\lambda,i}\, V_{\lambda}(\vec{k},t)\,e^{-i \vec{k}\cdot
\vec{x}}\right] \; .
\label{psiexmajo}
\ee
In the case of Majorana fields the     Lagrangian density, and the energy momentum tensor  must be multiplied by a factor $1/2$ since a Majorana field has half the number of degrees of freedom of the Dirac field. Furthermore, one can take linear combinations of the Weyl spinors $\chi_{1,2};\varphi_{1,2}$ and construct helicity eigenstates. The steps leading to the main results are the same as for the Dirac case, with the only difference being a factor $2$ instead of the factor $4$ because for a Majorana field particles are the same as antiparticles therefore $n_k=\overline{n}_k$, thereby halving the number of degrees of freedom. In the expansion of the state $\ket{\Phi,0_F}$ in terms of the asymptotic adiabatic states (see equation (\ref{excist})) the operator $\widetilde{d}^\dagger_{-\vk,s} \rightarrow \widetilde{b}^\dagger_{-\vk,s}$. In conclusion, Majorana instead of Dirac fermions only change the results quantitatively by a factor 2 but there are no major conceptual differences.

\section{Equation of motion up to second adiabatic order:}\label{app:eom}
For $f_k(t)$ given by  the WKB ansatz (\ref{wkb}), we find
\be    V^\dagger_{-\vk,s}(t)\gamma^0 V_{-\vk,s}(t) = N^2_k |f_k(t)|^2 \, \Big[k^2-\big(\Om^*_k(t)+m_F(t) \big)\big(\Om_k(t)+m_F(t) \big)  \Big] \,,\ee and using the result (\ref{newnorm}) yields
\be   V^\dagger_{-\vk,s}(t)\gamma^0 V_{-\vk,s}(t) =  \frac{\Big[k^2-\big(\Om^*_k(t)+m_F(t) \big)\big(\Om_k(t)+m_F(t) \big)  \Big]}{\Big[k^2+\big(\Om^*_k(t)+m_F(t) \big)\big(\Om_k(t)+m_F(t) \big)  \Big]}\,.\label{VV}  \ee The numerator becomes
\bea \textrm{numerator} & = &  -\Big[ 2m_F(t)\big(E_k(t)+m_F(t)\big)+ \Big(\big(\Om^2_{Rk}(t)-E^2_k(t)\big)+\Om^2_{Ik}(t)+ 2m_F(t)\big( \Om_{Rk}(t)-E_k(t) \Big)\Big] \nonumber \\  & = & -2m_F(t)\big(E_k(t)+m_F(t)\big)\, \Bigg[ 1+ \underbrace{\frac{\Big(\big(\Om^2_{Rk}(t)-E^2_k(t)\big)+\Om^2_{Ik}(t)+ 2m_F(t)\big( \Om_{Rk}(t)-E_k(t) \Big)}{2m_F(t)\big(E_k(t)+m_F(t)\big)}}_{ 2^{nd}~order +\cdots}\Bigg] \nonumber \eea and the denominator becomes

\bea \textrm{denominator} & = &   \Big[ 2E_k(t)\big(E_k(t)+m_F(t)\big)+ \Big(\big(\Om^2_{Rk}(t)-E^2_k(t)\big)+\Om^2_{Ik}(t)+ 2m_F(t)\big( \Om_{Rk}(t)-E_k(t) \Big)\Big] \nonumber \\  & = &  2E_k(t)\big(E_k(t)+m_F(t)\big)\, \Bigg[ 1+ \underbrace{\frac{\Big(\big(\Om^2_{Rk}(t)-E^2_k(t)\big)+\Om^2_{Ik}(t)+ 2m_F(t)\big( \Om_{Rk}(t)-E_k(t) \Big)}{ 2E_k(t)\big(E_k(t)+m_F(t)\big)}}_{ 2^{nd}~order +\cdots}\Bigg] \,. \nonumber \eea Keeping terms up to second adiabatic order, the relations (\ref{realOm},\ref{imagOm}) yields

\be \sum_{s=\pm1}\, V^\dagger_{-\vk,s}(t)\gamma^0 V_{-\vk,s}(t)  =  -2 \,\frac{m_f(t)}{E_k(t)}\Bigg[1+ \frac{\mathcal{R}(t)}{2m_F(t)E_k(t)}\,\Big[\frac{E_k(t)-m_F(t)}{E_k(t)+m_F(t)}  \Big]  \Bigg] \,,\ee where up to second order
\be \mathcal{R}(t)  = 2\,\Om^2_{Ik}(t)\,\Big(1+ \frac{m_f(t)}{2E_k(t)} \Big)+ \dot{\Om}_{Ik}(t)\, \Big(1+ \frac{m_f(t)}{E_k(t)} \Big)\,, \ee
with
\be \Om^2_{Ik}(t) \simeq \frac{\dot{m}^2_F(t)}{4E^2_k(t)}+ \cdots~~;~~ \dot{\Om}_{Ik}(t) \simeq - \frac{\ddot{m}_F(t)}{2E_k(t)}+ \cdots \,\label{apx}\ee  The dots stand for terms that do not yield ultraviolet logarithmic divergences, which are completely determined by the second term in (\ref{apx}), yielding for the equation of motion (\ref{eomNf})
\be \ddot{\varphi}(t)+ \underbrace{v'(\varphi(t))-2 y^2_0 \varphi(t)\int \frac{1}{E_k(t)} \,\frac{d^3k}{(2\pi)^3}}_{\frac{d}{d\varphi}\, v_{eff}(\varphi(t))} +\frac{ y^2_0}{2} \ddot{\varphi}(t) \,\int \frac{k^2}{E^5_k(t)}\,\frac{d^3k}{(2\pi)^3} + \cdots =0  \,,\label{appeom}\ee where the dots stand for higher order terms that do not contribute to the logarithmic ultraviolet divergence and $v_{eff}$ is given by (\ref{finbVeff}).
Carrying out the integral of the last term with an ultraviolet cutoff $\Lambda$, we find
\be \ddot{\varphi}(t)\, \Big[1 + \frac{y^2_0}{4\pi^2}\,\ln\Big( \frac{2\Lambda}{|m_f(t)|}\Big) \Big]   + \frac{d}{d\varphi}\,v_{eff}(\varphi) +\cdots     = 0\,,\label{secderfi} \ee  where again the dots stand for finite terms. Carrying out the same steps leading up to equation (\ref{Esecordfin}) and using the renormalization conditions (\ref{rentotfin}) we finally obtain the equation (\ref{eomZ}) quoted in the main text.

\section{Calculation of $f_{2k}(t)$ in the adiabatic approximation:}\label{app:Icalc}

Let us define
\be I(t) \equiv \int^t_0  e^{2i \int^{t'}_0 \Om_k(t'')dt''}\, {dt'} \,,\label{Int}\ee  where using equation (\ref{imagOm})
\be I(t) = \frac{1}{\Om_{Rk}(0)} \, \int^t_0 \underbrace{ \Om_{Rk}(t')\, e^{2i \int^{t'}_0 \Om_{Rk}(t'')dt''}}_{\frac{1}{2i}\,\frac{d}{dt'}\,\Big(e^{2i \int^{t'}_0 \Om_{Rk}(t'')dt''}\Big)}\,\,   e^{\int^{t'}_0 \frac{\dot{m}_F(t'')}{\Om_{Rk}(t'')}dt''}\, {dt'} \,,\label{Int2}\ee integrating by parts
\bea I(t)  &  =  &  \frac{1}{2i \Om_{Rk}(0)}\,  \Bigg\{   e^{2i \int^{t}_0 \Om_{Rk}(t')dt'}\, e^{\int^{t}_0 \frac{\dot{m}_F(t')}{\Om_{Rk}(t')}dt'} -1  \nonumber \\ &  -  & \int^t_0\,\underbrace{ \Om_{Rk}(t')\, e^{2i \int^{t'}_0 \Om_{Rk}(t'')dt''}}_{\frac{1}{2i}\,\frac{d}{dt'}\,\Big(e^{2i \int^{t'}_0 \Om_{Rk}(t'')dt''}\Big)}\,\,   e^{\int^{t'}_0 \frac{\dot{m}_F(t'')}{\Om_{Rk}(t'')}dt''}\,  \,\Big(\frac{\dot{m}_F(t')}{\Om^2_{Rk}(t')}\Big) \,{dt'}  \Bigg\} \,. \label{int3}\eea Integrating again by parts and using $\dot{m}_F(0) =0$ as the initial condition for the condensate, we find
\be I(t) =  \frac{1}{2i \Om_{Rk}(0)}\,  \Bigg\{-1+    e^{2i \int^{t}_0 \Om_{Rk}(t')dt'}\, e^{\int^{t}_0 \frac{\dot{m}_F(t')}{\Om_{Rk}(t')}dt'}\Big[1 + i \frac{\dot{m}_F(t)}{2\Om^2_{Rk}(t)}+ \cdots \Big]\Bigg\} \,\label{intc}\ee where the dots stand for higher adiabatic orders that can be obtained by subsequent integration by parts following the above strategy.

\section{Bogoliubov transformation for Fermionic fields}\label{app:bogofer}

In this section we summarize the general framework for Bogoliubov transformations of fermion fields.
Since the Bogoliubov transformations (\ref{btil},\ref{ddtil}) are the same for all flavors,   we will suppress the flavor label ``i'' in the annihilation and creation operators, writing these Bogoliubov transformations simply as

\bea \widetilde{b}_{\vec{k},s} & = &   {b}_{\vec{k},s} A_{k,s} - {d}^{\dagger}_{-\vec{k},s} B^*_{k,s} \label{btilapp} \\
\widetilde{d}^{\,\dagger}_{-\vec{k},s} & = &   {d}^{\dagger}_{-\vec{k},s} A^*_{k,s} + {b}_{\vec{k},s} B_{k,s} \,, \label{ddtilapp} \eea The Bogoliubov coefficients fulfill the unitarity condition (\ref{unitary}), therefore  we parametrize them as

\be A_{k,s} = \cos(\theta_k) \,e^{i(\varphi_+ + \varphi_-)}~~;~~ B_{k,s} = \sin(\theta_k) \,e^{i(\varphi_+ - \varphi_-)} \label{ABdefs}\ee  for the case corresponding to the Bogoliubov coefficients (\ref{ACbogen},\ref{BDbogen}), the angle $\theta_k$ does not depend on helicity (s) and the $k,s$  arguments of the phases are implicit. We now absorb the phases into a redefinition (canonical transformation) of the various operators,
\bea  \widetilde{b}_{\vec{k},s}\,e^{-i\varphi_-} \rightarrow  \widetilde{b}_{\vec{k},s}  ~~;~~ \widetilde{d}^{\,\dagger}_{-\vec{k},s}\,e^{i\varphi_-} \rightarrow \widetilde{d}^{\,\dagger}_{-\vec{k},s}  \nonumber \\ {b}_{\vec{k},s}\,e^{i\varphi_+} \rightarrow {b}_{\vec{k},s}   ~~;~~ {d}^{\,\dagger}_{-\vec{k},s}\,e^{-i\varphi_+} \rightarrow  {d}^{\,\dagger}_{-\vec{k},s}   \,.  \label{bdphases}\eea In terms of these redefinitions the Bogoliubov transformations (\ref{btil},\ref{ddtil}) read
\bea \widetilde{b}_{\vec{k},s} & = &   {b}_{\vec{k},s} \cos(\theta_k) - {d}^{\dagger}_{-\vec{k},s} \sin(\theta_k) \label{btil2} \\
\widetilde{d}^{\,\dagger}_{-\vec{k},s} & = &   {d}^{\dagger}_{-\vec{k},s} \cos(\theta_k) + {b}_{\vec{k},s} \sin(\theta_k) \,. \label{ddtil2} \eea The inverse transformation is
\bea  {b}_{\vec{k},s} & = &   \widetilde{b}_{\vec{k},s} \cos(\theta_k) + \widetilde{d}^{\dagger}_{-\vec{k},s} \sin(\theta_k) \label{btil2inv} \\
 {d}^{\,\dagger}_{-\vec{k},s} & = &   \widetilde{d}^{\dagger}_{-\vec{k},s} \cos(\theta_k) - \widetilde{b}_{\vec{k},s} \sin(\theta_k) \,. \label{ddtil2inv} \eea
  It is convenient to define the antihermitian operator
\be \gamma_{\vk,s} = \widetilde{b}^\dagger_{\vec{k},s}\,\widetilde{d}^{\dagger}_{-\vec{k},s} - \widetilde{d}_{-\vec{k},s}\,\widetilde{b}_{\vec{k},s}~~;~~ \gamma^\dagger_{\vk,s} = - \gamma_{\vk,s}\,, \label{gam}\ee in terms of which, this inverse transformation is generated by the unitary operator
 \be S[\theta_k] = \exp\big\{-\theta_k\,\gamma_{\vk,s}  \big\} \,,  \label{Sferop}\ee namely
 \bea {b}_{\vec{k},s} & = &  S[\theta_k]\,\widetilde{b}_{\vec{k},s}\,S^{-1}[\theta_k]\label{bofteta}\\
{d}^{\,\dagger}_{-\vec{k},s} & = &   S[\theta_k]\,\widetilde{d}^\dagger_{-\vec{k},s}\,S^{-1}[\theta_k]\,. \label{ddofteta}\eea To see that this is the case, consider the definitions
 \bea \alpha(\theta_k)  & = &  S[\theta_k]\,\widetilde{b}_{\vec{k},s}\,S^{-1}[\theta_k]\label{alfa}\\
\beta(\theta_k) & = &   S[\theta_k ]\,\widetilde{d}^\dagger_{-\vec{k},s}\,S^{-1}[\theta_k]\,. \label{beta}\eea Using the anticommutation relations we find
\bea \frac{d\alpha(\theta_k)}{d\theta_k} & = & \beta(\theta_k) \label{dalfa}\\
\frac{d\beta(\theta_k)}{d\theta_k} & = & -\alpha(\theta_k)\,, \label{dbeta}\eea
with the ``initial conditions''
\bea \alpha(0) & = & \widetilde{b}_{\vec{k},s}~~;~~ \frac{d\alpha(\theta_k)}{d\theta_k}\Big|_{\theta_k=0} = \beta(0) = \widetilde{d}^\dagger_{-\vec{k},s} \label{inialfa}\\
\beta(0) & = & \widetilde{d}^\dagger_{-\vec{k},s}~~;~~ \frac{d\beta(\theta_k)}{d\theta_k}\Big|_{\theta=0} = -\alpha(0) = -\widetilde{b}_{\vec{k},s} \label{inibeta} \,. \eea The solutions of equations (\ref{dalfa},\ref{dbeta}) with the initial conditions (\ref{inialfa},\ref{inibeta}) are given by
\bea \alpha(\theta_k)  & = &  \widetilde{b}_{\vec{k},s} \,\cos(\theta_k) + \widetilde{d}^\dagger_{-\vec{k},s}\,\sin(\theta_k) = {b}_{\vec{k},s} \label{alfasolu}\\
\beta(\theta_k)  & = &  \widetilde{d}^\dagger_{-\vec{k},s} \,\cos(\theta) - \widetilde{b}_{\vec{k},s}\,\sin(\theta) = {d}^{\,\dagger}_{-\vec{k},s}\,,  \label{betasolu} \eea hence, confirming the relations (\ref{bofteta},\ref{ddofteta}).

A simple expression for the unitary operator $S[\theta]$ is recognized from its expansion,

\be S[\theta_k] = 1-\theta_k\,\gamma_{\vk,s} + \frac{1}{2!}\,\theta^2_k\,\gamma^2_{\vk,s} + \frac{1}{3!}\,\theta^3_k\,\gamma^3_{\vk,s} + \cdots \label{sexp}\ee Using the canonical anticommutation relations we find
\be \gamma^2_{\vk,s} = - \Big[ \widetilde{b}^\dagger_{\vec{k},s}\, \widetilde{b}_{\vec{k},s}\, \widetilde{d}^\dagger_{-\vec{k},s}\, \widetilde{d}_{-\vec{k},s}+ \widetilde{d}_{-\vec{k},s}\, \widetilde{d}^\dagger_{-\vec{k},s}\,\widetilde{b}_{\vec{k},s}\, \widetilde{b}^\dagger_{\vec{k},s}  \Big] = - \mathbf{P}_{\vk,s}\,. \label{gam2}\ee $\mathbf{P}_{\vk}$ is a projection operator, which in terms of \be \widetilde{b}^\dagger_{\vec{k},s}\, \widetilde{b}_{\vec{k},s} = \widehat{n}_{\vk,s}~~;~~\widetilde{d}^\dagger_{-\vec{k},s}\, \widetilde{d}_{-\vec{k},s} = \widehat{\overline{n}}_{-\vk,s}\,, \label{ns}\ee
  may also be written as
\be \mathbf{P}_{\vk} = \widehat{n}_{\vk,s}\,\,\,\widehat{\overline{n}}_{-\vk,s}+ (1-\widehat{n}_{\vk,s})\,(1-\widehat{\overline{n}}_{-\vk,s})~~;~~ \mathbf{P}^2_{\vk,s} = \mathbf{P}_{\vk,s}\,. \label{projop}\ee Again using the anticommutation relations we find
\be \gamma_{\vk,s}\,\mathbf{P}_{\vk,s} = \mathbf{P}_{\vk,s}\,\gamma_{\vk,s} = \gamma_{\vk,s}\,, \label{itergam}\ee iterating yields
\be \gamma^3_{\vk,s} = - \gamma_{\vk,s}~~;~~ \gamma^4_{\vk,s} = \mathbf{P}_{\vk,s} ~~;~~ \gamma^5_{\vk,s} = \gamma_{\vk,s}\,\mathbf{P}_{\vk,s} =  \gamma_{\vk,s} \cdots \label{series}\ee Combining these results we finally find
\be S[\theta_k] = 1- \mathbf{P}_{\vk,s} + \mathbf{P}_{\vk,s}\,\cos(\theta_k) - \gamma_{\vk,s}\,\sin(\theta_k)\,. \label{Sffin}\ee Since the operators $\gamma_{\vk}$ commute  for different values of $\vk$ it follows that the full unitary transformation is
\be S[\theta] = \Pi_{\vk} S[\theta_k]\,.  \label{fullSf}\ee

Let us define as $\ket{0}$ as the state annihilated by  $ {b}_{\vec{k},s},{d}_{\vec{k},s}$ and as  $\ket{\widetilde{0}}$ as the state annihilated by $\widetilde{b}_{\vec{k},s},\widetilde{d}_{-\vec{k},s}$ for all $\vk,s$, namely

 \bea   {b}_{\vec{k},s}\,\ket{ {0}} =0  & ~~;~~ &  {d}_{-\vec{k},s}\,\ket{ {0}} =0  \,,\label{invac} \\  \widetilde{b}_{\vec{k},s}\,\ket{\widetilde{0}} =0 & ~~;~~ & \widetilde{d}_{-\vec{k},s}\,\ket{\widetilde{0}} =0\,.  \label{outvac}\eea Pre-multiplying $ \widetilde{b}_{\vec{k},s}; \widetilde{d}_{-\vec{k},s}$  by $S[\theta]$ and inserting $S^{-1}[\theta]\,S[\theta]=1$ into equations (\ref{outvac})  yields
 \be \underbrace{\Big( S[\theta]\,\widetilde{b}_{\vec{k},s}\,S^{-1}[\theta]\Big)}_{b_{\vk,s}}\,\underbrace{\Big(S[\theta]\,\ket{\widetilde{0}}\Big)}_{\ket{0}} =0 ~~;~~\underbrace{\Big( S[\theta]\,\widetilde{d}_{-\vec{k},s}\,S^{-1}[\theta]\Big)}_{d^\dagger_{-\vk}}\,\underbrace{\Big(S[\theta]\,\ket{\widetilde{0}}\Big)}_{\ket{0}} =0\,.   \label{Toutvacfer}\ee

Therefore, the vacua $\ket{0},\ket{\widetilde{0}}$ are related by
\be \ket{0}= S[\theta]\ket{\widetilde{0}} = \Pi_{\vk,s}\Big[\cos(\theta_k) - e^{2i\,\varphi_-}\,\sin(\theta_k)\,\widetilde{b}^\dagger_{\vec{k},s}\,\widetilde{d}^\dagger_{-\vec{k},s}\Big]\ket{\widetilde{0}}\,, \label{Sonvac}\ee where we restored the phases as per equation (\ref{bdphases}). It proves convenient to write this result as
\be \ket{0} = \Pi_{\vk,s}\Bigg\{\Big[\cos(\theta_k)\Big]~ \,\sum_{\widetilde{n}_{\vk,s}=0}^{1} \Bigg(-e^{2i\varphi_-(k)}\,\tan(\theta_k) \Bigg)^{\widetilde{n}_{\vk,s}} \ket{n_{\vk,s};\overline{n}_{-\vk,s}}\Bigg\} \,, \label{vacin}\ee where the fermionic   particle and antiparticle states are given by
\be  \ket{n_{\vk,s};\overline{n}_{-\vk,s}}  = \frac{\Big(\widetilde{b}^\dagger_{\vec{k},s}\Big)^{\widetilde{n}_{\vk,s}}}{\sqrt{\widetilde{n}_{\vk,s} !}}~
\frac{\Big(\widetilde{d}^\dagger_{-\vec{k},s}\Big)^{\widetilde{n}_{\vk,s}}}{\sqrt{\widetilde{n}_{\vk,s} !}}\, \ket{\widetilde{0}} ~~;~~ \widetilde{n}_{\vk,s} = 0, 1\,. \label{noutsapp}\ee Unitarity of the transformation is confirmed by obtaining
\be \langle 0|0\rangle= \Pi_{\vk,s}\Bigg\{\cos^2(\theta_k)\Big[1+ \tan^2(\theta_k) \Big]  \Bigg\} =1 \,. \label{normaI}\ee Furthermore, we find
\be  \langle 0|\widetilde{b}^\dagger_{\vec{k},s}\,\widetilde{b}_{\vec{k},s} |0\rangle = \langle 0|\widetilde{d}^\dagger_{\vec{k},s}\,\widetilde{d}_{\vec{k},s} |0\rangle = \sin^2(\theta_k) = |B_{k,s}|^2   \,.\label{exvalN}\ee



\end{document}